\documentclass[aps,pof,reprint,twocolumn,longbibliography]{revtex4-2}

\usepackage[utf8]{inputenc}
\usepackage[T1]{fontenc}
\usepackage{mathptmx}
\usepackage{courier}
\usepackage{etoolbox}

\usepackage{graphicx}
\usepackage{orcidlink}

\usepackage[letterpaper,
            top=2.1cm,
            bottom=2.0cm,
            left=1.8cm,
            right=1.8cm,
            columnsep=0.6cm
            ]{geometry}

\usepackage[font=small,justification=justified]{caption}

\usepackage{lmodern} %
\rmfamily %
\DeclareFontShape{T1}{lmr}{b}{sc}{<->ssub*cmr/bx/sc}{}
\DeclareFontShape{T1}{lmr}{bx}{sc}{<->ssub*cmr/bx/sc}{}

\usepackage{amsmath,amssymb}
\usepackage{array}
\usepackage{tabularray}
\usepackage{units}
\usepackage{subfig}

\usepackage{mathtools,nccmath}
\usepackage{bm}
\usepackage{setspace}

\usepackage{xargs}
\usepackage[usestackEOL]{stackengine}

\usepackage{xstring}
\usepackage{tikz}
\usepackage{scalerel}
\usepackage{wasysym}

\usepackage{pgffor}
\usepackage{cancel}
\usepackage{hyperref}
\definecolor{myblue}{rgb}{0.0, 0.0, 0.85}
\hypersetup{
    colorlinks=true,
    linkcolor=myblue,
    filecolor=magenta,
    urlcolor=myblue,
    citecolor=myblue
}

\usepackage[capitalise,nameinlink]{cleveref}
\crefname{section}{Sec.}{Secs.}
\Crefname{section}{Sec.}{Secs.}

\usepackage{xfrac}

\usepackage{upgreek}

\usepackage{microtype}

\newcommand{\pp}{\partial}

\newcommand{\dd}{\mathrm{d}}

\newcommand*{\IsInteger}[3]{%
    \IfStrEq{#1}{ }{%
        #3%
    }{%
        \IfInteger{#1}{#2}{#3}%
    }%
}%

    \newcommandx{\indexSequence}[4][1=1,2=4,3={},4=\alpha]{
        {
        \ifnum\the\numexpr#2-#1\relax>-1
            \ifnum #2=#1
                #3{#4_{#1}}
            \else
                \ifnum\the\numexpr#2-#1\relax<3
                    \foreach \i in {#1,...,#2}
                    {
                    #3{#4_{\i}}
                    }
                \else
                    {
                    #3{#4_{#1}}..#3{#4_{#2}}
                    }
                \fi
            \fi
        \else
            #3{#4_{\text{ !!error (#2<#1)!! }}}
        \fi
        }
    }

    \newcommandx{\Tensor}[6][1=T,2=1,3=1,4={},5=\alpha,6={}]{
        \def\literal{#1_{#6#4{#5_{#2}}..#4{#5_{#3}}}}
        \IsInteger{#2}{
            \IsInteger{#3}{
                #1_{#6\indexSequence[#2][#3][#4][#5]}
                }{
                \literal
            }
        }{
            \literal
        }
    }
    \newcommandx{\ifstrnotempty}[3]{\ifstrempty{#1}{#3}{#2}}
    \newcommandx{\sIndex}[2][2=\alpha]{\IsInteger{#1}{#2_{#1}}{#1}}
    \newcommandx{\MTensor}[6][1=T,2=1,3={},4={},5={},6={}]{
        \ifstrnotempty{#6}{
            #1_{{\sIndex{#2}}{\sIndex{#3}}{\sIndex{#4}}{\sIndex{#5}}{\sIndex{#6}}}
        }
        {
            \ifstrnotempty{#5}{
                #1_{{\sIndex{#2}}{\sIndex{#3}}{\sIndex{#4}}{\sIndex{#5}}}
            }
            {
                \ifstrnotempty{#4}{
                    #1_{{\sIndex{#2}}{\sIndex{#3}}{\sIndex{#4}}}
                }
                {
                    \ifstrnotempty{#3}{
                        #1_{{\sIndex{#2}}{\sIndex{#3}}}
                    }
                    {
                        #1_{{\sIndex{#2}}}
                    }
                }
            }
        }
    }

    \let\T\Tensor
    \let\MT\MTensor

\newcommand\Wtilde[1]{\ThisStyle{%
  \setbox0=\hbox{$\SavedStyle#1$}%
  \stackengine{-.1\LMpt}{$\SavedStyle#1$}{%
    \stretchto{\scaleto{\SavedStyle\mkern.2mu\AC}{.5150\wd0}}{.3\ht0}%
  }{O}{c}{F}{T}{S}%
}}

\newcommand{\zero}{{(0)}}
\newcommand{\one}{{(1)}}
\newcommand{\two}{{(2)}}

\newcommand{\Neq}{{\rm neq}}
\newcommand{\cNeq}{{\rm cneq}}
\newcommand{\Eq}{{\rm eq}}

\newcommand{\sgs}{{\rm sgs}}

\renewcommand\Re{\mathrm{Re}}
\newcommand\Ma{\mathrm{Ma}}

\newcommand\Kn{\mathrm{Kn}}

\newcommand\mcL{\mathcal{L}}
\newcommand\mcT{\mathcal{T}}
\newcommand\mcU{\mathcal{U}}
\newcommand\Ccc{\mathcal{C}_{\rm cc}}
\newcommand\CRey{\mathcal{C}_{\rm Rey}}
\newcommand\CLeo{\mathcal{C}_{\rm Leo}}
\newcommand\CCross{\mathcal{C}_{\rm Cross}}
\newcommand\Cchaos{\mathcal{C}_{\rm chaos}}
\newcommand\Ecc{\mathcal{E}_{\rm cc}}

\newcommand\vartau{\uptau}
\newcommand\varnu{\upnu}

\newcommand\labelAndRemember[2]
{\expandafter\gdef\csname labeled:#1\endcsname{#2}%
\label{#1}#2}

\newcommand\recallEqAndTag[1]
{\csname labeled:#1\endcsname\tag{\ref{#1}}}

\newenvironment{alignedEq}[1][\unskip]{\equation #1 \aligned}{\endaligned\endequation}

\newenvironment{alignatEq}[2][\unskip]{\equation #1 \alignedat{ #2 }}{\endalignedat\endequation}

\makeatletter
\def\@email#1#2{%
 \endgroup
 \patchcmd{\titleblock@produce}
  {\frontmatter@RRAPformat}
  {\frontmatter@RRAPformat{\produce@RRAP{*#1\href{mailto:#2}{#2}}}\frontmatter@RRAPformat}
  {}{}
}%

\def\underbracex#1#2{\mathop{\vtop{\m@th\ialign{##\crcr
   $\hfil\displaystyle{#2}\hfil$\crcr
   \noalign{\kern3\p@\nointerlineskip}%
   #1\crcr\noalign{\kern3\p@}}}}\limits}

\def\underbracea{\underbracex\upbracefilla}

\def\upbracefilla{$\m@th \setbox\z@\hbox{$\braceld$}%
  \bracelu\leaders\vrule \@height\ht\z@ \@depth\z@\hfill
\kern\p@\vrule \@width\p@\kern\p@\vrule \@width\p@\kern\p@\vrule \@width\p@
$}

\def\underbraceb{\underbracex\upbracefillb}

\def\upbracefillb{$\m@th \setbox\z@\hbox{$\braceld$}%
\vrule \@width\p@\kern\p@\vrule \@width\p@\kern\p@\vrule \@width\p@\kern\p@
 \leaders\vrule \@height\ht\z@ \@depth\z@\hfill\bracerd
  \braceld\leaders\vrule \@height\ht\z@ \@depth\z@\hfill
\kern\p@\vrule \@width\p@\kern\p@\vrule \@width\p@\kern\p@\vrule \@width\p@
$}

\def\underbraced{\underbracex\upbracefilld}
\def\upbracefilld{$\m@th \setbox\z@\hbox{$\braceld$}%
\vrule \@width\p@\kern\p@\vrule \@width\p@\kern\p@\vrule \@width\p@\kern\p@
 \leaders\vrule \@height\ht\z@ \@depth\z@\hfill\braceru$}

\def\underbracebd{\underbracex\upbracefillbd}
\def\upbracefillbd{$\m@th \setbox\z@\hbox{$\braceld$}%
\vrule \@width\p@\kern\p@\vrule \@width\p@\kern\p@\vrule \@width\p@\kern\p@
\bracerd\braceld
 \leaders\vrule \@height\ht\z@ \@depth\z@\hfill\braceru$}
\newcommand{\pushright}[1]{\ifmeasuring@#1\else\omit\hfill$\displaystyle{#1}$\fi\ignorespaces}

\makeatother

\newcounter{comparisoncell}
\makeatletter
\newcommand{\comparisoncelllabel}[3]{%
  \refstepcounter{comparisoncell}%
  \begingroup
    \def\@currentlabel{\ensuremath{#1_{\mathrm{#2}}}}%
    \label{#3}%
  \endgroup
}
\makeatother
\newcommand{\comparisoncelltag}[2]{\textup{(\ensuremath{#1_{\mathrm{#2}}})}\enspace}
\newcommand{\compref}[1]{\textup{(\ref{#1})}}

\crefformat{paragraph}{section~#2#1#3}
\crefname{paragraph}{section}{sections}
\Crefname{paragraph}{Section}{Sections}

\begin{document}

\title{Kinetic closure of turbulence: collision-side modeling beyond the filtered Boltzmann equation}
\author{Francesco Marson$^{1,2,*}$\,\orcidlink{0000-0002-0900-4193} and Orestis Malaspinas$^{2}$\,\orcidlink{0000-0001-9427-6849}}
\email{francesco.marson@proton.me}
\affiliation{$^{1}$Department of Computer Science, University of Geneva}%
\affiliation{%
  $^{2}$HEPIA, University of Applied Sciences and Arts of Western Switzerland%
}

\date{September 23, 2026}

\begin{abstract}
This article extends a recently introduced kinetic closure of turbulence by developing its theoretical framework, operational realizations, and validation. In contrast with filtered Navier--Stokes formulations, filtering the Boltzmann equation retains subgrid transport under the linear streaming operator, so that unresolved physics is concentrated on the collision side. We show that in dilute-gas large-eddy simulation, the main limitation of Bhatnagar--Gross--Krook (BGK)-type collision models is not the breakdown of molecular chaos, but the retention of a Markovian collision at a scale where filtering can induce finite temporal correlations in the collision product. In a BGK-type framework, the closure problem is dual: one must infer the filtered fine-grained equilibrium, not computable from filtered moments alone, and model the non-Markovian collision dynamics generated by the collision-product covariance. The present framework makes this dual structure explicit and represents the resulting collision-covariance source term through a BGK-like closure built from the equilibrium commutation residual, with the turbulent relaxation frequency given by a first phenomenological realization. The framework relies on a Chapman--Enskog analysis organized by the reference timescale ratio emerging from the nondimensionalization of the kinetic equation, performed in the classical sense, thereby avoiding artificial turbulent scale separations. We show that the Chapman--Enskog structure is not a pure one-parameter Knudsen scaling: the primary ordering is set by the kinetic-to-macroscopic timescale ratio, while higher moments retain an additional Mach dependence through the mixed scaling of particle velocity. The resulting kinetic closures are validated through lattice Boltzmann simulations and compared with the Smagorinsky model and regularization-based collision models.
\end{abstract}

\maketitle

\newcommand{\cpartial}{\pp^{\flat}}
\section{Introduction}\label{sec:intro}
Turbulent flows involve a broad range of dynamically active scales. Resolving all of them in direct numerical simulations (DNS) is generally infeasible because the number of degrees of freedom grows steeply with the Reynolds number. This computational constraint motivates coarse-grained descriptions such as Reynolds-averaged Navier--Stokes (RANS) and large-eddy simulation (LES). In both cases, averaging or filtering the nonlinear advective term in the Navier--Stokes equations (NSE) introduces unclosed advective correlations, viz. Reynolds stresses in RANS and subgrid-scale (SGS) stresses in LES, which must be modeled to represent the influence of unresolved scales of motion on the resolved (mean) flow \cite{pope_turbulent_2000}.

The NSE provides a macroscopic fluid description, whereas the Boltzmann equation (BE) describes transport and dissipation in phase space through particle distribution functions. Because the BE recovers the NSE in its hydrodynamic limit \cite{chapman_mathematical_1953}, several works applied filtering directly to the BE and imported NSE-based eddy-viscosity ideas into the kinetic setting (\cref{sec:review}). Yet this step changes the closure structure. In the filtered NSE, SGS velocity fluctuations are not transported explicitly, and the missing momentum flux must therefore be modeled. By contrast, in the filtered BE, advection is linear and commutes with filtering, so SGS transport remains embedded in the filtered distribution. As proposed in~\cite{marson_kinetic_2025} and developed here, the kinetic closure problem is therefore distinct from the macroscopic SGS-transport problem: the collision step must act on a distribution that still carries SGS momentum transport.

\begin{table*}[t]
\caption{Structural comparison between the macroscopic Filtered Navier--Stokes (FNSE) and the filtered Boltzmann equation (FBE), highlighting where SGS information is carried, how it is transported, and which contribution requires modeling. The detailed comparison in \cref{tab:deep_comparison} specializes this structure to the BGK closure using the formal notation developed in \cref{CE}.}
\label{tab:comparison}
\comparisoncelllabel{a}{NSE}{tab:comparison:a-nse}
\comparisoncelllabel{a}{BE}{tab:comparison:a-be}
\comparisoncelllabel{b}{NSE}{tab:comparison:b-nse}
\comparisoncelllabel{b}{BE}{tab:comparison:b-be}
\comparisoncelllabel{c}{NSE}{tab:comparison:c-nse}
\comparisoncelllabel{c}{BE}{tab:comparison:c-be}
\comparisoncelllabel{d}{NSE}{tab:comparison:d-nse}
\comparisoncelllabel{d^1}{BE}{tab:comparison:d-be1}
\comparisoncelllabel{d^2}{BE}{tab:comparison:d-be2}
\comparisoncelllabel{d}{BE}{tab:comparison:d-be}
\begin{tblr}{
  width=\linewidth,
  colspec={lXX},
  hline{1,2,6}={solid, 0.08em},
}
Feature & Filtered Navier--Stokes (FNSE) & Filtered Boltzmann equation (FBE) \\
\hline
Dependent variable & \comparisoncelltag{a}{NSE}Filtered velocity and density & \comparisoncelltag{a}{BE}Filtered distribution function $\overline{F}$ \\
Advection mechanism & \comparisoncelltag{b}{NSE}Non-linear coupling (velocity--velocity) & \comparisoncelltag{b}{BE}Linear translation at microscopic velocity \\
SGS advection & \comparisoncelltag{c}{NSE}Unresolved and represented by the advective SGS stress tensor & \comparisoncelltag{c}{BE} Unknown but exactly transported by the linear streaming of $\overline{F}$, including its advective SGS stress moments \\
Required modeled contribution
& \comparisoncelltag{d}{NSE}Modeled advective SGS stress tensor, often through an eddy-viscosity/Boussinesq closure
 & \comparisoncelltag{d^1}{BE}SGS equilibrium commutation residual and \comparisoncelltag{d^2}{BE} collision-covariance source term for collision-side SGS dissipation
 \\
\end{tblr}
\end{table*}

Table~\ref{tab:comparison} summarizes where SGS information is stored, how it is transported, and which contribution remains to be modeled in each framework.
The difference appears directly in the advection-mechanism entries \compref{tab:comparison:b-nse} and \compref{tab:comparison:b-be}, and in the modeled-contribution entries \compref{tab:comparison:d-nse} and \compref{tab:comparison:d-be}. While the filtered NSE (FNSE) closure targets an unresolved advective flux, the filtered BE (FBE) transports SGS information through the linear streaming of the filtered distribution function $\overline{F}$. Therefore, the kinetic closure addresses the collision side~\cite{marson_kinetic_2025}.

A recurrent strand of kinetic-turbulence theory approaches closure by going beyond Boltzmann's molecular-chaos factorization through bilocal or higher-order descriptions~\cite{krieger_molecular_1961,zhigulev_equations_1965,tsuge_approach_1974,chliamovitch_kinetic_2017,chliamovitch_turbulence_2018}. That line of work can suggest that molecular chaos is the main source of the closure problem (\cref{sec:review}).
The present framework, however, does not tie the filtered kinetic closure problem to the breakdown of molecular chaos, but to temporal correlations generated in the collision product by finite-time averaging over the coarse-graining observation window. Through free-streaming characteristics, this time filter also samples a velocity-dependent spatial extent (\cref{sec:filtered_recorrelated_BGK_Boltzmann_equation}). Thus, the LES-type spatial-filter width should not be smaller than the induced kinetic spatial extent and may be chosen of the same order to define a consistent space--time coarse-graining scale.
The associated collision-product covariance defines, through the collision integral, the \emph{collision-covariance source term}. This covariance source term makes the one-particle collision process non-Markovian. While the standard derivation of the BE neglects this non-Markovianity over the kinetic coarse-graining interval, in the filtered hydrodynamic setting considered here, this kinetic hypothesis must be reconsidered, because subfilter \emph{advective} variance \compref{tab:comparison:c-be}, i.e. the SGS stress tensor precursor, can induce temporal collision correlations at the filter scale.

For an operative collision-model reduction of the BE, such as the BGK--BE \cite{bhatnagar_model_1954}, the collision-side closure problem \compref{tab:comparison:d-be} is therefore dual. First, one must distinguish the filtered fine-grained equilibrium from the equilibrium built from filtered moments alone \compref{tab:comparison:d-be1}. Second, one must model the non-Markovian relaxation process generated at the filter scale. We refer to the BGK equation with modeled non-Markovianity as the recorrelated BGK--BE (\cref{sec:filtered_recorrelated_BGK_Boltzmann_equation}).

To formalize these points, following~\cite{marson_kinetic_2025}, we develop a Chapman--Enskog (CE) analysis that uses the ratio of the kinetic mean-free-time to the reference macroscopic transport time as the expansion parameter (\cref{CE}). Thus, we do not use the Knudsen number in the CE as an isolated length-scale ratio. This time-scale small parameter follows directly from the nondimensionalization of the BE and remains valid in the filtered setting considered here. This choice, however, does not make the asymptotic expansion one-parameter: after moment projection, the mixed convective-thermal scaling of the particle velocity leaves an independent Mach-number dependence in the higher order moments of the one-particle distribution function. Therefore, contrary to some previous studies, in the turbulent regime considered here we perform the CE expansion in the classical sense: introducing \textit{ad-hoc} smallness parameters based on macroscopic turbulence scales would impose an artificial and unphysical scale separation.

The CE analysis gives the hydrodynamic limit of the recorrelated BGK--BE, namely the NSE-equivalent momentum equation implied by this kinetic model (\cref{CE}). From this limit, we obtain the conditions for convergence to the FNSE and the asymptotic order of the non-Markovian term. Finally, the same calculation identifies the advective precursors used in the kinetic closure proposed in \cref{sec:modeling_Ecc}.

In summary, the remainder of this paper is structured as follows. \Cref{sec:review} reviews kinetic approaches to turbulence. \Cref{sec:filtered_recorrelated_BGK_Boltzmann_equation} formalizes the filtered recorrelated BGK--BE, whose hydrodynamic limit is derived in \cref{CE}.
The physical modeling of the collision-covariance source term is discussed in \cref{sec:modeling_Ecc}, while specific operational kinetic closures are formulated in \cref{sec:kinetic_closures}.
The numerical validation via three-dimensional test cases is presented in \cref{sec:test_cases}.
Finally, the Appendices follow the order of their technical dependencies, covering collision-product covariance analysis (Appendix~\ref{app:klimontovich_covariance}), collision-side-only scaling assumptions in filter-based kinetic closures (Appendix~\ref{app:ansumali_scaling}), the relative-frame formulation (Appendix~\ref{app:girimaji}), the comparison between the present kinetic closure and the Klimontovich formulation of Chen et al. (Appendix~\ref{app:chen_comparison}), Hermite representations (Appendix~\ref{app:hermite_representation}), and lattice--Boltzmann discretization with operational implementation details and the trapezoidal reconstruction of the two-rate collision (Appendix~\ref{app:lbm_discretization} and Appendices~\ref{app:omega_t_operational}--\ref{app:trapezoidal_reconstruction}).

\section{Review of kinetic approaches to turbulence}\label{sec:review}
Over the past six decades, a broad literature has developed kinetic approaches to turbulence. The review below follows six lines of work: foundational correlation theories, renormalization-inspired approaches, filter-based approaches, macroscopic-consistent engineering models, advanced collision models and implicit filtering, and turbulence quasi-particle approaches.

\paragraph{Foundational kinetic theories of correlations}
One route to kinetic turbulence starts from a limitation already present in the BE: the molecular-chaos factorization removes explicit two-particle correlations. Early theoretical efforts therefore relaxed the strict molecular chaos assumption (\emph{Stosszahlansatz}) to account for unresolved spatial and interparticle correlations. Foundational bilocal formulations by Krieger, Zhigulev, and Tsug\'e \cite{krieger_molecular_1961,zhigulev_equations_1965,tsuge_breakdown_1971, tsuge_approach_1974, tsuge_new_1975} retain the joint two-particle distribution $\mathcal F_2$, thereby avoiding the factorization $\mathcal F_2 \approx \mathcal F \mathcal F$ ($\mathcal F$ being the one-particle distribution). Contemporary efforts \cite{chliamovitch_truncation_2015,chliamovitch_kinetic_2017,chliamovitch_turbulence_2018} continue this philosophy by attempting to truncate the exact Bogoliubov--Born--Green--Kirkwood--Yvon (BBGKY) hierarchy~\cite{bogoliubov1946problems,bogoliubov1962problems,born1946general,kirkwood1946statistical,yvon1935theorie} at the two-particle level via Maximum Entropy principles.

However, evolving exact multi-point distributions is practically impossible due to the overwhelming dimensionality of the governing equations. These theories correctly identify two-particle correlations omitted by the Boltzmann closure, but they address a different approximation from the one treated here. As detailed in Section \ref{sec:filtered_recorrelated_BGK_Boltzmann_equation} and Appendix \ref{app:klimontovich_covariance}, the SGS closure term modeled here does not stem primarily from molecular-chaos failure ($\mathcal F_2 \neq \mathcal F \mathcal F$), but from the temporal covariance of the one-particle collision product under finite-time filtering ($\langle \mathcal F \mathcal F \rangle_{\Delta_t} \neq \langle \mathcal F \rangle_{\Delta_t} \langle \mathcal F \rangle_{\Delta_t}$, with $\langle\cdot\rangle_{\Delta_t}$ denoting the kinetic time filter). Two-particle correlations omitted by molecular chaos and the time-filtered collision-product covariance are therefore distinct objects. Only the latter is modeled in the present framework.

\paragraph{Renormalization-inspired kinetic approaches}
Following these fundamental derivations, practical modeling efforts sought to represent turbulence within the tractability of the single-particle Boltzmann framework. Inspired by the renormalization-group (RNG) analysis~\cite{yakhot_renormalization_1986,yakhot_renormalization-group_1986}, these models incorporate turbulent fluctuations by introducing an effective, scale-dependent relaxation rate into the collision operator~\cite{chen_analysis_1999,chen_extended_2003,succi_towards_2002}. This ``renormalized'' collision rate is not a simple macroscopic dissipative timescale. It represents the scale-coupling dynamics of the SGS fluctuations. In this framework, the primary modeling effort shifts to determining this renormalized relaxation time---often loosely equated to an eddy-viscosity---using inertial-range scaling arguments~\cite{chen_expanded_2004}.

Because these formulations can recover Smagorinsky-type scalings when projected onto the hydrodynamic limit, an implicit interpretive bias emerged in the literature: the assumption that the filtered BGK--BE and the FNSE require the modeling of the same unresolved physics. As discussed in~\cite{marson_kinetic_2025}, this conflates the macroscopic transport closure \compref{tab:comparison:c-nse} and \compref{tab:comparison:d-nse} with the collision-side SGS dissipation \compref{tab:comparison:d-be} in \cref{tab:comparison}. By equating the kinetic renormalized relaxation rate with a macroscopic eddy-viscosity, one implicitly frames the kinetic closure problem as a model for unclosed macroscopic \emph{transport} \compref{tab:comparison:d-nse}, rather than recognizing it as a model for collision-side SGS \emph{dissipation} associated with the SGS equilibrium commutation residual and the collision-covariance source term \compref{tab:comparison:d-be}~\cite{marson_kinetic_2025}. The constitutive difference between the effective relaxation rate in the filtered BGK--BE and the standard NSE turbulent viscosity was analyzed in~\cite{malaspinas_consistent_2012}.

Compounding this interpretive difficulty is the use of scale-separation arguments to justify or study the resulting macroscopic closures. Some formulations perform a CE or CE-like expansion using either the kinetic Knudsen number with filter-width scaling or an effective turbulent Knudsen number~\cite{ansumali_kinetic_2004,chen_expanded_2004}. This parallels earlier analytical strategies, such as Yen's use in~\cite{yen_kinetic_1972} of a ``turbulent energy number'' as an expansion parameter to derive a closed set of RANS equations.
However, for a multiscale expansion like the CE, it is not sufficient that the small parameter of the expansion remain below unity. It must define a genuine asymptotic separation between fast and slow dynamics, and for that the small parameter should be asymptotically small. This is not the case for these turbulence parameters, because, physically, the unresolved fluctuations remain dynamically coupled to the resolved flow rather than forming an independent dynamic scale.
Given that the turbulent fluctuations neither decouple from the resolved scale nor satisfy an independent kinetic evolution equation, these macroscopic flow parameters add a modeling assumption rather than defining an asymptotic limit.

\paragraph{Filter-based kinetic approaches} \label{filter_kinetic_models}
In a distinct category of ``pure'' kinetic models, the SGS closure is not imported from macroscopic mechanics but is derived from the kinetic theory itself or the algebraic properties of the filter kernel.

Early attempts~\cite{ansumali_kinetic_2004}, as well as inverse filtering techniques such as the Approximate Deconvolution Method (ADM)~\cite{sagaut_toward_2010, malaspinas_advanced_2011}, derived LES--BE closures from filter noncommuting collision-side residuals \compref{tab:comparison:d-be1} rather than from a modeled advective commutation term \compref{tab:comparison:d-nse} as in LES--NSE.
Their analysis correctly places the SGS contribution in the collision residual, but neglects the emergence of the advective SGS contribution when taking the moments of the FBE (please see Appendix~\ref{app:ansumali_scaling} for the details). At the macroscopic moment level this choice keeps the resulting Euler-level equations in resolved laminar form, whereas the well-known filtered Euler equations contain the $O(1)$ SGS stress in the advective flux \compref{tab:comparison:c-nse}.
These models can still work in practice in a lattice Boltzmann method (LBM) framework because the exact streaming step transports the full filtered populations, including SGS advection not represented in their hydrodynamic limit. However, even if the SGS advection \compref{tab:comparison:c-be} is still present in these models, the closure is not complete. The collision model still acts mainly on the filtered fine-grained equilibrium and leaves the non-Markovian collision relaxation \compref{tab:comparison:d-be2} unresolved.
A detailed mathematical analysis of these collision-side-only scaling assumptions and their consequences on the recovered macroscopic equations is provided in Appendix~\ref{app:ansumali_scaling}.

A later work treated the lack of explicit advective SGS terms in the filtered Boltzmann equation as a defect~\cite{girimaji_boltzmann_2007}. An earlier RANS-oriented formulation by Yen~\cite{yen_kinetic_1972} already used the resolved BGK collision model~\cite{bhatnagar_model_1954} by expanding the distribution function around a Maxwellian moving at the time-averaged macroscopic velocity. This construction leads to a Reynolds stress tensor at the moment level. In contrast, the relative-frame formulation for RANS and LES~\cite{girimaji_boltzmann_2007} shifts the velocity coordinate to a non-inertial reference frame tied to unresolved fluctuations. Unresolved transport then appears as transformed transport and fictitious acceleration terms. As detailed in Appendix~\ref{app:girimaji}, this shift constitutes a representation change rather than a closure. The inertial filtered equation already carries these velocity correlations through the filtered distribution, while the transformation neither closes the collision-covariance source term nor derives a Navier--Stokes-order constitutive law via the Chapman--Enskog expansion.

\paragraph{Macroscopic-consistent engineering models}
A separate engineering line uses the kinetic solver as a host for macroscopic turbulence models. The works of Premnath et al. are an example. In~\cite{premnath_generalized_2009}, a multiple-relaxation-time formulation improves stability and enables local strain-rate evaluation for Smagorinsky-type LES.
This architectural connection links these engineering models to implicit LES (ILES) strategies discussed below, where advanced collision methods, such as the entropic model~\cite{ansumali_stabilization_2000}, essentially act as implicit SGS closures.
In~\cite{premnath_three_2011}, the central-moment formulation addresses Galilean-invariance errors associated with raw lattice-frame moments.
In~\cite{premnath_inertial_2012}, Reynolds or SGS stresses are embedded through the relative-frame forcing representation of~\cite{girimaji_boltzmann_2007}. This is precisely the ambiguity highlighted here: the transformation externalizes an advective SGS contribution already contained in the filtered distribution and then turns it into a separate closure target.

Similarly, Righi's gas-kinetic scheme~\cite{righi_gas-kinetic_2016} couples a BGK-based finite-volume flux solver to a standard $k$--$\omega$ RANS model, while Asinari et al.~\cite{asinari_kinetic_2016} recast the $k$--$\varepsilon$ RANS model as coupled BGK-like equations with forcing.

\paragraph{Advanced collision models and implicit filtering} Independently of these developments, the LBM community devoted significant effort to refining kinetic collision operators for improved stability and accuracy.
Representative examples include multiple-relaxation-time (MRT) models \cite{dhumieres_multiplerelaxationtime_2002}, projective (PR) \cite{latt_lattice_2006}, recursive (RR) \cite{malaspinas_increasing_2015} and hybrid recursive regularized (HRR) \cite{jacob_new_2018} collisions, entropic formulations \cite{ansumali_stabilization_2000}, and cumulant-based schemes \cite{geier_cumulant_2015}, which explicitly distinguish between slow hydrodynamic moments and rapidly relaxing non-hydrodynamic modes.
By assigning different relaxation rates to different moments, these models enhance numerical robustness and extend the range of accessible flow regimes. Within these methods, fast non-hydrodynamic modes are primarily manipulated as auxiliary degrees of freedom whose rapid relaxation suppresses lattice artifacts and stabilizes the computation, without being associated with turbulent fluctuations or scale-dependent energy transfer.
In under-resolved turbulent simulations, these advanced collision models can act as ILES mechanisms: they control higher-order kinetic content and add implicit dissipation without prescribing a separate SGS stress model.
In their standard forms, these models are stabilizations of the unfiltered kinetic equation. They tune, relax, or reconstruct higher-order kinetic modes without an explicit SGS carrier. The recursive regularization reconstruction~\cite{malaspinas_increasing_2015} gives a clear example. It is derived for an unfiltered, fully resolved Chapman--Enskog description. In a fully resolved laminar regime, the distribution remains close to the corresponding manifold, so the components replaced by RR are predominantly ghost modes and higher-order discretization errors. However, applying the same reconstruction to a turbulent filtered distribution replaces higher-order nonequilibrium departures that also carry physical SGS content. RR then removes the SGS content carried by those departures together with numerical artifacts. For comparison, Smag.\ Naive follows a different route. It retains raw population content and damps it through eddy-viscosity collision relaxation. Neither operation, however, separates numerical artifacts from physical SGS fluctuations.

\paragraph{Macroscopic PDF and fluid-element kinetic formulations}
A common RANS-oriented top-down family represents macroscopic turbulence through one-point probability distributions of velocity and, in some formulations, scalar variables. Lundgren introduced a hierarchy of turbulent velocity-distribution equations and later proposed a model equation in which the unclosed pressure-fluctuation term is represented by a phenomenological relaxation ansatz~\cite{lundgren_distribution_1967,lundgren_model_1969}. This relaxation is a closure at the macroscopic PDF level rather than a molecular collision operator. Pope subsequently derived a joint velocity--scalar PDF transport equation from the macroscopic Navier--Stokes and scalar equations~\cite{pope_transport_1981,pope_pdf_1985}. Its second moment gives the Reynolds stresses, while the pressure--strain closure is constructed consistently with Reynolds-stress models and includes a linear return-to-isotropy contribution of Rotta type.

Degond and Lemou introduced a more explicit kinetic representation of statistical fluid structures, with an isotropization operator that leads to $k$--$\varepsilon$-type macroscopic models~\cite{degond_turbulence_2002}. This formulation retains the macroscopic PDF interpretation but introduces explicit kinetic transport and isotropization.

More recently, Chen et al. revived this route with a Klimontovich-type kinetic equation for fluid elements, later extended by Xin et al.; related kinetic variants include Saveliev's Boltzmann-based turbulent-fluid-quasiparticle model and the Fokker--Planck construction of Luan et al. \cite{chen_average_2023,chen_average_2024,saveliev_kinetic_2024,xin_model_2026,luan_constructing_2025}. In Chen's formulation the Klimontovich equation is used as a top-down kinetic lifting of the macroscopic Navier--Stokes dynamics, rather than as a bottom-up molecular kinetic derivation. This line should also be distinguished from earlier kinetic-analogy work \cite{chen_expanded_2004}, which already introduced a BGK-like turbulent Gaussian and a Chapman--Enskog-like expansion for Reynolds-stress modeling.
The formulation introduced by Chen et al. in~\cite{chen_average_2023} obtains a collision-like covariance term in velocity space. Its advective moments recover the Reynolds stress through the second order velocity moment of the averaged fluid-element distribution. The main difference with the present framework lies in the closure step. Chen's formulation closes this covariance through a BGK relaxation toward a turbulent Gaussian and an additional dissipation-rate model. In the present work, the corresponding unresolved term is identified from the time-filtered collision dynamics and closed through the SGS carrier and a constitutive relaxation rate. Appendix~\ref{app:chen_comparison} provides the detailed mathematical comparison.

Subsequent extensions \cite{xin_model_2026,luan_constructing_2025} refine the relaxation-time modeling, develop near-wall treatments, and formulate related Fokker--Planck descriptions that lead to eddy-viscosity models through Chapman--Enskog expansion.

\section{The filtered recorrelated BGK--BE}\label{sec:filtered_recorrelated_BGK_Boltzmann_equation}
Fluid motion is described by the evolution of the one-representative-particle mass probability density function (PDF) $\mathcal{F}(\boldsymbol{x},\boldsymbol{\xi},t) \equiv m \, n(\boldsymbol{x},\boldsymbol{\xi},t)$, where $\boldsymbol{x}$ denotes position, $\boldsymbol{\xi}$ microscopic particle velocity, $t$ time, $m$ the particle mass, and $n$ is the one-particle PDF. Unless stated otherwise, $\alpha_r\in\{x,y,z\}$ denotes the $r$-th Cartesian tensor index, with $r$ a nonsummed tensor-slot label. Einstein summation applies only when the same Cartesian index occurs twice within one product; the unnumbered index $\alpha$ follows the same convention. The evolution of $\mathcal{F}$ is governed by the first equation of the BBGKY hierarchy~\cite{bogoliubov1946problems,bogoliubov1962problems,born1946general,kirkwood1946statistical,yvon1935theorie}. Neglecting external body forces, it reads:
\begin{equation}\label{BBGKY1}
\partial_t \mathcal{F} + \xi_\alpha\,\partial_\alpha \mathcal{F} = \mathcal{I}_{12}(\mathcal{F}_2) \equiv \frac{1}{m} \int \partial_{\alpha} \Phi(|\boldsymbol{x} - \boldsymbol{x}_{i\!i}|) \, \partial_{\xi_\alpha} \mathcal{F}_2 \, \mathrm{d}\boldsymbol{z}_{i\!i},
\end{equation}
where $\boldsymbol{z}_{i\!i} = (\boldsymbol{x}_{i\!i}, \boldsymbol{\xi}_{i\!i})$ denotes the phase coordinates of a second particle PDF and $\Phi$ is the intermolecular potential energy per unit mass. The interaction integral $\mathcal{I}_{12}$ depends functionally on the two-particle distribution $\mathcal{F}_2=m^2n_2$ (with $n_2$ the two-particle number density for distinct particles), highlighting the fundamental closure problem of the hierarchy \cite{yvon1935theorie,bogoliubov1946problems,bogoliubov1962problems}.

The kinetic equations used in practice, most commonly the BE or BGK--BE~\cite{cercignani_boltzmann_1988,kremer_introduction_2010,bhatnagar_model_1954}, result from successive reductions of the BBGKY hierarchy.
Since the present work revisits them in a filtered setting, the hypotheses behind each reduction must be made explicit.

The derivations in the herein article combine several filters, tensor conventions, and distinct CE and Hermite-rank labels.
Table~\ref{tab:notation} provides a compact reference for the notation used throughout the manuscript but does not replace the definitions given when each quantity is first introduced.

\begin{table*}[t]
\caption{Central notation and conventions used in the filtered kinetic framework. The table records notation used across sections or assigned a nonstandard role. Formal definitions are supplied at their first use.}
\label{tab:notation}
\centering
\small
\begin{tblr}{
  width=\linewidth,
  colspec={Q[l,wd=0.31\linewidth]X},
  hline{1,2}={solid, 0.08em},
}
Notation & Meaning or convention \\
\hline
\SetCell[c=2]{l}{\emph{Filtering and distribution levels}} \\
$\mathcal F \xrightarrow{\langle\cdot\rangle_{\Delta_t}} F \mapsto f \xrightarrow{\overline{(\cdot)}} \overline f$
  & Instantaneous mass PDF, kinetic-time-filtered dimensional distribution, its nondimensional form, and its homogeneous LES-filtered form. \\
$\langle\cdot\rangle_{\rm ens}$, $\langle\cdot\rangle_t$, $\overline{(\cdot)}$, $\tilde a$, $\langle\cdot\rangle$,
  & Ensemble average, time-filter, homogeneous LES space-filter, Favre filter, post-processing time average. These operations are distinct. \\
$\varrho$, $\rho=\varrho/\varrho_R$, $\bm u^\flat$, $\theta^\sharp$
  & Dimensional density, normalized density, convectively scaled velocity, and kinetically scaled temperature. \\
$f^\zero$, $\overline{f^\zero}$
  & Resolved equilibrium and its filtered fine-grained counterpart. In the CE hierarchy, $\overline{f^\zero}=f^\zero+f_{\sgs}$ is the leading filtered coefficient. \\
$f^\cNeq$, $\overline{f^\Neq}$, $f_{\sgs}$
  & Coarse-nonequilibrium departure, filtered fine-grained nonequilibrium, and SGS equilibrium commutation residual. \Cref{fig:entropic_descent} illustrates their geometric relation in a simplified Hilbert space. \\
$\Cchaos$, $\Ccc$, $\Ecc^\sharp$
  & Molecular-chaos defect, collision-product covariance, and its nondimensional collision-side source. The two covariances are distinct. \\
\hline
\SetCell[c=2]{l}{\emph{Tensor and discrete-index conventions}} \\
$\bm x$, $\bm\xi$; $\xi_{\alpha_r}$, $\xi_{i\alpha_r}$
  & Position, continuous particle velocity, its Cartesian component, and the corresponding component of a discrete velocity $\bm\xi_i$. \\
$\alpha_r\in\{x,y,z\}$, $r=1,\ldots,n$; $\delta_{\alpha_1\alpha_2}$
  & Cartesian tensor index, tensor-slot label, and Cartesian identity tensor. A repeated Cartesian index is summed only when it occurs twice within one product. \\
$i=0,\ldots,Q-1$
  & Discrete velocity-population index in the LBM sections. It is distinct from the $i\!i$ collision-partner label used in the continuous collision integral. \\
\hline
\SetCell[c=2]{l}{\emph{CE and Hermite-rank conventions}} \\
$g^{(k)}$, $O(\epsilon^k)$
  & CE coefficient multiplying $\epsilon^k$ in the expansion of $g$, and its formal CE order. \\
$H_{\alpha_1\dots\alpha_n}$, $m_{\alpha_1\dots\alpha_n}$, $a_{\alpha_1\dots\alpha_n}$
  & $H_{\alpha_1\dots\alpha_n}$ is a rank-$n$ tensor-valued Hermite polynomial of total degree $n$. $m_{\alpha_1\dots\alpha_n}$ and $a_{\alpha_1\dots\alpha_n}$ are the corresponding rank-$n$ raw moment and Hermite coefficient, respectively. Unless a $\sharp$ superscript is written explicitly, these continuum quantities use convective $\flat$ scaling; the $\flat$ superscript is omitted when that default is unambiguous. \\
$\overline{a^{\sgs(k)}_{\alpha_1\dots\alpha_n}}$
  & Composite reading rule: overbar for LES filtering, roman label for physical role, parenthesized number for CE order, and lower Cartesian indices for tensor rank. \\
\hline
\SetCell[c=2]{l}{\emph{Reference scales and dimensionless groups}} \\
$\flat$, $\sharp$
  & Convective scaling by $\{\mathcal L,\mathcal T,\mathcal U\}$ and kinetic/diffusive scaling by $\{\ell,\tau_{\rm mft},\sqrt{\theta_R}\}$. \\
$\epsilon$, $\Kn$, $\Ma$, $\Re_\ell$, $\Re$
  & CE timescale ratio, Knudsen number, Mach number, kinetic Reynolds-like number, and hydrodynamic Reynolds number. They satisfy $\epsilon=\Ma\Kn=\Ma^2/\Re_\ell$, $\Re_\ell=\Ma/\Kn$, and $\Re=\mathcal U\mathcal L/\nu=\Re_\ell/(\tau^\sharp\tilde\theta^\sharp)$. \\
\hline
\end{tblr}
\end{table*}

\paragraph{The Boltzmann equation.}
To render \cref{BBGKY1} tractable, the interaction integral $\mathcal{I}_{12}$ is replaced by a modeled collision operator, the Boltzmann integral $\mathcal{Q}_{\mathrm{Boltz}}$~\cite{boltzmann1872weitere}. This reduction involves a hierarchy of hypotheses:

\begin{enumerate}
    \item[(i)] \textit{Stosszahlansatz (Molecular Chaos):} The closure assumes that particles entering a collision are statistically uncorrelated. The error introduced by this assumption is defined as:
    \begin{equation}
        \Cchaos \equiv \mathcal{F}_2(\boldsymbol{z}, \boldsymbol{z}_{i\!i}) - \mathcal{F}(\boldsymbol{z})\mathcal{F}_{i\!i}(\boldsymbol{z}_{i\!i}).
    \end{equation}
    The Boltzmann integral assumes $\Cchaos \approx 0$~\cite{boltzmann1872weitere,cercignani_boltzmann_1988,kremer_introduction_2010}.

    \item[(ii)] \textit{Dilute Gas Approximation (Binary, Point-Particle Interactions):} The derivation assumes the gas is sufficiently dilute such that three-body collisions are negligible and interactions are binary. Furthermore, the variation of the distribution function over the molecular diameter is assumed negligible~\cite{kremer_introduction_2010}.

 \item[(iii)] \textit{Temporal filtering and collision-product covariance:} The derivation uses a short time average~\cite{kremer_introduction_2010}, represented here as a kinetic time filter of width $\Delta_t$, with $\tau_c \ll \Delta_t \ll \tau_{\rm mft}$, where  $\tau_c \sim d/\sqrt{\theta_R}$ denotes the collision duration (with $d$ the molecular diameter) and $\tau_{\rm mft}$ is the mean-free-time scale, i.e.\ the characteristic time between successive binary collisions.

 The lower bound $\tau_c \ll \Delta_t$ allows one to introduce the pre/post-collisional scattering map characterizing the BE collision integral~\cite{kremer_introduction_2010} and allows approximating $\mathcal I_{12}$ as
 \begin{alignedEq}\label{preBE}
\mathcal I_{12} &\approx \frac{1}{m} \int \left(\langle \mathcal F' \mathcal F'_{i\!i}\rangle_{\Delta_t} - \langle \mathcal F \mathcal F_{i\!i}\rangle_{\Delta_t} \right)  |\boldsymbol{\xi} - \boldsymbol{\xi}_{i\!i}| \, \sigma(\Omega) \, \mathrm{d}\Omega \, \mathrm{d}\boldsymbol{\xi}_{i\!i}\\
&\qquad\equiv \mathcal{Q}_{\rm Boltz}(F,F_{i\!i}) + \Ecc,
\end{alignedEq}
with $\langle\cdot\rangle_{\Delta_t}$ denoting this kinetic time filter, and where $\mathcal F' \equiv \mathcal F(\boldsymbol{z}') =  \mathcal F(\boldsymbol{x}, \boldsymbol{\xi}')$ and $\mathcal F'_{i\!i} \equiv \mathcal F(\boldsymbol{z}'_{i\!i}) = \mathcal F(\boldsymbol{x}, \boldsymbol{\xi}'_{i\!i})$, with $\boldsymbol{z}' = (\boldsymbol{x}, \boldsymbol{\xi}')$ and $\boldsymbol{z}'_{i\!i} = (\boldsymbol{x}, \boldsymbol{\xi}'_{i\!i})$. The prime superscript changes the velocity argument through the binary scattering map at fixed $\boldsymbol{x}$ (it does not denote a later time). Also, $\sigma$ represents the differential cross section and $\Omega$ the solid angle~\cite{boltzmann1872weitere,kremer_introduction_2010}.

 The upper bound $\Delta_t \ll \tau_{\rm mft}$ lets the ensemble one-particle distribution vary negligibly over the filter window. Let $F \equiv \langle \mathcal{F} \rangle_{\Delta_t}$ denote the filtered one-point distribution. In this kinetic limit the time-filtered collision-product covariance $\Ccc\left(\mathcal{F}\mathcal{F}_{i\!i}\right) \equiv \langle \mathcal{F}\mathcal{F}_{i\!i} \rangle_{\Delta_t} - F F_{i\!i}$ vanishes asymptotically, so the associated gain-minus-loss collision-covariance source
 \begin{alignedEq}\label{Ecc_def}
    \Ecc = \frac{1}{m} \int \left[ \Ccc\left(\mathcal{F}'\mathcal{F}'_{i\!i}\right) - \Ccc\left(\mathcal{F}\mathcal{F}_{i\!i}\right) \right] |\boldsymbol{\xi} - \boldsymbol{\xi}_{i\!i}| \, \sigma(\Omega) \, \mathrm{d}\Omega \, \mathrm{d}\boldsymbol{\xi}_{i\!i}.
\end{alignedEq}
 is neglected. In this limit we get the local Boltzmann integral that describes the ensemble rate of scattering events:
\begin{alignedEq}\label{BE}
\mathcal{Q}_{\rm Boltz}(F,F_{i\!i}) = \frac{1}{m} \int \left[ F' F'_{i\!i} - F F_{i\!i} \right]   \cdot |\boldsymbol{\xi} - \boldsymbol{\xi}_{i\!i}| \, \sigma(\Omega) \, \mathrm{d}\Omega \, \mathrm{d}\boldsymbol{\xi}_{i\!i}&.
\end{alignedEq}

\end{enumerate}

Assumption (iii) differs from the other statistical hypotheses (i) and (ii) because it depends on the temporal covariance of the collision product $\mathcal F\mathcal F_{i\!i}$ and scales explicitly with the chosen observation window $\Delta_t$. In contrast, molecular chaos and the dilute gas approximation do not scale with \(\Delta_t\); they concern pre-collisional pair correlations, gas density, molecular size, and the variation of the distribution over the interaction range (see the detailed comparison in Appendix~\ref{app:klimontovich_covariance}).

In the classical Boltzmann construction, this time-averaging window satisfies $\tau_c\ll\Delta_t\ll\tau_{\rm mft}$. In this limit $\Ccc \to 0$ and $\Ecc\to 0$ asymptotically, and the operator \cref{preBE} reduces to the local Boltzmann integral~\eqref{BE}. For temporal filters with $\tau_c\ll\Delta_t$ but $\Delta_t/\tau_{\rm mft}$ not asymptotically small, the Boltzmann argument no longer justifies discarding $\Ecc$. Therefore, we treat $\Ecc$ as an unclosed collision-side contribution: its asymptotic order after the homogeneous LES-type spatial filter can be determined via the CE (see \cref{CE}).
This finite time filter also has a kinematic interpretation. Although the collision integral is local in space, the streaming part transports particles during the filter window. Thus a time filter of \(\mathcal F\) samples a velocity-dependent spatial segment of size \(O(|\boldsymbol{\xi}|\Delta_t)\). This characteristic-induced spatial extent is not a homogeneous isotropic LES-type filter. The precise relation with the homogeneous LES-type spatial filter used below is stated in \cref{sec:filtered_nondimensional_rbgk_be}.

\paragraph{The BGK equation.}
For computational efficiency, the quadratic Boltzmann integral is frequently approximated by the BGK model \cite{bhatnagar_model_1954}. This assumes collisions drive the system toward a local equilibrium $F^{(0)}$ over a relaxation time $\tau$ and a constant collision frequency $\omega=1/\tau$:
\begin{equation}
    \mathcal{Q}_{\mathrm{BGK}}(F) \equiv - \omega (F - F^{(0)}).
\end{equation}
In practice, $\tau$ is chosen so that a Chapman--Enskog expansion recovers the desired macroscopic hydrodynamic limit, typically through the target viscosity.
The BGK operator approximates the Boltzmann integral~\eqref{BE} and inherits its strict Markovianity: the target equilibrium $F^{(0)}$ is constructed from the \textit{instantaneous} macroscopic moments, representing an adiabatic relaxation process with no subfilter memory.

This operator introduces three additional structural hypotheses regarding the equilibrium state and the relaxation pathway:

\begin{enumerate}
    \item[(iv)] \textit{Maximum Entropy Equilibrium:} The target state $F^{(0)}$ is the maximum–entropy state under fixed mass, momentum, and energy:
    \begin{alignedEq}\label{equilibrium}
        F^{(0)} &= \arg \max _F\Big\{-\int_\Xi F \ln F \,\mathrm{d} \bm{\xi} \Big|  \\
        &\hphantom{ \arg \max _F\Big\{--}\int_\Xi F \, \bm{\phi} \mathrm{d} \bm{\xi} = \left(\varrho, \varrho u_\alpha, \frac{3}{2}\varrho\theta\right),\\
          &\hphantom{ \arg \max _F\Big\{--}\bm{\phi} = \left(1, \xi_\alpha, \frac{1}{2}\zeta_{\alpha} \zeta_{\alpha}\right)\Big\}, \\
        &= \frac{\varrho}{(2\pi \theta)^\frac{3}{2}}e^{-\frac{\zeta_{\alpha} \zeta_{\alpha}}{2\theta}},
    \end{alignedEq}
    with
    \begin{gather}\label{rho_u_theta}
        \varrho = \int_\Xi F \,\mathrm{d} \bm{\xi},\quad
        u_\alpha = \frac{1}{\varrho}\int_\Xi F \xi_\alpha \mathrm{d} \bm{\xi},\quad
        \theta = \frac{1}{3\varrho}\int_\Xi F \zeta_{\alpha} \zeta_{\alpha} \mathrm{d} \bm{\xi},
    \end{gather}
    where $\zeta_\alpha=\xi_\alpha-u_\alpha$ and $\Xi$ denotes velocity space. It is important to stress that in \cref{rho_u_theta}, the moments are computed with respect to the time-filtered distribution $F$. If we denote by $\varrho^{\mathcal F}$ and $v_\alpha$ the instantaneous density and velocity associated with $\mathcal F$, then
\begin{equation}
    u_\alpha = \frac{\langle \varrho^{\mathcal F} v_\alpha\rangle_{\Delta_t}}{\langle \varrho^{\mathcal F}\rangle_{\Delta_t}},
\end{equation}
that is, the filtered velocity is the density-weighted temporal average of the instantaneous velocity (i.e. a Favre filtered variable~\cite{favre_equations_1965}).
\item[(v)] \textit{Linearity:} The BGK operator replaces \cref{BE} with a term linear in the departure from equilibrium, $F - F^{(0)}$, reducing the relaxation to a first-order decay. The operator remains globally non-linear through the dependence of $F^{(0)}$ on the macroscopic moments of $F$.

    \item[(vi)] \textit{Isotropy:} The relaxation frequency $\omega$ is a scalar independent of $\boldsymbol{\xi}$ (single relaxation time), fixing the Prandtl number to unity. This limitation is addressed, e.g., by the ES-BGK model~\cite{holway1966new}, which uses an anisotropic Gaussian target with a temperature tensor, and by the Shakhov model~\cite{shakhov1968generalization}, which modifies the target equilibrium to decouple the heat flux relaxation.

\end{enumerate}

The fundamental limitation of the BGK model in LES is inherited from the Boltzmann integral and it is the structural inability to capture the collision-covariance source term $\Ecc$. For temporal filters with $\tau_c\ll\Delta_t$ but $\Delta_t/\tau_{\rm mft}$ not asymptotically small, the temporal filter can retain memory related either to microscopic fluctuations or to unresolved macroscopic advective fluctuations. The unresolved coherent structures make the macroscopic statistical state no longer Markovian over the filter width. Therefore, even if the linear relaxation $-\omega(F - F^{(0)})$ approximates the Boltzmann integral $\mathcal{Q}_{\rm Boltz}$, it fails to account for the collision-product covariance.

In the following, we retain the collision-covariance source term $\Ecc$. In this way we break the inherited Markovianity of the kinetic equation: the collision-product covariance $\Ccc$ acts as a macroscopic memory that actively perturbs the local relaxation trajectory. We define the sum of the BGK collision operator and the collision-covariance source term as the \emph{recorrelated BGK collision model}~\cite{marson_kinetic_2025}:
\begin{equation}\label{RBGK}
    \mathcal{Q}_{\mathrm{RBGK}} \equiv \mathcal{Q}_{\mathrm{BGK}} + \Ecc\,.
\end{equation}
The specific modeling and closure of $\Ecc$ is deferred to \cref{sec:modeling_Ecc}, after the multiscale analysis has established a way to disentangle the SGS advection from the main flow.

\subsection{Nondimensional recorrelated BGK--BE}
We refer to the equation
\begin{equation}
    \partial_t {F} + \xi_\alpha\,\partial_\alpha {F} = \mathcal{Q}_{\rm RBGK}
\end{equation}
as the \emph{recorrelated BGK--BE} (RBGK--BE). Its nondimensionalization proceeds by introducing suitable reference scales so that all terms in the governing equations become \(O(1)\).
This nondimensionalization concentrates the physical information into a small set of nondimensional numbers (Reynolds, Mach, Knudsen), which express the relative strengths of the underlying mechanisms each one playing a different asymptotic role.
Such a formulation avoids artificial scale imbalances and clarifies the hierarchy of effects, enabling consistent asymptotic analysis.

In the specific case of the BE, nondimensionalization must explicitly reflect the presence of two physically distinct scales.
The first is the macroscopic ``convective scale'', characterized by a reference length \(\mathcal L\), time \(\mathcal T\), and velocity \(\mathcal U = \mathcal L / \mathcal T\).
These define the advective transport of the distribution function.
The second is the microscopic ``diffusive scale'', defined by the mean free path \(\ell\), the microscopic reference velocity \(\sqrt{\theta_R}\), and the mean-free-time scale \(\tau_{\rm mft}\), related via \begin{equation}
    \tau_{\rm mft}\equiv\ell /\sqrt{\theta_R}\,.
\end{equation}
These quantities govern the relaxation dynamics of \(F\) toward equilibrium.
Here \(\theta_R\) is the reference temperature defined from a reference or average PDF of the fluid $F_R$ as
\begin{equation}
    \theta_R\equiv  \frac{1}{3\varrho_R}\int_\Xi F_R \zeta_{R\alpha} \zeta_{R\alpha}\,\mathrm{d}\bm{\xi},
\end{equation}
where $\varrho_R$ and $\zeta_{R\alpha}\equiv\xi_\alpha-u_{R\alpha}$ are defined from $F_R$ as in \cref{rho_u_theta}.
The root-mean-square (RMS) thermal speed \(c_t\) is related to $\theta_R$ by \( c_t = \sqrt{3\theta_R}\).
For an ideal gas, the speed of sound is related to the reference temperature by \(c_s = \sqrt{\gamma \theta_R}\), where \(\gamma\) is the adiabatic index.
In this work, we adopt \(\sqrt{\theta_R}\) as the characteristic microscopic velocity scale.

Because the convective and collisional processes operate on different characteristic times, nondimensionalization must keep both scales visible: the ratio of their timescales supplies the primary Chapman--Enskog ordering, while the Mach number controls the mixed convective-thermal content of the particle velocity and therefore the relative size of the higher moments.
Identifying \(\{\mathcal L, \mathcal T, \mathcal U\}\) and \(\{\ell, \tau_{\rm mft}, \sqrt{\theta_R} \}\) is therefore essential for a consistent scaling analysis and for understanding how macroscopic fluid equations emerge from the kinetic description.

In practice, the convectively scaled one-particle distribution $f \equiv f^\flat \equiv \mcU^3 F/\varrho_R$ satisfies the nondimensional RBGK--BE, expressed in terms of the Mach number, $\Ma\equiv\mcU/ \sqrt{\theta_R} $, and the kinetic Reynolds-like number, $\Re_\ell \equiv \sfrac{\mcU\mcL}{  \sqrt{\theta_R} \ell}$, as
\begin{alignedEq}
    \label{eq:BEdimless2}
    \cpartial_t  f + \xi^{\flat}_{\alpha}\cpartial_{\alpha}f=&-\frac{\Re_{\ell}}{{\Ma^2}}\,{\omega^\sharp}\left({f-f^\zero}\right)+\frac{\Re_{\ell}}{{\Ma^2}}{\Ecc^\sharp}\,.
\end{alignedEq}
The superscripts $\flat$ and $\sharp$ denote convective and diffusive scaling by \(\{\mcL,\mcT,\mcU\}\) and \(\{\ell,\tau_{\rm mft},\sqrt{\theta_R}\}\), respectively. Therefore, \cref{eq:BEdimless2} uses convective scaling for the streaming variables and the one-particle distribution \(f\), whereas the collision frequency and thermal variables retain diffusive scaling.
The nondimensional variables are
    \begingroup \allowdisplaybreaks
    \begin{alignat*}{2}
        &\cpartial_t \equiv \mcT\partial_t,\   \cpartial_\alpha \equiv \mcL\partial_\alpha &\text{(time/space derivatives)}\\
        &\xi_\alpha^\flat \equiv \xi_\alpha/\mcU,\ {\zeta_\alpha^\sharp\equiv\zeta_\alpha/ \sqrt{\theta_R} }\ &\text{(particle/peculiar vel.)}\\
        &\rho \equiv \varrho/\varrho_R,\ \mathrlap{u_\alpha^\flat \equiv u_\alpha/\mcU}\quad &\text{(density/velocity)}
        \\
        &\theta^\sharp \equiv \theta/\theta_R,\ \omega^\sharp \equiv \omega\,\ell/ \sqrt{\theta_R} \quad &\text{(temperature/collision freq.)}\\
        &f^\zero \equiv\frac{\rho\Ma^3}{(2\pi \theta^\sharp)^\frac{3}{2}}e^{-\frac{\zeta^{\sharp 2}}{2\theta^\sharp}}
        \ &\text{(Maxwellian)},
    \end{alignat*}
\endgroup
Henceforth, \(f\equiv f^\flat\) with \(\rho=\int_\Xi f^\flat\,d\bm\xi^\flat\).
The nondimensionalization of \(f\) is local in \(\bm{x}\) and \(t\) and no factor \(\mcL^3\) appears because hydrodynamic moments integrate over velocity space rather than the full phase space. If \(f^\sharp\) denotes the same local distribution written with the diffusive velocity measure, then \(f^\flat d\bm\xi^\flat=f^\sharp d\bm\xi^\sharp\). Since \(d\bm\xi^\sharp=\Ma^3 d\bm\xi^\flat\), one has \(f^\flat=\Ma^3 f^\sharp\). The Maxwellian above is the corresponding convectively scaled equilibrium distribution.
The collision-covariance source and the timescale ratio are nondimensionalized as
\begin{equation}\label{Re2Ma}
    \Ecc^\sharp \equiv \frac{\tau_{\rm mft}\,\mcU^3\Ecc}{\varrho_R} \,, \qquad \frac{\mcT}{\tau_{\rm mft}}=\frac{\Re_{\ell}}{\Ma^2} = \frac{1}{\Ma\Kn} \,,
\end{equation}
with $\Kn \equiv \ell/\mcL$ being the Knudsen number. The parameter $\Re_\ell=\Ma/\Kn$ is not the classic hydrodynamic Reynolds number \(\Re\), which is introduced in \cref{subsec:navier_stokes_limit} through \cref{eq:Re_tau_relation} after the collision time is specified.
Crucially, although $\xi_\alpha^\flat \equiv \xi_\alpha/\mcU$, the particle velocity is not purely advective: using $\xi_\alpha=u_\alpha+\zeta_\alpha$ and $\Ma=\mcU/\sqrt{\theta_R}$, one has
\begin{equation}\label{xi_mixed_scales}
    \xi_\alpha^\flat = u_\alpha^\flat + \Ma^{-1}\zeta_\alpha^\sharp \,.
\end{equation}
The notation $\xi_\alpha^\flat$ is thus compact, but it hides an intrinsically mixed convective-thermal scaling.

Also the collision-covariance source term $\Ecc^\sharp$ has a mixed normalization: it uses the convective-scale reference velocity $\mathcal U$ and the diffusive-scale kinetic collision mean-free-time $\tau_{\rm mft}$. The superscript $\sharp$ records the latter because this source is ordered through the collision side of the CE hierarchy. The CE hierarchy following in \cref{CE} fixes the order at which the retained filtered source contributes after the application of the homogeneous LES-type spatial filter.

\subsection{Filtered nondimensional recorrelated BGK--BE}
\label{sec:filtered_nondimensional_rbgk_be}
We now introduce the homogeneous LES filtering used in the rest of the manuscript. The kinetic time filter maps the instantaneous one-particle distribution \(\mathcal F\) to \(F=\langle\mathcal F\rangle_{\Delta_t}\). A homogeneous LES-type spatial filter, denoted by \(\overline{(\cdot)}\), then acts on \(F\). The resulting filtering sequence is
\[
\mathcal F
\xrightarrow{\langle\cdot\rangle_{\Delta_t}}
F
\xrightarrow{\overline{(\cdot)}}
\overline F .
\]
After nondimensionalization, \(F\) is written as \(f\), and \(\overline F\) as \(\overline f\).
Therefore, \(\overline f\) is the LES-filtered form of an already time-filtered one-particle distribution.

The first operation is not purely temporal in physical content.
For a temporal kernel \(G_{\Delta_t}\),
\begin{equation}\label{eq:kinetic_time_filter}
F(\boldsymbol{x},\boldsymbol{\xi},t)
=
\int G_{\Delta_t}(s)\,
\mathcal F(\boldsymbol{x},\boldsymbol{\xi},t-s)\,\mathrm{d}s .
\end{equation}
For the free-streaming part of the BE~\cite{cercignani_boltzmann_1988,kremer_introduction_2010},
\begin{equation}\label{eq:free_streaming_characteristic}
    \partial_t\mathcal F+\xi_\alpha\partial_\alpha\mathcal F=0
    \quad\Rightarrow\quad
    \mathcal F(\boldsymbol{x},\boldsymbol{\xi},t-s)
    =
    \mathcal F(\boldsymbol{x}+\boldsymbol{\xi}s,\boldsymbol{\xi},t).
\end{equation}
Consequently, along the streaming characteristics, \cref{eq:kinetic_time_filter} reads
\begin{equation}\label{eq:time_filter_spatial_reach}
F(\boldsymbol{x},\boldsymbol{\xi},t)
\simeq
\int G_{\Delta_t}(s)\,
\mathcal F(\boldsymbol{x}+\boldsymbol{\xi}s,\boldsymbol{\xi},t)\,\mathrm{d}s .
\end{equation}
Therefore, the induced kinetic spatial extent is velocity dependent,
\begin{equation}\label{eq:kinetic_filter_width}
    \Delta_x^{\rm kin}(\boldsymbol{\xi})\sim |\boldsymbol{\xi}|\Delta_t,
    \qquad
    \Delta_x^{\rm kin}\sim c_t\Delta_t,
\end{equation}
where the second estimate uses the RMS thermal speed \(c_t=\sqrt{3\theta_R}\) defined above.
A consistent LES-type filtering interpretation therefore chooses the homogeneous spatial-filter width \(\Delta_x\) no smaller than \(c_t\Delta_t\), with a proportionality factor set by the temporal-kernel definition.
Taking \(\Delta_x=O(c_t\Delta_t)\) then fixes a consistent space--time coarse-graining scale connected by the RMS thermal speed.
Here \(O(\cdot)\) denotes a common characteristic scale, not a kernel-specific equality or a prescription for the spatial-filter shape.
The temporal kernel and particle velocities determine the induced extent, whereas the spatial-filter kernel determines how that extent maps to its nominal width. The purpose of the relation is to identify the correlation between the time and spatial coarse-graining scales, not to prescribe this kernel-dependent mapping.
For the LBM realization considered later, the same scale matching is recovered when \(\Delta_t\) and \(\Delta_x\) are identified with the time step and lattice spacing, respectively.
Appendix~\ref{app:lbm_discretization} gives the discrete derivation.
When \(\Delta_x=O(c_t\Delta_t)\), the homogeneous LES-type spatial filter and the kinetic time filter remain distinct operations, but their widths represent the same physical coarse-graining scale in space and time.
Thus, the consistency condition \(\Delta_x=O(c_t\Delta_t)\) fixes the interpretation of the filtered state but does not introduce an additional asymptotic parameter.

Following the standard LES formalism~\cite{sagaut_large_2006}, a homogeneous spatial filter with uniform width commutes with spatial and temporal differentiation, i.e., \(\overline{\partial\phi}=\partial\overline{\phi}\).
Applying the filter to the nondimensional RBGK--BE \eqref{eq:BEdimless2} gives the filtered RBGK--BE (FRBGK--BE):
\begin{gather}\label{fbe_bgk}
    \cpartial_t  \overline f + \xi_\alpha^\flat\cpartial_{\alpha} \overline f
    =  -\frac{\Re_{\ell}}{{\Ma^2}}\omega^\sharp\left({\overline f- \overline{f^{\zero }}}\right)
    +\frac{\Re_{\ell}}{{\Ma^2}}\overline{\Ecc^\sharp}\,.
\end{gather}

In \cref{fbe_bgk}, the filtered fine-grained equilibrium is
\begin{equation}\label{f0bar_full}
    \overline{f^\zero(f)}
    =
    \overline{
    \frac{\rho\Ma^3}{(2\pi\theta^\sharp)^{3/2}}
    \exp\left(
    -\frac{\Ma^2(\xi_\alpha^\flat-u_\alpha^\flat)(\xi_\alpha^\flat-u_\alpha^\flat)}
    {2\theta^\sharp}
    \right)
    }.
\end{equation}
Here \(\rho(f)\), \(u_\alpha^\flat(f)\), and \(\theta^\sharp(f)\) are computed from \(f\) before applying the homogeneous LES-type spatial filter.
Here, ``fine-grained'' means unfiltered only with respect to the homogeneous LES-type spatial filter, since \(f\) has already undergone the kinetic time filter \(\langle\cdot\rangle_{\Delta_t}\).
Hence, \(\overline{f^\zero(f)}\) is not directly computable from $\overline f$, and requires knowledge of the LES-unfiltered state \(f\).
Consequently, \cref{fbe_bgk} is not closed in \(\overline f\), because filtering the Maxwellian evaluated from the fine-grained fields is not equivalent to evaluating the Maxwellian from the filtered fields. The difference defines the SGS equilibrium commutation residual, made explicit by the decomposition
\begin{align}\label{f0bar}
    \overline{f^\zero(f)}
    &= f^\zero\left(\bar\rho,\tilde u^\flat,\tilde\theta^\sharp\right)+ f_{\sgs}(f)\,.
\end{align}
The resolved Maxwellian in \cref{f0bar} is built from the filtered thermodynamic fields:
\begin{equation}\label{f0resolved}
    f^\zero\left(\bar\rho,\tilde u^\flat,\tilde\theta^\sharp\right)
    \equiv
    \frac{\bar\rho\Ma^3}{(2\pi\tilde\theta^\sharp)^{3/2}}
    \exp\left(
    -\frac{\Ma^2(\xi_\alpha^\flat-\tilde u_\alpha^\flat)(\xi_\alpha^\flat-\tilde u_\alpha^\flat)}
    {2\tilde\theta^\sharp}
    \right),
\end{equation}
where we introduced the Favre-filtered fields
\(\MT[\tilde u^\flat][1] \equiv \sfrac{\overline{\rho \MT[u^\flat][1]}}{\bar \rho }\) and
\(\tilde \theta^\sharp \equiv \sfrac{\overline{\rho \theta^\sharp}}{\bar \rho }\,\)~\cite{favre_equations_1965,favre_turbulence_1983}.
The superscript zero denotes the equilibrium map \(f^{(0)}\) and its argument specifies the thermodynamic fields supplied to the map. Below, we use \(f^\zero\) as a shortcut for \(f^\zero(\bar\rho,\tilde u^\flat,\tilde\theta^\sharp)\) and \(\overline{f^\zero}\) for \(\overline{f^\zero(f)}\). With this notation, the SGS equilibrium commutation residual appearing in \cref{f0bar} is defined by
\begin{equation}\label{fsgs_def}
    f_{\sgs}(f)\equiv \overline{f^\zero} - f^\zero.
\end{equation}

\paragraph{Information content of the SGS equilibrium commutation residual $f_\sgs$}
A common misconception in the kinetic modeling of turbulence is the conflation of the commuting linear transport of the BE with the velocity advective covariances \(\Wtilde{\MT[u^\flat][1]\MT[u^\flat][2]} - \MT[\tilde u^\flat][1]\MT[\tilde u^\flat][2] \) generating the advective SGS stress tensor at the moment transport level.
To clarify this point, we can first use \cref{f0bar} to decompose $\overline f$ as
\begin{equation}\label{barf}
    \overline f = f^\zero + f_{\sgs} + \overline{f^\Neq}, \qquad \overline{f^\Neq}\equiv \overline f - \overline{f^\zero},
\end{equation}
where $\overline{f^\Neq}$ is the filtered fine-grained nonequilibrium.
\Cref{barf} clarifies that the lhs of \cref{fbe_bgk} linearly transports the SGS information carried by $f_{\sgs}$ that is embedded through \cref{barf} in the filtered distribution function $\overline f$. This also explains why in \cref{tab:comparison} the SGS advection \compref{tab:comparison:c-be}, represented here by the advection of $f_\sgs$, is described as ``exactly transported but not closed by the filtered moments.''

Second, we can clarify the connection of $f_\sgs$ with the macroscopic SGS stress tensor \compref{tab:comparison:c-nse} by expanding it in Hermite series (see Appendix~\ref{app:hermite_representation} for the details about the expansion).
For an isothermal flow with prescribed $\theta^\sharp=\tilde\theta^\sharp=1$, the representation reduces to
\begin{align}\label{f_sgs_hermite}
    f_{\sgs}
    &= \frac{w^\flat}{\rho_R}\,\bar\rho\left[
        \frac{\Ma^2}{2!}H^\sharp_{\alpha_1\alpha_2}\left(\Wtilde{\MT[u^\flat][1]\MT[u^\flat][2]} - \MT[\tilde u^\flat][1]\MT[\tilde u^\flat][2]\right) \right.\nonumber\\
    &\left.\quad+ \frac{\Ma^3}{3!}H^\sharp_{\alpha_1\alpha_2\alpha_3}\left(\Wtilde{\MT[u^\flat][1]\MT[u^\flat][2]\MT[u^\flat][3]} - \MT[\tilde u^\flat][1]\MT[\tilde u^\flat][2]\MT[\tilde u^\flat][3]\right) + \cdots\right],
\end{align}
where $\Wtilde{\MT[u^\flat][1]\MT[u^\flat][2]} \equiv \overline{\rho \MT[u^\flat][1]\MT[u^\flat][2]}/\bar\rho$ are the Favre-averaged second-order velocity correlations.
\Cref{f_sgs_hermite} shows that the leading isothermal contribution to $f_\sgs$ contains the same velocity covariance \(\Wtilde{\MT[u^\flat][1]\MT[u^\flat][2]} - \MT[\tilde u^\flat][1]\MT[\tilde u^\flat][2] \) that defines the macroscopic SGS stress tensor. In fact, taking moments of $f_\sgs$ up to second order shows that the zeroth and first raw moments vanish,
\begin{align}
    \int_\Xi f_{\sgs}\,\dd\bm\xi^\flat &= 0, \label{f_sgs_zero_moment}\\
    \int_\Xi f_{\sgs}\,\xi^\flat_{\alpha_1}\,\dd\bm\xi^\flat &= 0, \label{f_sgs_first_moment}
\end{align}
whereas the second raw moment gives the SGS stress tensor
\begin{equation}\labelAndRemember{m_sgs_2}{
    \MT[m^{\sgs}][1][2]
    \equiv \int_\Xi f_{\sgs}\,\xi^\flat_{\alpha_1}\xi^\flat_{\alpha_2}\,\dd\bm\xi^\flat
    = \bar{\rho} \left( \Wtilde{\MT[u^\flat][1]\MT[u^\flat][2]} - \MT[\tilde u^\flat][1]\MT[\tilde u^\flat][2] \right)}.
\end{equation}
Thus $f_\sgs$ is transported by the linear BE streaming operator because it is part of $\overline f$, but it remains unknown because its evaluation requires fine-grained correlations not determined by the filtered moments alone.

The modeling problem in \cref{f0bar} underlies several kinetic turbulence approaches~\cite{succi_towards_2002,girimaji_boltzmann_2007,ansumali_kinetic_2004}. In particular, the relative-frame construction in \cite{girimaji_boltzmann_2007} makes advective second order SGS terms embedded in \cref{f_sgs_hermite} explicit by a change of velocity variables in a non-inertial reference frame (Appendix~\ref{app:girimaji}). In the present interpretation, this reformulates content already present in the inertial filtered distribution: the velocity SGS covariance appears in $f_{\sgs}$ through \cref{f_sgs_hermite} and it is already transported by \cref{fbe_bgk} through \cref{barf}. Also, the transformation in \cite{girimaji_boltzmann_2007} neither supplies a collision-side closure nor represents a Navier--Stokes-order constitutive law via a Chapman--Enskog expansion in the relative variables (see Appendix~\ref{app:girimaji}).

\section{Hydrodynamic limit of the filtered recorrelated BGK--BE}
\label{CE}

The present derivation of the hydrodynamic limit of the FRBGK--BE \eqref{fbe_bgk} follows the classical CE multiscale expansion~\cite{chapman_mathematical_1953}, with the expansion parameter identified explicitly and unambiguously through the nondimensionalization.
The expansion parameter must compare fast collisional relaxation with slow macroscopic transport. In \cref{eq:BEdimless2}, the BGK prefactor ${\Re_{\ell}}/{\Ma^2}$ gives the ratio of the molecular relaxation rate to the macroscopic transport rate~\cite{marson_kinetic_2025}. Its reciprocal equals $\Ma\,\Kn$ via \cref{Re2Ma}. The kinetic description therefore contains two physically distinct scale ratios: $\Kn$ compares the mean free path with the macroscopic length, whereas $\Ma\,\Kn$ compares the reference kinetic and macroscopic times. Their ratio is $\Ma$, which compares the convective and thermal velocity scales. The mixed nondimensionalization retains these distinct spatial, temporal, and velocity-scale ratios; it does not introduce them. The natural CE small parameter is therefore the reciprocal of the prefactor in \cref{eq:BEdimless2},
\begin{equation}\label{CE_eps}
    \epsilon \equiv \frac{\Ma^2}{\Re_{\ell}} \overset{\eqref{Re2Ma}}{=} \frac{\tau_{\rm mft}}{\mcT} = \Ma\,\Kn \ll 1\,.
\end{equation}
Thus, $\epsilon$ follows from the nondimensionalized BE rather than from an imposed turbulent scale separation. Its smallness is necessary for the formal CE hierarchy, but $\epsilon$ alone does not specify the hydrodynamic distinguished limit: the projected moment magnitudes also require the independent spatial ratio $\Kn$, or equivalently $\Ma$ and $\Re_\ell$. Since \(\Re_\ell/\Ma^2=1/\epsilon\), premultiplying \cref{fbe_bgk} by \(\epsilon\) gives
\begin{equation}\label{fbe_bgk2}
    \epsilon\,\cpartial_t \overline f + \epsilon\,\xi_\alpha^\flat\cpartial_{\alpha} \overline f
    =  -\omega^\sharp\left({\overline f- \overline{f^{\zero }}}\right)
    +\overline{\Ecc^\sharp}\,.
\end{equation}
This premultiplied form shows that the transport terms carry an explicit \(\epsilon\) prefactor, whereas the BGK operator and the retained source are written in microscopic normalization.

Following~\cite{marson_kinetic_2025}, the CE procedure expands the filtered distribution function in successive powers of this parameter,
\begin{equation}\label{CE_f}
    \overline{f} = \sum_{k=0}^{\infty} \epsilon^k\, \overline{f^{(k)}}\,.
\end{equation}
In the CE expansion, $g^{(k)}$ denotes the coefficient multiplying $\epsilon^k$. The tensor and scaling conventions are collected in \cref{tab:notation}. The leading filtered coefficient is $\overline{f^\zero}=f^\zero+f_{\sgs}$ by \cref{f0bar}, rather than the resolved equilibrium $f^\zero$ alone.
The nondimensional space and time derivative operators are decomposed separately by asymptotic order in $\epsilon$.
The explicit factor in the premultiplied balance \cref{fbe_bgk2} can be integrated in the derivative expansion, making them appear starting from the $O(\epsilon)$ term, i.e.\ from $k=1$:
\begin{equation}\label{CE_deriv}
    \epsilon\,\cpartial_\alpha = \sum_{k=1}^{\infty} \epsilon^{k}\, \partial_\alpha^{(k)}\,,\qquad
    \epsilon\,\cpartial_t    = \sum_{k=1}^{\infty} \epsilon^{k}\, \partial_t^{(k)}\,.
\end{equation}
The first resolved transport derivatives therefore enter the premultiplied hierarchy at \(O(\epsilon)\).

Because the LES-type spatial filter is homogeneous, the commutation property stated in \cref{fbe_bgk} applies equally to each multiscale derivative operator appearing in \cref{CE_deriv}. The CE hierarchy derived below is therefore restricted to uniform filter widths, for which filtering and multiscale differentiation commute by the standard convolution property.

With this convention, the order-by-order BGK contribution is represented as
\begin{gather}\label{CE_Q}
    \overline{\mathcal{Q}_{\mathrm{BGK}}}
    = \sum_{k=0}^{\infty} \epsilon^k\, \overline{\mathcal{Q}_{\mathrm{BGK}}^{(k)}}\,,\\
    \overline{\mathcal{Q}_{\mathrm{BGK}}^{(0)}}
    =0,\qquad
    \overline{\mathcal{Q}_{\mathrm{BGK}}^{(k)}}
    = -\omega^\sharp\overline{f^{(k)}}\,,\qquad k\ge 1\,.\label{Qbgk}
\end{gather}
The filtered collision-covariance source \(\overline{\Ecc^\sharp}\) is likewise expanded by CE order:
\begin{equation}\label{CE_Ecc}
    \overline{\Ecc^\sharp}
    =
    \sum_{k=0}^{\infty}
    \epsilon^k\overline{\Ecc^{\sharp(k)}}.
\end{equation}
The leading equilibrium balance imposes
\begin{equation}\label{fce0}
    \overline{\Ecc^{\sharp(0)}}=0.
\end{equation}
The first retained coefficient, \(\overline{\Ecc^{\sharp(1)}}\), enters the \(O(\epsilon)\) kinetic balance and contributes to the Navier--Stokes-order momentum equation through its second moment. This is a CE-order placement of the retained LES collision-side source, not a filter-width estimate. Substituting \cref{CE_f,CE_deriv,CE_Q,CE_Ecc} into \cref{fbe_bgk2} and collecting powers of $\epsilon$ gives the kinetic hierarchy:
\begin{equation}
    \epsilon\, \partial_t^{(1)} \overline {f^{(0)}}+\xi_{\alpha}^\flat \epsilon\, \partial_\alpha^{(1)} \overline {f^{(0)}} = \epsilon\,\overline{\mathcal{Q}_{\mathrm{BGK}}^{(1)}} +\epsilon\,\overline{\Ecc^{\sharp(1)}}\,, \label{fce1}
\end{equation}
\begin{alignat}{2}
    \epsilon^2 \partial_t^{(1)} \overline {f^{(1)}}+ \epsilon^2 \partial_t^{(2)} \overline {f^{(0)}}+&\nonumber\\
    \xi_{\alpha}^\flat \epsilon^2 \partial_\alpha^{(1)} \overline {f^{(1)}}+\xi_{\alpha}^\flat \epsilon^2 \partial_\alpha^{(2)} \overline{f^{(0)}} && & \label{fce2}\\
    &= \epsilon^2\,\overline{\mathcal{Q}_{\mathrm{BGK}}^{(2)}} + \epsilon^2\,\overline{\Ecc^{\sharp(2)}}\,.\nonumber
\end{alignat}
Since \(\overline{f^\zero} = f^\zero(\overline f) + f_\sgs\)  (\cref{f0bar}), filtering creates two distinct equilibrium constructions: the filtered fine-grained equilibrium \(\overline{f^\zero}\) (\cref{f0bar_full}) and the equilibrium reconstructed from the filtered moments \(f^\zero(\overline f)\) (\cref{f0resolved}). Therefore, we define the coarse-level nonequilibrium
\begin{equation}
    f^\cNeq \equiv \overline f - f^\zero,
\end{equation}
where the superscript ``\(\cNeq\)'' denotes departure from the equilibrium reconstructed from filtered moments and does not imply purely nonequilibrium content. On the contrary, the filtered fine-grained nonequilibrium
\begin{alignedEq}
    \overline{f^\Neq} \equiv \overline f - \overline{f^\zero}&=\overline f - f^\zero -f_\sgs=f^\cNeq - f_\sgs\\
    &=\sum_{k=1}^\infty \epsilon^k \overline{f^{(k)}} \approx \epsilon \overline{f^\one}.
\end{alignedEq}
contains purely filtered non equilibrium content with terms premultiplied by \(\epsilon^{k\ge 1}\) as a direct consequence of the filtered CE hierarchy \cref{Qbgk,fce1,fce2}.
Accordingly, the expression of \(\overline{f^\one}\) differs from that of the unfiltered case, i.e.:
\begin{alignedEq}\label{approx1}
    \epsilon \overline{f^\one} &= \overline{f^\Neq} - \sum_{k=2}^\infty  \epsilon^k \overline{f^{(k)}} \\
    &= f^\cNeq - f_\sgs - \sum_{k=2}^\infty  \epsilon^k \overline{f^{(k)}} \\
    &\approx f^\cNeq - f_\sgs\,.
\end{alignedEq}
Thus \(f^\cNeq\) is \(O(1)\) because it contains the SGS equilibrium commutation residual \(f_\sgs\), which is not premultiplied by \(\epsilon\). The asymptotically small CE remainder is instead \(\overline{f^\Neq} = \epsilon \overline{f^\one} + O(\epsilon^2)\).
If one further expands \(f_\sgs=f_\sgs^\zero+\epsilon f_\sgs^\one+O(\epsilon^2)\), then
\begin{equation}\label{f1_def}
    \epsilon \overline{f^\one}
    = \underbrace{f^\cNeq - f_\sgs^\zero}_{\epsilon f^\one}
    - \epsilon f_\sgs^\one + O(\epsilon^2)\,,
\end{equation}
This matches the unbarred \(f^\one\) variable used in~\cite{marson_kinetic_2025}.
The key distinction is therefore between the coarse-level nonequilibrium \(f^\cNeq\), which contains the \(O(1)\) contribution from the leading order of the SGS equilibrium commutation residual \(f_\sgs^\zero\), and the filtered fine-grained nonequilibrium \(\overline{f^\Neq}\), which is \(O(\epsilon)\).
The same distinction fixes the ordering of $f_{\sgs}$ in \cref{approx1}. Since $f_{\sgs}$ carries the unresolved second moment, it must retain an $O(1)$ component when the filter width is a macroscopic LES scale. A collision-side-only scaling such as $f_{\sgs} = \epsilon f_{\sgs}^\one$, with $f_{\sgs}^\zero = 0$, places the SGS contribution only at viscous order and removes $\MT[m^{\sgs}][1][2]$ from the Euler limit. Appendix~\ref{app:ansumali_scaling} shows how this restriction appears in the filtered formulation of~\cite{ansumali_kinetic_2004} and in the approximate deconvolution method (ADM) in~\cite{sagaut_toward_2010,malaspinas_advanced_2011}.
In coefficient-level CE balances below, an unsuperscripted SGS moment denotes its leading coefficient unless a CE superscript is shown explicitly.

The CE hierarchy assumes no additional scale separation within hydrodynamic turbulence. Energy-containing, inertial-range, and Kolmogorov-scale motions remain coupled through nonlinear energy transfer.
Unlike some earlier kinetic-turbulence formulations (e.g.,~\cite{chen_expanded_2004}), which use a CE-like expansion in an effective turbulent Knudsen number, the present analysis uses the kinetic timescale ratio supplied by the nondimensionalization, $\epsilon = \Ma^2/\Re_\ell$, to order the derivative hierarchy.
The Mach number nevertheless remains an independent parameter in the projected hydrodynamic equations representing the unconstrained presence of the space-scale ratio $\Kn$ after the scale separation ordered by the time-scale ratio \(\epsilon\). Through \cref{xi_mixed_scales}, the thermal component of $\xi^\flat$ introduces inverse powers of $\Ma$, beginning with $\Ma^{-2}$ in the second moment. Consequently, moment projection can map an $O(\epsilon)$ kinetic correction to an $O(\epsilon\Ma^{-2})=O(1/\Re_\ell)$ macroscopic contribution.

\subsection{Euler level limit}

Taking the zeroth- and first-order velocity moments (multiplication by $1$ and $\xi_{\alpha_1}^\flat$, respectively, and integration over $\dd\bm\xi^\flat$) of the $O(\epsilon)$ equation \cref{fce1}, and using \(\overline{f^{(0)}}=f^{(0)}+f_{\sgs}\), yields the macroscopic continuity and Euler momentum equations at the convective time scale:
\begin{align}
    \epsilon \partial_t^{(1)} \bar{\rho} + \epsilon \partial_{\alpha_1}^{(1)} (\bar{\rho} \MT[\tilde{u}^\flat][1]) &= 0 \label{mass_moments}, \\
    \epsilon \partial_t^{(1)} (\bar{\rho} \MT[\tilde{u}^\flat][1]) + \epsilon \partial_{\alpha_2}^{(1)} \left( \bar{\rho} \MT[\tilde{u}^\flat][1] \MT[\tilde{u}^\flat][2] + \bar{p}^\flat \MT[\delta][1][2] + \MT[m^{\sgs}][1][2] \right) &= 0 \label{momentum_moments},
\end{align}
where $\bar{p}^\flat \equiv \bar{\rho} \tilde{\theta}^\sharp \Ma^{-2}$ is the resolved convectively nondimensional thermodynamic pressure.
For constant $\tilde\theta^\sharp$, this pressure separates into
\[
\bar p^\flat
=
\underbrace{\frac{\rho_R\tilde\theta^\sharp}{\Ma^2}}_{\text{uniform background}}
+
\underbrace{\frac{(\bar\rho-\rho_R)\tilde\theta^\sharp}{\Ma^2}}_{\text{pressure variation}}.
\]
In the usual isothermal incompressible low-Mach scaling, $\bar\rho-\rho_R=O(\Ma^2)$, so the pressure variation remains $O(1)$.

No collision term contributes to \cref{mass_moments,momentum_moments} for two reasons. First, the BGK term $\overline{\mathcal{Q}_{\mathrm{BGK}}}$ has zero conserved moments because the equilibrium and the distribution share the same mass and momentum. Second, $\overline{\Ecc^{\sharp(1)}}$ is the filtered gain-minus-loss difference of collision operators that conserve the same quantities. Therefore its zeroth and first velocity moments vanish. The operative closure in \cref{bgk_closure} preserves this constraint.

Therefore, the vanishing zeroth and first SGS moments in \cref{f_sgs_zero_moment,f_sgs_first_moment} leave the conserved balances unchanged, while the second-moment identity \[\recallEqAndTag{m_sgs_2}\] supplies the standard macroscopic SGS stress tensor in \cref{momentum_moments}. In contrast to~\cite{ansumali_kinetic_2004,sagaut_toward_2010,malaspinas_advanced_2011}, this demonstrates that the unclosed macroscopic SGS stress tensor
$ \MT[m^{\sgs}][1][2]$ emerges naturally and exactly at the Euler level solely as a consequence of the unresolved velocity variance inherently present within the filtered thermodynamic state $\overline{f^{(0)}}$.
Contrary to~\cite{girimaji_boltzmann_2007}, we treat this fact as an advantage, not as a problem.

\subsection{Navier--Stokes limit}\label{subsec:navier_stokes_limit}
By taking the first-order moment of \cref{fce2} after expressing $\overline{f^\one}$ using \cref{Qbgk,CE_Ecc,fce1}, we obtain:
\begin{alignatEq}{2}
    \label{filtered_second_moments}
    0 =\ &
    \T[\pp^\two][2][2][] \T[m^{\zero }][1][2][] +
    \pp_t^\two \T[m^{\zero }][1][1][]+\T[\pp^\two][2][2][] \T[m^{\sgs}][1][2][] \\
    &\underbracea{-
        \tau^\sharp \T[\pp^{(1)}][2][3][] \T[m^{\zero}][1][3][]
        -\tau^\sharp\pp_{\alpha_2}^{(1)}
        \pp_t^{(1)}\T[m^\zero ][1][2]}\\
        &
        \pushright{\underbracebd{-\tau^\sharp \T[\pp^{(1)}][2][3][] \T[m^{\sgs}][1][3][]
        -\tau^\sharp\pp_{\alpha_2}^{(1)}\pp_t^{(1)}\T[m^{\sgs}][1][2]}_{\phantom{..}\pp_{\alpha_2}^{(1)}(\bullet)\hfill}}\\
        &+\tau^\sharp\T[\pp^\one][2][2][] E^{{\rm cc},(1)}_{\alpha_1\alpha_2}\,.
\end{alignatEq}
Here we consider $\tau^\sharp = 1/\omega^\sharp$ uniform and constant and we defined the second-order moment of $\overline{\Ecc^{\sharp(1)}}$ as
\begin{equation}\label{Ecc_second_moment}
    E^{{\rm cc},(1)}_{\alpha_1\alpha_2} \equiv \int_\Xi \xi_{\alpha_1}^\flat\xi_{\alpha_2}^\flat \overline{\Ecc^{\sharp(1)}}\,d\bm\xi^\flat\,.
\end{equation}
The last line in \cref{filtered_second_moments} is therefore proportional to the divergence of the filtered second order raw moment $E^{{\rm cc},(1)}_{\alpha_1\alpha_2}$ of the collision-covariance source term in \cref{fce1}, namely $\overline{\Ecc^{\sharp(1)}}$, and, as will become evident, must be retained to recover the filtered NSE exactly through the $O(\epsilon^2)$ CE balance.
The underbraced term $\pp_{\alpha_2}^{(1)}(\bullet)$ can be rewritten in two different ways~\cite{marson_kinetic_2025}.

\paragraph{The first way} uses the balances obtained from the zeroth and first velocity moments of \cref{fce1} to eliminate the time derivatives of the resolved equilibrium moments in $(\bullet)$ for an isothermal and incompressible flow.
This procedure follows Appendix A.2.2 of \cite{kruger_lattice_2017}, with an additional level of complexity due to the subfilter-scale moments~\cite{marson_kinetic_2025}.
The resulting $O(\epsilon^2)$ momentum balance is then combined with the $O(\epsilon)$ Euler momentum balance obtained from \cref{fce1}. Recombining the CE derivative scales yields the hydrodynamic limit of the FBE. In the resulting macroscopic equations, we drop the CE superscript on derivative operators:
\begin{equation}\begin{aligned}\label{HLBE}
    0 =
\T[\pp][2][2][] \left(
    \bar \rho
        \,\MT[\tilde u^\flat][1] \MT[\tilde u^\flat][2]\right)
        +
        \T[\pp][1][1][]\bar p^\flat
+
\pp_t \left(\bar \rho   \MT[\tilde u^\flat][1]\right)
\\+
\T[\pp][2][2]\left[\bar\rho(\Wtilde{\MT[u^\flat][1]\MT[u^\flat][2]}-\MT[\tilde u^\flat][1]\MT[\tilde u^\flat][2])
\right]
\\
\underbracea{- \frac{1}{\Re}\,
    \MT[\pp][2]
\left[\bar \rho
\left(
    \MT[\pp][1]\MT[\tilde u^\flat][2]
    +\MT[\pp][2]\MT[\tilde u^\flat][1]
\right)
\right]}        \\
        \underbraceb{+\Big[\MT[\pp][2]\big(\MT[\tilde u^\flat][1] \MT[\pp][3]\MT[m^{\sgs}][2][3]\big)
        +\MT[\pp][2]\big(\MT[\tilde u^\flat][2]\MT[\pp][3]\MT[m^{\sgs}][1][3]\big)}
        \\
        \underbraced{-  \T[\pp][2][3][] \T[m^{\sgs}][1][3][]
        - \pp_{\alpha_2}\pp_t\T[m^{\sgs}][1][2]}
        \\
        +\T[\pp][2][2][] E^{{\rm cc},(1)}_{\alpha_1\alpha_2}\Big]\frac{\Ma^2}{\Re\,\tilde\theta^\sharp}
\,,
\end{aligned}
\end{equation}
where the hydrodynamic Reynolds number follows from the BGK viscosity relation:
\begin{equation}\label{eq:Re_tau_relation}
    \nu=\tau\tilde{\theta},
    \qquad
    \tau^\sharp\tilde{\theta}^\sharp
    =
    \frac{\tau\tilde{\theta}}{\ell\sqrt{\theta_R}},
    \qquad
    \Re
    \equiv
    \frac{\mcU\mcL}{\nu}
    =
    \frac{\Re_{\ell}}{\tau^\sharp\tilde{\theta}^\sharp}.
\end{equation}
In \cref{HLBE}, the underbraced terms are the recombined form of the term $\pp_{\alpha_2}^{(1)}(\bullet)$ in \cref{filtered_second_moments}, while the last line corresponds to the macroscopic effect of $\overline{\Ecc^{\sharp(1)}}$~\cite{marson_kinetic_2025}.

The prefactor $\sfrac{\Ma^2}{\Re\,\tilde\theta^\sharp\tau^\sharp}=\sfrac{\Ma^2}{\Re_{\ell}}$ arises from the recombination of the $O(\epsilon)$ and $O(\epsilon^2)$ equations.

Its explicit $\Ma^2$ factor is compensated by the $O(\Ma^{-2})$ part of the filtered fine-grained nonequilibrium moment $\overline{m^{(1)}_{\alpha_1\alpha_2}}$, yielding the Newtonian stress at $O(1/\Re_{\ell})$.
In a non-isothermal flow, the third-order SGS moment can contain $O(\Ma^{-2})$ temperature--velocity covariances that compensate this prefactor and promote the corresponding terms to $O(1/\Re_{\ell})$.
In the isothermal case considered here, these thermal contributions cancel from the second- and third-order SGS equilibrium commutation moments.
However, this cancellation does not establish uniform $O(1)$ bounds on the differentiated SGS moments, particularly the time derivative of the SGS stress, or on $E^{{\rm cc},(1)}_{\alpha_1\alpha_2}$ and its divergence. For generality, we therefore retain all terms in the final bracket of \cref{HLBE} at the stated CE order in the following analysis.

\paragraph{The second way} to rewrite $(\bullet)$ proceeds by considering the second-order raw moment of \cref{fce1}:
\begin{alignedEq}
	    \overline{\T[m^\one][1][2]} - \tau^\sharp E^{{\rm cc},(1)}_{\alpha_1\alpha_2} &=\\
	    &\underbracea{-\tau^\sharp\MT[\pp^{(1)}][3]\MT[m^{\zero}][1][2][3]-
	    \tau^\sharp\pp_{t}^{(1)} \MT[m^{\zero}][1][2]}
	    \\
	    &
	    \underbracebd{-\tau^\sharp\MT[\pp^{(1)}][3]\MT[m^{\sgs}][1][2][3]-
	        \tau^\sharp\MT[\pp^{(1)}][t]\MT[m^{\sgs}][1][2]
    }_{\phantom{..}(\bullet)\hfill}.
\end{alignedEq}

Therefore, by comparison with \cref{filtered_second_moments} and the underbraced terms in \cref{HLBE}, we can write:
\begin{equation}\label{approx2}
\begin{aligned}
	    \overline{\T[m^\one][1][2]} - \tau^\sharp E^{{\rm cc},(1)}_{\alpha_1\alpha_2} = &-\frac{\tau^\sharp{\tilde\theta^\sharp}}{\Ma^2}\left[\bar \rho
	    \left(
	    \MT[\pp^{(1)}][1]\MT[\tilde u^\flat][2]
	    +\MT[\pp^{(1)}][2]\MT[\tilde u^\flat][1]
	    \right)
	    \right]\\&+\tau^\sharp\MT[\tilde u^\flat][1] \MT[\pp^\one][3]\MT[m^{\sgs}][2][3]
	        \\&+\tau^\sharp\MT[\tilde u^\flat][2]\MT[\pp^\one][3]\MT[m^{\sgs}][1][3]
	        \\
	        &-  \tau^\sharp\T[\pp^{(1)}][3][3][] \T[m^{\sgs}][1][3][]
	        \\&- \tau^\sharp\pp_t^{(1)}\T[m^{\sgs}][1][2].
\end{aligned}
\end{equation}
Here we retain the CE superscript on derivative operators because \cref{approx2,aneq,msgs} still refer to the $O(\epsilon)$ hierarchy, not to the recombined macroscopic equations.

\subsection{Analysis of the hydrodynamic limit and convergence to FNSE}
We now disentangle the terms in \cref{approx2}.
A natural separation is the following:
\begin{align}
    \overbrace{\omega^\sharp{\T[m^\one][1][2]}- \omega^\sharp{\T[m^{\sgs\one}][1][2]}}^{\omega^\sharp\overline{\T[m^\one][1][2]}}\approx &-\frac{\tilde\theta^\sharp}{\Ma^2}\left[\bar \rho
    \left(
    \MT[\pp^{(1)}][1]\MT[\tilde u^\flat][2]
    +\MT[\pp^{(1)}][2]\MT[\tilde u^\flat][1]
    \right)
    \right]\label{aneq}
    \\
     E^{{\rm cc},(1)}_{\alpha_1\alpha_2} \approx &-\MT[\tilde u^\flat][1] \MT[\pp^\one][3]\MT[m^{\sgs}][2][3] \nonumber
        \\&-\MT[\tilde u^\flat][2]\MT[\pp^\one][3]\MT[m^{\sgs}][1][3]
        \label{msgs} \\
        &+  \T[\pp^{(1)}][3][3][] \T[m^{\sgs}][1][3][]
        + \pp_t^{(1)}\T[m^{\sgs}][1][2],\nonumber
\end{align}
which connects the source-term moment defined in \cref{Ecc_second_moment} to the advective evolution of the SGS stress tensor.
Here \(\T[m^{\sgs\one}][1][2]\) denotes the second raw moment of \(f_{\sgs}^{(1)}\) (see \cref{m_sgs_2,f1_def}).

The relation between the two filter lengths is important in the identity-filter limit.
The homogeneous LES-type spatial filter and the characteristic-induced spatial extent of the kinetic time filter do not need to be identical, but they collapse together in the DNS limit.
Since a consistent coarse graining requires \(\Delta_x\) not smaller than \(c_t\Delta_t\), taking \(\Delta_x\to0\) also sends the kinetic time-filter width to zero at the macroscopic scale.
Then \(m^{\sgs}_{\alpha_1\alpha_2}\to0\), the retained collision-product covariance vanishes, and \cref{msgs} reduces to \(0=0\).
The remaining relation \cref{aneq} gives the resolved Newtonian branch, so the hydrodynamic limit recovers the classical unfiltered NSE.
Here \(\Delta_x\to0\) is only the identity-filter limit, not a CE ordering such as \(\Delta_x/L\sim\sqrt{\Kn}\) in~\cite{ansumali_kinetic_2004}; the LES regime retains finite \(\Delta_x\), while \(\epsilon=\Ma^2/\Re_\ell\) remains the CE parameter.

Even if the separation of \cref{approx2} into \cref{aneq,msgs} is not unique, it is natural because it associates the viscous stress tensor with the molecular diffusivity associated with $\omega^\sharp$ and the added transport SGS terms with the induced covariance.
An alternative disentanglement that yields an explicit SGS dissipation term in the macroscopic equations within the $O(\epsilon^2)$ CE balance is discussed in the appendix of~\cite{marson_kinetic_2025}.

More importantly, \cref{aneq,msgs} are sufficient for the hydrodynamic limit of the FBE to exactly recover the filtered NSE through the $O(\epsilon^2)$ CE balance. The split becomes necessary once the constitutive gauge is fixed by requiring that no additional divergence-free stress term can be reassigned between its two parts.

In fact, if we assume \cref{msgs} holds and inject it into \cref{HLBE}, we obtain:
\begin{equation}
\labelAndRemember{FNSE}{\begin{aligned}
    0 =
\T[\pp][2][2][] \left(
    \bar \rho
        \,\MT[\tilde u^\flat][1] \MT[\tilde u^\flat][2]\right)
        +
        \T[\pp][1][1][]\bar p^\flat
+
\pp_t \left(\bar \rho   \MT[\tilde u^\flat][1]\right)
\\+
\T[\pp][2][2]\left[\bar\rho(\Wtilde{\MT[u^\flat][1]\MT[u^\flat][2]}-\MT[\tilde u^\flat][1]\MT[\tilde u^\flat][2])
\right]
\\
- \frac{1}{\Re}\,
    \MT[\pp][2]
\left[\bar \rho
\left(
    \MT[\pp][1]\MT[\tilde u^\flat][2]
    +\MT[\pp][2]\MT[\tilde u^\flat][1]
\right)
\right]\mathrlap{,}
\end{aligned}}
\end{equation}
which is exactly the (isothermal, incompressible) FNSE, i.e.\ the low-Mach asymptotic form recovered from the present Favre-filtered kinetic description. Here, extending the derivation to the fully compressible case would require retaining the energy equation together with the associated compressibility terms.

\Cref{aneq,msgs} constitute a central analytical result of this work, as they provide the foundation for disentangling filter-scale from subfilter-scale effects.
Formally, the validity of \cref{aneq,msgs} implies that the Newtonian constitutive laws for the stress tensor hold even in the filtered case.
\Cref{msgs} constrains the second moment of the collision-covariance source term: it shows that any physically consistent model for $\overline{\Ecc^\sharp}$ should reproduce the advective evolution of the SGS stress tensor $\MT[m^\sgs][1][2]$. The specific modeling of $\overline{\Ecc^\sharp}$ that satisfies this constraint is developed in \cref{sec:modeling_Ecc}.

\begin{table*}[t]
\caption{Detailed structural comparison between the Filtered Navier--Stokes Equations (FNSE, \cref{FNSE}) and the Filtered Recorrelated BGK (FRBGK--BE, \cref{fbe_bgk,filtered_RBGK2}) framework. The advective closure hypothesis and the diffusive term hypothesis are listed as separate rows, reflecting conceptually independent modeling assumptions. In the FNSE, SGS transport requires an explicit advective closure (Boussinesq), whereas in the FRBGK--BE, the same information is exactly advected inside $f_{\sgs}$ and the closure is relocated entirely to the collision side.}
\label{tab:deep_comparison}
\renewcommand{\arraystretch}{1.4}
\begin{tblr}{
  width=\linewidth,
  colspec={lXX},
  hline{1,2}={solid, 0.08em},
}
Feature & Filtered Navier--Stokes (FNSE) & Filtered Recorrelated BGK (FRBGK--BE) \\
\hline
Equations
    & \cref{FNSE}
    & \cref{fbe_bgk,filtered_RBGK2} \\
Dependent variable
    & $\MT[\tilde u^\flat][1]$, $\bar\rho$
    & $\overline f$ \\
Resolved advection
    & Non-linear: $u_\alpha\partial_\alpha u_\beta$
    & Linear: $\xi_\alpha\partial_\alpha\overline f = \overline{\xi_\alpha\partial_\alpha f}$ \\
SGS advection
    & $\T[\pp][2][2]\left[\bar\rho(\Wtilde{\MT[u^\flat][1]\MT[u^\flat][2]}-\MT[\tilde u^\flat][1]\MT[\tilde u^\flat][2])
\right]$: unresolved by the filtered equation, must be fully modeled
    & $\xi_{\alpha_1}^\flat\MT[\pp][1]f_{\sgs} = \xi_{\alpha_1}^\flat\MT[\pp][1](\overline{f^\zero} -  f^\zero)$: exactly advected as part of $\overline{f}$ by the linear streaming operator \\
SGS stress tensor
    & $\bar\rho(\Wtilde{\MT[u^\flat][1]\MT[u^\flat][2]}-\MT[\tilde u^\flat][1]\MT[\tilde u^\flat][2])$: unresolved and fully unknown
    & $\MT[m^{\sgs}][1][2] = \int_\Xi f_{\sgs}\,\xi^\flat_{\alpha_1}\xi^\flat_{\alpha_2}\,\dd\bm\xi^\flat$: carried as a moment of $f_{\sgs}$ (\cref{m_sgs_2}), directly encoding the same macroscopic Reynolds stress, but not closed by the filtered moments alone \\
Advective closure hypothesis
    & Boussinesq: assumes local equilibrium between production and dissipation to model the advective commutation error as a diffusive term, $\bar\rho(\Wtilde{\MT[u^\flat][1]\MT[u^\flat][2]} - \MT[\tilde u^\flat][1]\MT[\tilde u^\flat][2]) \approx \frac{1}{3}m^{\sgs}_{\alpha_3\alpha_3}\delta_{\alpha_1\alpha_2}-\bar\rho\nu_t^\flat(\partial_{\alpha_1} \MT[\tilde u^\flat][2] + \partial_{\alpha_2}\MT[\tilde u^\flat][1])$, with $\nu_t^\flat\equiv\nu_t/(\mcU\mcL)$
    & Unnecessary: advection is linear and exact; the SGS information is already embedded in $f_{\sgs}$ and exactly streamed \\
Diffusive term hypothesis
    & Newton's constitutive law is assumed valid for the filtered flow; implicitly, the modeling error of the viscous stress does not scale with the filter width
    & Relaxation toward the resolved equilibrium $ f^\zero$ for the resolved BGK; more generally, the filtered equation is closed by the source model $\overline{\Ecc^\sharp}\approx-\omega_t^\sharp f_{\sgs}$, which appears in \cref{filtered_RBGK2} as $-\epsilon^{-1}\omega_t^\sharp f_{\sgs}$ and models the turbulent relaxation of SGS kinetic modes \\
Convergence to FNSE
    & By definition
    & Requires \cref{aneq,msgs}: sufficient conditions for the hydrodynamic limit to recover the filtered NSE \eqref{FNSE} \\
Diffusive term variable
    & $\MT[\tilde u^\flat][1]$, $\bar\rho$
    & $\MT[\tilde u^\flat][1]$, $\bar\rho$ (enter through the resolved equilibrium $ f^\zero$) \\
\hline \hline
\end{tblr}
\end{table*}

The analysis of the present section makes it possible to formalize the structural comparison between the macroscopic and kinetic frameworks introduced qualitatively in Table~\ref{tab:comparison}. Table~\ref{tab:deep_comparison} restates each feature using the notation developed above. In particular, the table distinguishes two conceptually independent modeling hypotheses: the \emph{advective closure}, which addresses the unclosed SGS transport term, and the \emph{diffusive hypothesis}, which concerns the constitutive hypothesis governing the dissipation of momentum.

The diffusive hypothesis enters the two formulations at different levels. In FNSE, one retains Newton's constitutive law after filtering: the resolved viscous stress uses the same local strain-rate relation as the unfiltered flow. The filter-width dependence is then assigned mainly to an advective SGS model. In FRBGK--BE, the CE split \cref{aneq,msgs} first recovers the same Newtonian resolved stress from the molecular BGK relaxation. Because the right-hand side of \cref{msgs} arises from the second-moment projection of kinetic streaming, it represents the transport and production of SGS stress rather than its irreversible dissipation. The source model $\overline{\Ecc^\sharp}\approx-\omega_t^\sharp f_{\sgs}$ addresses the remaining constraint on the collision-covariance source term in \cref{msgs}: its second moment must dissipate the advective SGS stress content carried by $f_{\sgs}=\overline{f^\zero}-f^\zero$. Thus the filtered kinetic closure still recovers the Newtonian resolved stress, but it assigns the unresolved filter-dependent contribution to collision-side relaxation rather than to an eddy-viscosity stress ansatz.

\section{Modeling the collision-covariance source term}
\label{sec:modeling_Ecc}

The hydrodynamic analysis established in \cref{CE} suggests that any physically consistent model for the collision-covariance source term $\overline{\Ecc^\sharp}$ should satisfy the macroscopic condition \cref{msgs}: its second moment, defined in \cref{Ecc_second_moment}, is tied to the advective evolution of the SGS stress tensor $\MT[m^{\sgs}][1][2]$. This result connects the collision-covariance source term with the advective SGS variance encoded by $f_{\sgs}$, motivating the closure strategy developed below.

The proposed framework diverges from classical filter-based kinetic LES techniques. Conventional approaches typically formulate the closure as a filter--collision commutation problem arising from the nonlinearity of the equilibrium distribution. In contrast, the present formulation focuses more holistically on modeling the physical collision-product covariance. Treating the commutation error in isolation limits SGS modeling to a mathematical reconstruction of the filtered collision term rather than a phenomenological representation of the unresolved collision dynamics.

\subsection{BGK closure for the collision-covariance source term}\label{bgk_closure}

The kinetic time filter already samples a characteristic-induced spatial extent (\cref{sec:filtered_nondimensional_rbgk_be}), while the width \(\Delta_x\) of the subsequent homogeneous LES-type spatial filter sets the macroscopic scale on which the retained collision-product covariance is interpreted.
Choosing \(\Delta_x=O(c_t\Delta_t)\) (\cref{eq:kinetic_filter_width}) makes the space and time coarse-graining scales consistent through the RMS thermal speed.
At the chosen coarse-graining widths, the collision-product covariance is retained rather than discarded.
It thereby enters the filtered equation as the unclosed source \(\overline{\Ecc^\sharp}\), which requires closure.
Under the split in \cref{aneq,msgs}, the Chapman--Enskog constraint specifies the first retained second moment of the modeled source:
it must supply the collision-side balance associated with the advective evolution of the SGS stress tensor.
Thus the source model must act on a carrier that contains this SGS stress content.

The filtered-equilibrium decomposition provides the exact SGS equilibrium commutation residual
\[
    f_\sgs=\overline{f^\zero}-f^\zero,
\]
defined in \cref{fsgs_def}.
The zeroth and first moments of $f_\sgs$ vanish by construction (\cref{f_sgs_zero_moment,f_sgs_first_moment}), so using $f_\sgs$ as the SGS carrier in a scalar relaxation does not add local conserved-moment source terms.
At the same time, its second and higher moments contain the unresolved SGS stress and higher-order kinetic content (\cref{f_sgs_hermite}).
This makes \(f_\sgs\) a natural admissible carrier for the CE-constrained second moment of an isothermal source. The scalar model also removes SGS energy as part of the isothermal reduction, unlike the exact elastic collision covariance in \cref{Ecc_def}.
We therefore follow the BGK idea and represent the unresolved collision action as a relaxation of this carrier.

Keeping the relaxation operator general at this stage gives the source-level model
\begin{equation}
    \overline{\Ecc^\sharp}
    \approx
    -\Omega_t^\sharp\circ f_{\sgs}.
\end{equation}
At the CE-coefficient level, this corresponds to
\begin{equation}
    \overline{\Ecc^{\sharp(1)}}
    \approx
    -\left(\Omega_t^\sharp\circ f_{\sgs}\right)^{(1)}.
\end{equation}
Here $\Omega_t^\sharp$ is the unresolved collision-side relaxation operator acting on $f_{\sgs}$.
It must preserve the zero mass and momentum moments of the source, but it need not be scalar.
In the present work we use the BGK-like scalar specialization \(\Omega_t^\sharp \circ f_{\sgs} \approx \omega_t^\sharp f_{\sgs}\), where \(\omega_t^\sharp\) is a constitutive turbulent relaxation frequency.
Unlike the molecular BGK rate \(\omega^\sharp\), it depends on the local SGS state and on the coarse-graining level.
Its operational model is specified later in \cref{turbulent_scaling}.
Accordingly, the source-level closure used in the present work is
\begin{equation}\label{Ebgk_approx}
    \overline{\Ecc^\sharp}
    \approx
    -\omega_t^\sharp f_{\sgs}\,.
\end{equation}
Equivalently, its first CE coefficient satisfies
\begin{equation}\label{Ebgk_coeff_approx}
    \overline{\Ecc^{\sharp(1)}}
    \approx
    -\left(\omega_t^\sharp f_{\sgs}\right)^{(1)}\,.
\end{equation}
Since $f_{\sgs}$ has vanishing zeroth and first velocity moments, the scalar BGK-like unresolved relaxation preserves mass and momentum. Thus, the proposed relaxation targets zero rather than a separate turbulent equilibrium.

Substituting the source closure \cref{Ebgk_approx} into the FRBGK--BE gives
\begin{equation}\labelAndRemember{filtered_RBGK2}{
    \cpartial_t \overline f + \xi^{\flat}_{\alpha}\cpartial_{\alpha} \overline f
    = -\frac{\omega^\sharp}{\epsilon}\,\overline{f^\Neq}
    -\frac{\omega_t^\sharp}{\epsilon}\, f_{\sgs}}\,,
\end{equation}
where the filtered fine-grained nonequilibrium is
\begin{equation}\label{eq:filtered_fneq_definition}
    \overline{f^\Neq}
    \equiv
    \overline{f} - \overline{f^{\zero}}
    =
    f^\cNeq - f_\sgs
    \approx
    \epsilon\overline{f^\one}.
\end{equation}
This quantity is distinct from the coarse-level nonequilibrium \(f^\cNeq = \overline f - f^\zero\).

At this stage, \(\omega_t^\sharp\) remains a constitutive input. Its operational closure, detailed in \cref{turbulent_scaling}, uses the diffusive-scaling ($\sharp$) SGS eddy-viscosity estimate \(\tilde \theta^\sharp / \omega_t^\sharp = \nu_t^\sharp \propto \Delta_x^\sharp\sqrt{k_\sgs^\sharp}\).
The dimensional lattice counterpart is given in Appendix~\ref{app:omega_t_operational}.

\paragraph{The resolved BGK limit.} \Cref{filtered_RBGK2} recovers the standard resolved-BGK collision when the SGS and molecular relaxation rates are set equal, \(\omega_t^\sharp=\omega^\sharp\). To see this, use \(f_\sgs=\overline{f^\zero}-f^\zero\). The collision term then reduces to
\begin{align}
    -\frac{\omega^\sharp}{\epsilon}\,\overline{f^\Neq}
    -\frac{\omega^\sharp}{\epsilon}\, f_{\sgs}
    &= -\frac{\omega^\sharp}{\epsilon}\left( \overline{f} - \overline{f^{\zero}} \right)
    -\frac{\omega^\sharp}{\epsilon}\left( \overline{f^{\zero}} - f^{\zero} \right) \nonumber\\
    &= -\frac{\omega^\sharp}{\epsilon}\left( \overline{f} - f^{\zero} \right) .
\end{align}
Thus the resolved BGK limit can be read as an implicit SGS closure~\cite{yen_kinetic_1972,marson_kinetic_2025}. Linear streaming carries the SGS advective content in \(\overline f\), while the BGK collision damps \(f_{\sgs}\) at the molecular rate. This interpretation also shows the limitation of the resolved-BGK limit. The SGS damping is fixed by the molecular BGK prefactor, not by the unresolved state. Therefore, this limit does not define a general SGS closure.

\begin{figure}
        \centering
        \includegraphics[width=.9\linewidth]{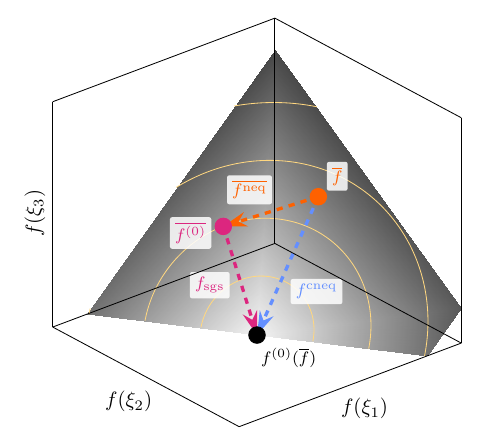}
        \caption{Visual qualitative representation of the combined equilibration and isothermal heat removal in a simplified 3-component Hilbert space. The gray plane depicts the manifold of conserved filtered moments ($\bar{\rho}, \tilde{u}^\flat$). The blue dashed line is the coarse-level nonequilibrium $f^{\cNeq} = \overline f - f^\zero$. The orange dashed line is the filtered fine-grained nonequilibrium $\overline{f^\Neq} = \overline f - \overline{f^\zero}$. The magenta dashed line is the SGS equilibrium commutation residual $f_\sgs = \overline{f^\zero} - f^\zero$.}
        \label{fig:entropic_descent}
    \end{figure}

\paragraph{MaxEnt interpretation.} The closure \cref{filtered_RBGK2} admits a direct information-theoretic interpretation~\cite{shannon_mathematical_1948,jaynes_information_1957} independent of the specific hydrodynamic regime (DNS, LES, RANS) based on the principle of maximum entropy (MaxEnt)~\cite{marson_kinetic_2025}. This interpretation is illustrated schematically in \cref{fig:entropic_descent}.

Let \(f_E^{(0)}\) be an auxiliary Maxwellian with the same mass, momentum, and total energy as \(\overline{f^{(0)}}\). Then
\[
    f_{\sgs}
    = \left(\overline{f^{(0)}}-f_E^{(0)}\right)
    + \left(f_E^{(0)}-f^{(0)}\right).
\]
Applying the homogeneous LES-type spatial filter to the time-filtered state \(f\) introduces two equilibrium constructions. The first, $f_E^{(0)}$, is the coarse-grained equilibrium: by \cref{equilibrium}, it is the Maxwellian that maximizes the same local Boltzmann--Shannon entropy functional, but under the filtered mass, momentum, and total-energy constraints. Filtering changes the constraints, not the MaxEnt functional. Being the least constrained state compatible with the filtered moments, it occupies the entropy maximum under these constraints. The second, $\overline{f^{(0)}(f)}$, is the filtered fine-grained equilibrium: it retains additional SGS information and is therefore generally not the entropy maximizer under the coarse constraints alone.

Geometrically, the blue segment in \cref{fig:entropic_descent} is the coarse-level nonequilibrium \(f^\cNeq = \overline f - f^\zero\). It splits into the orange filtered fine-grained nonequilibrium \(\overline{f^\Neq} = \overline f - \overline{f^\zero}\) and the magenta SGS equilibrium commutation residual \(f_\sgs = \overline{f^\zero} - f^\zero\). The energy-conserving contribution $-\epsilon^{-1}\omega_t^\sharp \left(\overline{f^{(0)}}-f_E^{(0)}\right)$ in \cref{filtered_RBGK2} therefore represents an entropic descent from the information-rich, lower-entropy fine-grained state $\overline{f^{(0)}}$ toward the higher-entropy coarse-grained state $f_E^{(0)}$. In the present isothermal case, the remaining contribution \(-\epsilon^{-1}\omega_t^\sharp(f_E^{(0)}-f^{(0)})\) represents the heat removal required to maintain \(\tilde\theta^\sharp\). Thus, \cref{fig:entropic_descent} shows the combined effect of equilibration and heat removal.

\paragraph{Relation to corrected BGK models.}
At the operator level, \cref{filtered_RBGK2} is structurally close to corrected BGK models that modify the relaxation path, such as Shakhov's heat-flux correction~\cite{shakhov1968generalization}. The analogy is only formal. Shakhov-type models alter molecular relaxation to recover gas-kinetic transport coefficients. Here, the SGS term \(-\epsilon^{-1}\omega_t^\sharp f_{\sgs}\) instead relaxes an SGS carrier with zero mass and momentum moments.
This also separates the present closure from Chen et al.'s Klimontovich RANS formulation~\cite{chen_average_2023,chen_average_2024}. There, the unresolved term relaxes toward a turbulent Gaussian and introduces a turbulent temperature. Here, \cref{filtered_RBGK2} dissipates SGS stress and higher-order SGS content without introducing a separate turbulent equilibrium; see Appendix~\ref{app:chen_comparison}.

\subsection{LES-regime specialization of the CE constraint}
\label{subsec:Ecc_CE_consistency}

The generic second-moment constraint is already fixed by \cref{msgs}. We now specialize the CE with the coefficient-level model \cref{Ebgk_coeff_approx}. The modeled projected contribution must satisfy the same zeroth-order compatibility condition \eqref{fce0}:
\begin{equation}\label{E_cc_0}
    E^{{\rm cc},(0)}_{\alpha_1\alpha_2}=0,
    \qquad
    -\left(\omega_t^\sharp \T[m^{\sgs}][1][2]\right)^{\zero}
    = -\omega_t^{\sharp\zero}\T[m^{\sgs\zero}][1][2]
    = 0\,.
\end{equation}
At first order, the modeled second-moment source is obtained by expanding the product \(\omega_t^\sharp \T[m^{\sgs}][1][2]\):
\begin{equation}\label{Edecomposition}
    E^{{\rm cc},(1)}_{\alpha_1\alpha_2} \approx -\left(\omega_t^\sharp \T[m^{\sgs}][1][2]\right)^{\one} \approx -\omega_t^{\sharp\zero} \T[m^{\sgs\one}][1][2] - \omega_t^{\sharp\one} \T[m^{\sgs\zero}][1][2]\,.
\end{equation}
The full CE contribution is therefore \(\epsilon E^{{\rm cc},(1)}_{\alpha_1\alpha_2}+O(\epsilon^2)\).
The zeroth-order condition \eqref{E_cc_0} leaves two admissible scalings:
\begin{enumerate}
    \item $\T[m^{\sgs\zero}][1][2] = 0,\ \omega_t^{\sharp\zero}\neq 0$: the leading-order SGS stress tensor vanishes in the filtered Euler equations; there is no leading-order advective SGS transport and the laminar Euler equation is recovered in this regime.
    \item $\T[m^{\sgs\zero}][1][2] \neq 0,\ \omega_t^{\sharp\zero}= 0$: the unresolved (RANS, LES) regime of primary interest. Here consistency with the CE hierarchy requires $\omega_t^{\sharp\zero}=0$, so \cref{Edecomposition} reduces to
\begin{equation}\label{Ei}
E^{{\rm cc},(1)}_{\alpha_1\alpha_2} \approx -\omega_t^{\sharp\one}\T[m^{\sgs\zero}][1][2]\,.
\end{equation}
\end{enumerate}

Substituting \cref{Ei} into \cref{msgs} gives the specialized LES constraint:
\begin{alignedEq}\label{msgs_modeled}
     -{\omega_t^{\sharp\one}}\MT[m^{\sgs\zero}][1][2] &\approx -\MT[\tilde u^\flat][1] \MT[\pp^\one][3]\MT[m^{\sgs\zero}][2][3]
        -\MT[\tilde u^\flat][2]\MT[\pp^\one][3]\MT[m^{\sgs\zero}][1][3]
        \\&\quad +  \T[\pp^{(1)}][3][3][] \T[m^{\sgs\zero}][1][3][]
        + \pp_t^{(1)}\T[m^{\sgs\zero}][1][2],
\end{alignedEq}
which identifies the admissible LES scaling of the model: the SGS carrier $f_{\sgs}$ supports the required second moment, while the constitutive relaxation enters through $\omega_t^{\sharp\one}$.

Within the present disentanglement, $\omega_t^{\sharp\one}$ governs the irreversible collision-side dissipation.
In the LES regime, the limit \(\omega_t^{\sharp\one}\to0\) removes this collision-side dissipation.
The SGS transport then approaches a reversible collisionless limit, so cascading energy is no longer efficiently thermalized and may accumulate spectrally~\cite{marson_kinetic_2025}.

\section{Kinetic closures of turbulence}
\label{sec:kinetic_closures}

Having established the source-term model in \cref{sec:modeling_Ecc}, we now turn
\[\recallEqAndTag{filtered_RBGK2} \]
into operational closures.
The resolved equilibrium \(f^\zero=f^\zero(\bar\rho,\tilde{\bm u},\tilde\theta)\) is built from the filtered moments, so the coarse-level nonequilibrium \(f^\cNeq=\overline f-f^\zero\) is directly available.
The molecular BGK term in \cref{filtered_RBGK2} also needs the filtered fine-grained nonequilibrium \(\overline{f^\Neq}=\overline f-\overline{f^\zero}\).
We approximate this filtered fine-grained nonequilibrium by its first-order CE contribution, \(\overline{f^\Neq}\approx\epsilon\overline{f^\one}\), as in \cref{eq:filtered_fneq_definition}.
Then \cref{approx1} gives the split
\begin{equation}\label{eq:coarse_residual_split}
    f^\cNeq = \epsilon\overline{f^\one} + f_\sgs + O(\epsilon^2).
\end{equation}
Under the constitutive split in \cref{aneq}, the second moment of \(\epsilon\overline{f^\one}\) gives the Newtonian viscous stress reconstructed from the gradients of the filtered velocity moments.
Multiplying \cref{aneq} by \(\epsilon=\Ma^2/\Re_\ell\) and using \cref{eq:Re_tau_relation} gives the shared moment estimate
\begin{equation}\label{eq:eps_m_one_newtonian}
    \epsilon \overline{m^\one_{\alpha_1 \alpha_2}}
    \approx
    -\frac{\bar{\rho}}{\Re}
    \left(\pp_{\alpha_2}^\flat\tilde u_{\alpha_1}^\flat
    +\pp_{\alpha_1}^\flat\tilde u_{\alpha_2}^\flat\right)
    =
    -\frac{2\bar{\rho}}{\Re}\MT[\tilde S^\flat][\alpha_1][\alpha_2],
\end{equation}
where $\MT[\tilde S^\flat][\alpha_1][\alpha_2] = \frac{1}{2}(\pp_{\alpha_2}^\flat \tilde u_{\alpha_1}^\flat + \pp_{\alpha_1}^\flat \tilde u_{\alpha_2}^\flat)$.
The closures below use this relation to estimate the filtered first-order CE contribution \(\epsilon\overline{f^\one}\).
The way this estimate is projected or completed distinguishes the different closure variants.
Once this CE contribution is estimated, \cref{eq:coarse_residual_split} gives the closure-specific SGS-carrier estimate, \(f_\sgs\approx f^\cNeq-\epsilon\overline{f^\one}\).
Its second and higher moments contain the unresolved SGS stress and higher-order SGS kinetic content (see \cref{f_sgs_hermite}).
Operational closure therefore requires two choices: a closure-specific approximation to \(\epsilon\overline{f^\one}\), which defines the corresponding estimate of \(f_\sgs\) through \cref{eq:coarse_residual_split}, and a turbulent relaxation coefficient \(\omega_t^\sharp\) for the collision-side source in \cref{Ebgk_approx,filtered_RBGK2}.

We first define the common relaxation-frequency estimate.
The subsequent subsections define three SGS-carrier approximations: the Residual-Based Closure (KC-RB), the Moment-Projected Closure (KC-MP), and the Recursive Kinetic Closure (KC-RR).
This section gives their continuum formulation.
Appendix~\ref{app:hermite_representation} summarizes the Hermite representation used below, and Appendix~\ref{app:lbm_discretization} gives the discrete formulas used in the validation cases.

\subsection{Turbulent relaxation frequency}
\label{turbulent_scaling}
The turbulent relaxation frequency $\omega_t^\sharp$ is the constitutive rate that damps the SGS carrier \(f_\sgs\) in \cref{filtered_RBGK2}.
We therefore need a local estimate for this rate.
The present closure does not introduce a separate turbulence transport equation, because the SGS information is already carried by \(\overline f\) and transported by the linear streaming operator.
Instead, we use a Prandtl-type eddy-viscosity argument: first extract a local SGS kinetic-energy amplitude from the evolving kinetic moments, then convert the resulting SGS viscosity into a relaxation frequency.

For the exact SGS stress tensor, the trace \(m^{\sgs}_{\alpha_1\alpha_1}\) is associated with the SGS kinetic-energy density.
In the reconstructed closures, however, this trace is not necessarily positive.
We therefore use only its scalar magnitude.
Since the SGS second moment in \cref{m_sgs_2} is a convective raw moment, the corresponding thermal-unit amplitude is
\begin{equation}\label{eq:ksgs_sharp_trace}
    k_\sgs^\sharp
    \approx
    \frac{\Ma^2}{2\bar\rho}
    \left|m^{\sgs}_{\alpha_1 \alpha_1}\right|.
\end{equation}

The corresponding eddy-viscosity estimate, written in the same diffusive \((\sharp)\) scaling as \(\omega_t^\sharp\), is
\begin{equation}\label{eq:nu_t_sharp_closure}
    \nu_t^\sharp = C_\nu \Delta_x^\sharp \sqrt{k_\sgs^\sharp},
\end{equation}
where \(\Delta_x^\sharp \equiv \Delta_x/\ell\) and \(C_\nu\) is the empirical closure coefficient.
The viscosity--relaxation relation in \cref{eq:Re_tau_relation}, applied to the SGS viscosity, then gives the source relaxation frequency
\begin{equation}\label{eq:omega_t_sharp_closure}
    \omega_t^\sharp = \frac{\tilde{\theta}^\sharp}{\nu_t^\sharp}\,.
\end{equation}

Finally, we bound the SGS relaxation by the molecular BGK relaxation and use \(\min(\omega^\sharp,\omega_t^\sharp)\) in the modeled equation.
This bound recovers the standard BGK model when \(f_{\sgs}\) and \(k_\sgs^\sharp\) vanish in the DNS limit.
The filtered kinetic evolution equation, obtained from \cref{fbe_bgk,filtered_RBGK2}, reads
\begin{equation}\label{eq:full_eff_bgk}
    \cpartial_t \overline f + \xi^{\flat}_{\alpha}\cpartial_{\alpha} \overline f
    = -\frac{\omega^\sharp}{\epsilon}\,\left({\overline f- \overline{f^{\zero }}}\right)
    -\frac{1}{\epsilon}\min\!\left(\omega^\sharp, \omega_t^\sharp\right)\, f_{\sgs}\,.
\end{equation}
where \(\overline{f^{\zero }} = f^\zero + f_\sgs\).

For implementation, the same closure can be written directly with dimensional derivatives and relaxation frequencies:
\begin{equation}
    \partial_t \overline f + \xi_{\alpha}\partial_{\alpha} \overline f
    = -\omega\,\left({\overline f- \overline{f^{\zero }}}\right)
    - \min\!\left(\omega, \omega_t\right)\, f_{\sgs}\,,
\end{equation}
where \(\omega_t = \tilde\theta/\nu_t\). The exact SGS equilibrium commutation residual \(f_\sgs = \overline{f^\zero}-f^\zero\) serves as the SGS carrier and is approximated through a closure-specific reconstruction.
The corresponding lattice implementation used in the simulations employed for the validations in \cref{sec:test_cases} is given in Appendix~\ref{app:omega_t_operational}.

\subsection{Residual-Based Closure (KC-RB)}
\label{subsec:KCRB_theory}

The KC-RB model, originally proposed in~\cite{marson_kinetic_2025,marson_kinetic_2025_1}, is the simplest realization of \cref{filtered_RBGK2} considered here.
It reconstructs the filtered first-order CE contribution $\epsilon\overline{f^\one}$ by projecting its viscous second-order moment onto the Hermite basis. The SGS carrier is then estimated from the remainder of the coarse-level nonequilibrium:
\begin{equation}\label{eq:f_sgs_KCRB}
    f_\sgs \approx \overline{f} - f^\zero - \epsilon\overline{f^\one},
\end{equation}
In detail, substituting the Newtonian estimate \cref{eq:eps_m_one_newtonian} into the truncated Hermite expansion \cref{eq:second_order_ce_reconstruction} (see Appendix~\ref{app:hermite_representation}) gives
\begin{equation}\label{eq:f1_KCRB}
    \epsilon\overline{f^\one} \approx -\frac{\Ma^2\bar{\rho}\, w^\flat}{\rho_R\theta_R^{\sharp 2}\Re} H^\sharp_{\alpha_1 \alpha_2} \MT[\tilde S^\flat][\alpha_1][\alpha_2],
\end{equation}
where $w^\flat$ is the weighting function defined in \cref{w}~\cite{shan_2006_hydrodynamics-navier-stokes}, and
\(
    \MT[\tilde S^\flat][\alpha_1][\alpha_2]
    =
    \frac{1}{2}
    \left(\pp_{\alpha_2}^\flat \tilde u_{\alpha_1}^\flat
    + \pp_{\alpha_1}^\flat \tilde u_{\alpha_2}^\flat\right).
\)
Thus the KC-RB estimate of \(\epsilon\overline{f^\one}\) is obtained by computing the gradients of the filtered velocity.
Under this construction, the KC-RB estimate of \(f_\sgs\) contains both the SGS equilibrium commutation residual and fine-grained nonequilibrium content not represented by the second-order Navier--Stokes reconstruction.
\paragraph{Implementation choice.}
The LBM implementation used for the validation in \cref{sec:test_cases} retains the raw incompressible second-order reconstruction from~\cite{marson_kinetic_2025}, rather than the Hermite-consistent form in \cref{eq:f1_KCRB}. Its weak-compressibility defect and numerical effect are discussed in \cref{subsec:tgv}.

\subsection{Moment-Projected Closure (KC-MP)}
\label{subsec:KCMP_theory}

The KC-MP model represents the SGS carrier \(f_\sgs\) through its second-order Hermite projection. Let $m^{\cNeq}_{\alpha_1 \alpha_2}$ denote the second-order raw moment of the coarse-level nonequilibrium \(f^\cNeq = \overline f - f^\zero\):
\begin{equation}\label{eq:m1_KCMP}
    m^{\cNeq}_{\alpha_1 \alpha_2} \equiv \int_{\Xi} (\overline{f} - f^\zero) \xi^\flat_{\alpha_1} \xi^\flat_{\alpha_2} \, \dd\bm{\xi}^\flat.
\end{equation}
Using the resolved viscous stress contribution in \cref{eq:eps_m_one_newtonian}, define the common reconstructed second-order CE reference stress as
\begin{equation}\label{eq:m_ce_reference}
    m^{\rm CE}_{\alpha_1\alpha_2}
    \equiv
    - \frac{2\bar{\rho}}{\Re} \MT[\tilde S^\flat][\alpha_1][\alpha_2]
    \approx
    \epsilon \overline{m^\one_{\alpha_1\alpha_2}}.
\end{equation}
Within the second-order truncation,
\(
    m^{\cNeq}_{\alpha_1\alpha_2}
    \approx
    m^{\rm CE}_{\alpha_1\alpha_2}
    +
    m^{\sgs}_{\alpha_1\alpha_2}
\),
the model obtains the SGS second-order moment by subtracting the CE reference stress from the second-order moment of the coarse-level nonequilibrium:
\begin{equation}\label{eq:m_sgs_KCMP}
    m^{\sgs}_{\alpha_1 \alpha_2}
    \approx
    m^{\cNeq}_{\alpha_1 \alpha_2}
    -
    m^{\rm CE}_{\alpha_1\alpha_2}.
\end{equation}
Since $f_\sgs$ has vanishing zeroth and first moments, its second-order Hermite reconstruction follows the same projection as \cref{eq:second_order_ce_reconstruction}, with $\epsilon\overline{m^\one_{\alpha_1\alpha_2}}$ replaced by $m^{\sgs}_{\alpha_1\alpha_2}$:
\begin{equation}\label{eq:f_sgs_KCMP}
    f_\sgs \approx \frac{\Ma^2 w^\flat}{2\rho_R\theta_R^{\sharp 2}}  H^\sharp_{\alpha_1 \alpha_2} m^{\sgs}_{\alpha_1 \alpha_2}.
\end{equation}
Under this construction, moments above second order do not contribute to the KC-MP reconstruction of $f_\sgs$.
Appendix~\ref{app:kcmp_operational} gives the explicit discrete implementation.

\subsection{Recursive Kinetic Closure (KC-RR)}
\label{subsec:KCRR_theory}

The KC-RR closure generalizes the projection philosophy of KC-MP by retaining the SGS second-order stress and adding higher-order departures from the recursive-regularized reference manifold. By design, the recursive regularization model (RR)~\cite{malaspinas_increasing_2015} reconstructs a reference set of convectively scaled nonequilibrium Hermite coefficients $a^{\rm rr}_{\alpha_1 \dots \alpha_n}$ from lower-order hydrodynamic moments. The recursive relations~\cite{malaspinas_increasing_2015} are derived for an unfiltered, fully resolved Chapman--Enskog description. In a fully resolved laminar regime, RR removes components outside the corresponding CE manifold, which are predominantly ghost modes and higher-order discretization errors. In a filtered turbulent distribution, however, the higher-order departures replaced by RR also carry SGS content. Standard RR has no independent SGS carrier and therefore cannot preserve that content.

In the present LES context, however, the filtered distribution $\overline{f}$ contains both this resolved macroscopic state and the SGS equilibrium commutation residual $f_\sgs$, whose moments encode unresolved fluctuations. The KC-RR approach repurposes the RR procedure: rather than using it simply to replace higher-order departures by their fully resolved reference values (as in classical regularization), it uses the RR reference manifold to actively partition the moments into a reference component and an estimated unresolved (fluctuating) component (possibly including reconstruction and discretization errors).
Throughout this subsection, \(a^{\cNeq}_{\alpha_1\dots\alpha_n}\) denotes the order-\(n\) convectively scaled Hermite coefficient (see Appendix~\ref{app:hermite_representation}) of the coarse-level nonequilibrium \(f^\cNeq\equiv\overline f-f^\zero\).
Because \(f^\cNeq\) has vanishing density and momentum moments, its second- and third-order Hermite coefficients coincide with the corresponding raw moments; in particular, \(a^{\cNeq}_{\alpha_1\alpha_2}=m^{\cNeq}_{\alpha_1\alpha_2}\).
This coincidence does not generally extend to fourth and higher orders.

For the second-order moment ($n=2$), KC-RR uses the same CE reference stress in \cref{eq:m_ce_reference}; its RR seed is the corresponding Hermite coefficient,
\begin{equation}
    a^{\rm rr}_{\alpha_1 \alpha_2}
    =
    m^{\rm CE}_{\alpha_1\alpha_2}.
\end{equation}
The reconstructed SGS coefficient is therefore
\begin{equation}
    a^{\sgs}_{\alpha_1 \alpha_2}
    \equiv
    a^{\cNeq}_{\alpha_1 \alpha_2}-a^{\rm rr}_{\alpha_1 \alpha_2}
    =
    m^{\cNeq}_{\alpha_1 \alpha_2}
    -
    m^{\rm CE}_{\alpha_1\alpha_2}
    \approx
    m^{\sgs}_{\alpha_1 \alpha_2}.
\end{equation}
For higher orders ($n \ge 3$), the RR reference coefficients are generated recursively via the isothermal RR relationship \cite{malaspinas_increasing_2015}:
\begin{equation}\labelAndRemember{eq:kcrr_rr}{
\begin{aligned}
    a^{\rm rr}_{\alpha_1 \dots \alpha_n}
    &= \tilde{u}^\flat_{\alpha_n} a^{\rm rr}_{\alpha_1 \dots \alpha_{n-1}} \\
    &\quad + \frac{1}{\bar{\rho}} \sum_{i=1}^{n-1}
    a^{\zero}_{\alpha_1 \dots \alpha_{i-1}\alpha_{i+1} \dots \alpha_{n-1}}
    a^{\rm rr}_{\alpha_i \alpha_n}
\end{aligned}
},
\end{equation}
where $a^{\zero}$ are the standard convectively scaled equilibrium Hermite coefficients. The coefficients $a^{\cNeq}_{\alpha_1 \dots \alpha_n}$ of the coarse-level nonequilibrium are obtained by projecting $f^\cNeq$ onto the Hermite basis.
The estimated higher-order SGS coefficients are then defined as the difference from the RR manifold:
\begin{equation}\label{eq:kcrr_sgs}
    a^{\sgs}_{\alpha_1 \dots \alpha_n} \equiv a^{\cNeq}_{\alpha_1 \dots \alpha_n} - a^{\rm rr}_{\alpha_1 \dots \alpha_n},
    \qquad n \ge 3.
\end{equation}
The KC-RR SGS-carrier estimate can then be reconstructed as
\begin{equation}\label{eq:f_sgs_KCRR}
    f_\sgs
    \approx
    \frac{w^\flat}{\rho_R}
    \left[
    \frac{\Ma^2}{2\theta_R^{\sharp 2}}
    H^\sharp_{\alpha_1\alpha_2}a^{\sgs}_{\alpha_1\alpha_2}
    +
    \sum_{n=3}^{\infty}
    \frac{\Ma^n}{n!\theta_R^{\sharp n}}
    H^\sharp_{\alpha_1\dots\alpha_n}
    a^{\sgs}_{\alpha_1\dots\alpha_n}
    \right].
\end{equation}
Appendix~\ref{app:kcrr_operational} gives the explicit discrete implementation.

\section{Test cases}
\label{sec:test_cases}

We assess the theoretical developments with lattice--Boltzmann simulations. The discrete realization of the closures, including the turbulent relaxation rate and the KC-RB, KC-MP, and KC-RR algorithms, is given in Appendix~\ref{app:lbm_discretization} and Appendices~\ref{app:omega_t_operational}--\ref{app:kcrr_operational}. The numerical study considers three problems of increasing complexity: the three-dimensional Taylor--Green vortex (TGV)~\cite{brachet_small-scale_1983}, the lid-driven cavity flow~\cite{leriche_direct_2000,Hegele_2018}, and the flow past a circular cylinder~\cite{williamson_vortex_1996,parnaudeau_experimental_2008}. The TGV and cavity provide quantitative benchmarks for unforced turbulent transition and wall-bounded recirculation, respectively. The cylinder case instead provides an exploratory consistency test of an obstacle-generated separated turbulent wake. Its wall resolution is insufficient for a grid-converged assessment of boundary-layer separation and force prediction, so this case demonstrates applicability to a more complex flow rather than comparative accuracy.

The closure variants are first prototype realizations of the proposed framework. The simulations provide an initial numerical assessment of the framework through its discrete realizations, not a systematic accuracy or performance ranking of fully calibrated and optimized closures. Dedicated studies of coefficient calibration, grid and boundary sensitivity, performance, and numerical optimization are left to future work.

We compare the proposed closures with the naive LBM implementation of the Smagorinsky model~\cite{malaspinas_consistent_2012} (Smag.\ Naive), which reconstructs the strain-rate magnitude locally from the deviatoric second-order moment \(m^{\cNeq}_{\alpha_1\alpha_2}\) of the coarse-level nonequilibrium, a Smagorinsky variant that uses a finite-difference stencil to compute the strain-rate tensor (Smag.\ FD),
and a simple resolved BGK model (BGK).
In the LBM literature, the resolved BGK collision constructed from the resolved macroscopic fields is commonly labeled simply BGK, thus we retain this conventional label in the figures.
Within the present framework, however, the same resolved-BGK dynamics is interpreted as the simplest implicit, non-dynamic SGS model: it relaxes the SGS carrier $f_\sgs$ at the molecular relaxation frequency rather than at an independently modeled turbulent frequency (see the resolved-BGK limit in \cref{bgk_closure}).
We also compare with the recursive regularized collision model~\cite{malaspinas_increasing_2015} (RR) and its hybrid recursive regularized version with regularization parameter $\sigma = 0.97$~\cite{jacob_new_2018} (HRR).

The KC-RB curves retain the legacy raw reconstruction in \cref{eq:kcrb_operational_raw_fone}, proposed in the original KC-RB implementation~\cite{marson_kinetic_2025}. Appendix~\ref{app:kcrb_reconstruction_sensitivity} isolates the raw-versus-Hermite reconstruction choice in a matched $N=64$ TGV calculation. Therefore, the direct KC-RB/KC-RR comparisons in the cavity and cylinder combine the raw KC-RB implementation with the different KC-RR recursive construction.

In the present validations, the coefficients $C_\nu$ and $C_s$ are empirical stability and dissipation parameters. Following the stability-limit procedure used in~\cite{marson_kinetic_2025}, KC-RB and Smag.\ FD were first brought close to their minimum stable values on the under-resolved $32\times32\times32$ TGV by decreasing $C_\nu$ and $C_s$ in steps of \(\Delta C_{\nu}=\Delta C_{s}=0.005\). The reported values were then set slightly above that limit: KC-RB uses $C_\nu=0.02$ instead of the earlier $C_\nu=0.015$ in~\cite{marson_kinetic_2025}, KC-RR is kept at the same $C_\nu=0.02$, and the Smagorinsky comparisons use $C_s=0.11$ instead of the earlier $C_s=0.105$ in~\cite{marson_kinetic_2025}. KC-MP uses $C_\nu=0.035$, chosen to give roughly similar dissipative behavior to KC-RB in the TGV.
 These values are then kept fixed for the lid-driven cavity and cylinder cases.
Systematic calibration, sensitivity studies, and a kinetic determination of these coefficients are left to future work.

As distinguished in \cref{tab:notation}, angle brackets in the reported statistics denote post-processing time averages over the stated sampling windows. They are used only for statistical diagnostics and are distinct from the model-defining filter operator $\langle \cdot \rangle_{\Delta_t}$.
In this section we drop the theoretical diffusive/convective superscripts $\sharp$ and $\flat$ and use the standard benchmark dimensional notation. Quantities are normalized explicitly by the reference scales of each case whenever a normalized form is reported. Within each benchmark, $u_0$ serves as the operational reference velocity. Quantities reported without explicit normalization are given in lattice units.

\subsection{Taylor--Green vortex}
\label{subsec:tgv}
The Taylor--Green vortex (TGV) is a standard benchmark for assessing numerical dissipation in an under-resolved periodic flow without physical boundaries \cite{brachet_small-scale_1983}. It consists of a periodic domain initialized with an anisotropic, deterministic flow field containing large-scale vortical structures. Following the standard weakly compressible formulation, the initial velocity field $(u_x, u_y, u_z)$ is prescribed in normalized form as:
\begin{align}
    \frac{u_x}{u_0} &= \sin X \cos Y \cos Z \,, \\
    \frac{u_y}{u_0} &= - \cos X \sin Y \cos Z \,, \\
    \frac{u_z}{u_0} &= 0 \,,
\end{align}
where $X = 2\pi x/L_{\rm box}, Y = 2\pi y/L_{\rm box}, Z = 2\pi z/L_{\rm box}$, and $L_{\rm box}$ is the periodic box length. The solver uses $L_{\rm TGV} = L_{\rm box}/(2\pi)$ as the reference length, so $\Re = u_0 L_{\rm TGV}/\nu$. Here $u_0$ is the characteristic velocity amplitude used to normalize the case.
The initial density is initialized from the standard TGV pressure fluctuation
$p_{\rm fluc}\propto [\cos(2X)+\cos(2Y)][\cos(2Z)+2]$, which balances the nonlinear advective acceleration of the prescribed velocity field at $t=0$.
 Driven by non-linear advection and vortex stretching, the flow undergoes unforced transition and subsequently decays viscously. The time evolution of the kinetic energy and enstrophy then provides a convenient basis for comparing the dissipative behavior of the different closures.

The post-processing uses the volume averages
\begin{equation}
    \begin{aligned}
        E_k &= \frac{1}{V}\int_{\Omega_{\rm TGV}} \frac{1}{2}\rho\,u_\alpha u_\alpha\,\dd\bm{x}, \\
        \mathcal{Z} &= \frac{1}{V}\int_{\Omega_{\rm TGV}} \frac{1}{2}\omega_\alpha\omega_\alpha\,\dd\bm{x},
        \qquad \bm{\omega} = \nabla \times \bm{u},
    \end{aligned}
    \label{eq:tgv_energy_enstrophy}
\end{equation}
where $V = |\Omega_{\rm TGV}|$ is the volume of the computational domain $\Omega_{\rm TGV}$. In the discrete diagnostics, these integrals are arithmetic averages over fluid lattice nodes. The vorticity is computed with a sixth-order centered finite-difference stencil, and the reported quantities are $E_k/u_0^2$ and $\mathcal{Z}/(u_0/L_{\rm TGV})^2$.

\begin{figure*}
    \includegraphics[width=\textwidth]{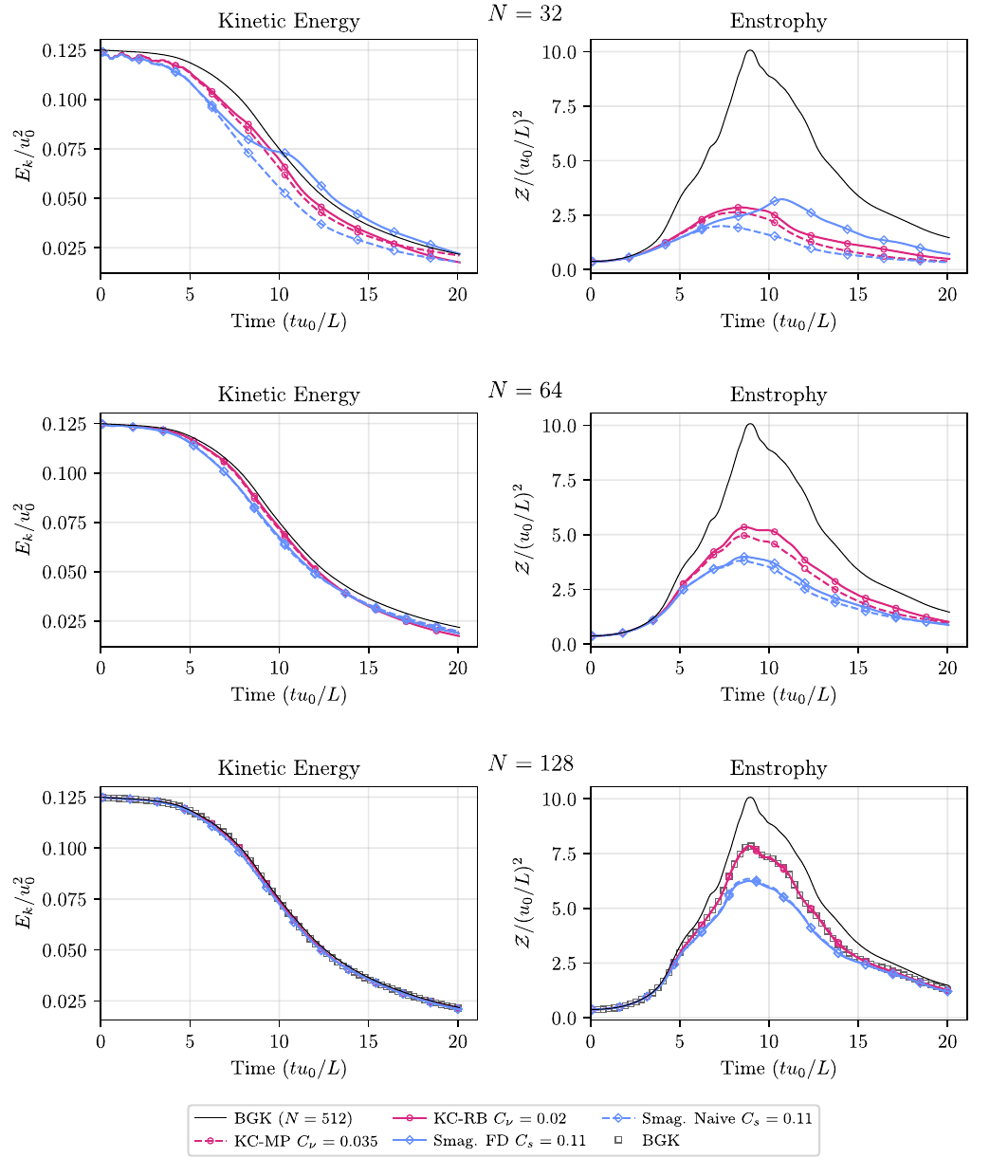}
    \caption{Time evolution of the normalized kinetic energy $E_k/u_0^2$ (left) and enstrophy $\mathcal{Z}/(u_0/L_{\rm TGV})^2$ (right) for the Taylor--Green vortex at $\Re = 1600$ and $\Ma = 0.2$. Comparison between the high-resolution resolved-BGK reference simulation (labeled BGK, $N=512$), the residual-based closures (KC-RB and KC-MP), and the classical Smagorinsky models at varying resolutions.}
    \label{fig:tgv-RB}
\end{figure*}

\begin{figure*}
    \includegraphics[width=\textwidth]{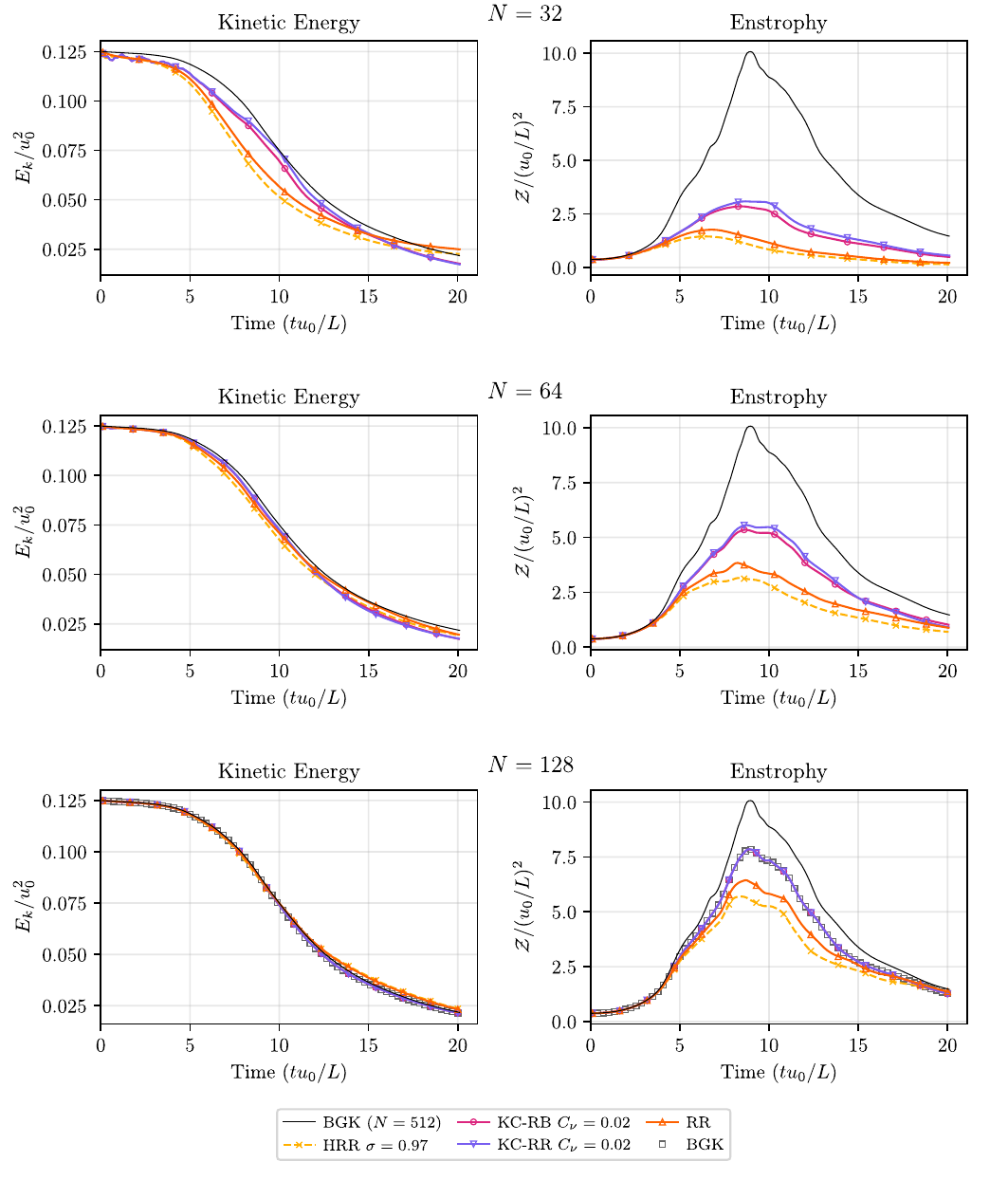}
    \caption{Time evolution of the normalized kinetic energy and enstrophy for the Taylor--Green vortex ($\Re = 1600$). Comparison between the recursive kinetic closure (KC-RR) and the standard recursive regularization (RR) and hybrid recursive regularization (HRR) models.}
    \label{fig:tgv-RR}
\end{figure*}

\begin{figure*}
    \includegraphics[width=\textwidth]{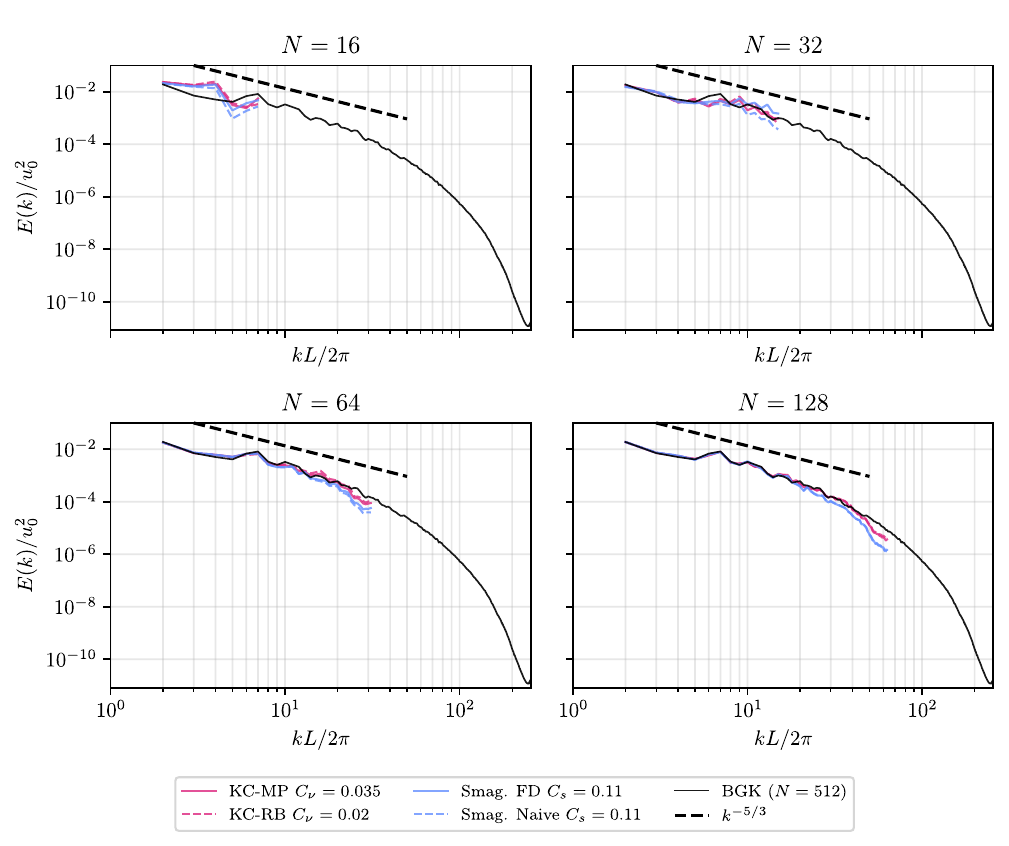}
    \caption{Turbulent kinetic energy spectra $E(k)$ for the Taylor--Green vortex at \(t u_0/L_{\rm TGV}=10\). Comparison between the residual-based closures (KC-RB and KC-MP), the Smagorinsky models, and the reference solution.}
    \label{fig:tgv-spectrum-RB}
\end{figure*}
\begin{figure*}
    \includegraphics[width=\textwidth]{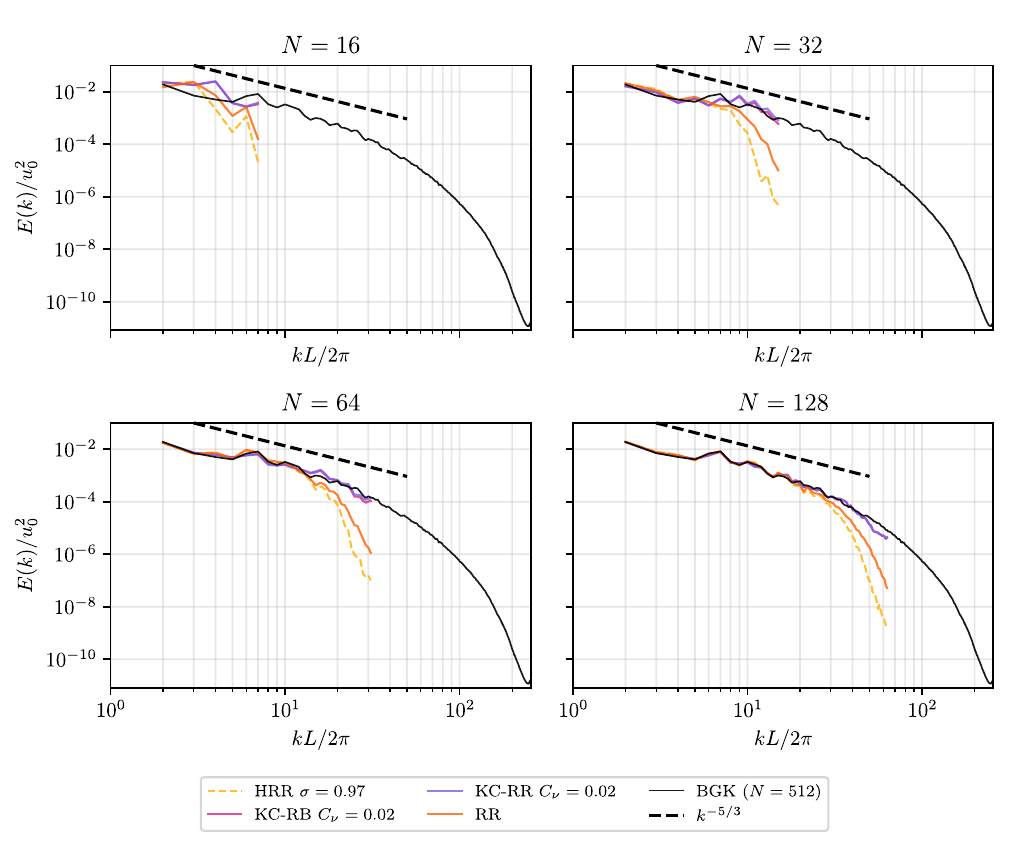}
    \caption{Turbulent kinetic energy spectra for the comparison between KC-RR, RR, HRR, and KC-RB at \(t u_0/L_{\rm TGV}=10\).}
    \label{fig:tgv-spectrum-RR}
\end{figure*}

\begin{figure*}
    \includegraphics[width=\textwidth]{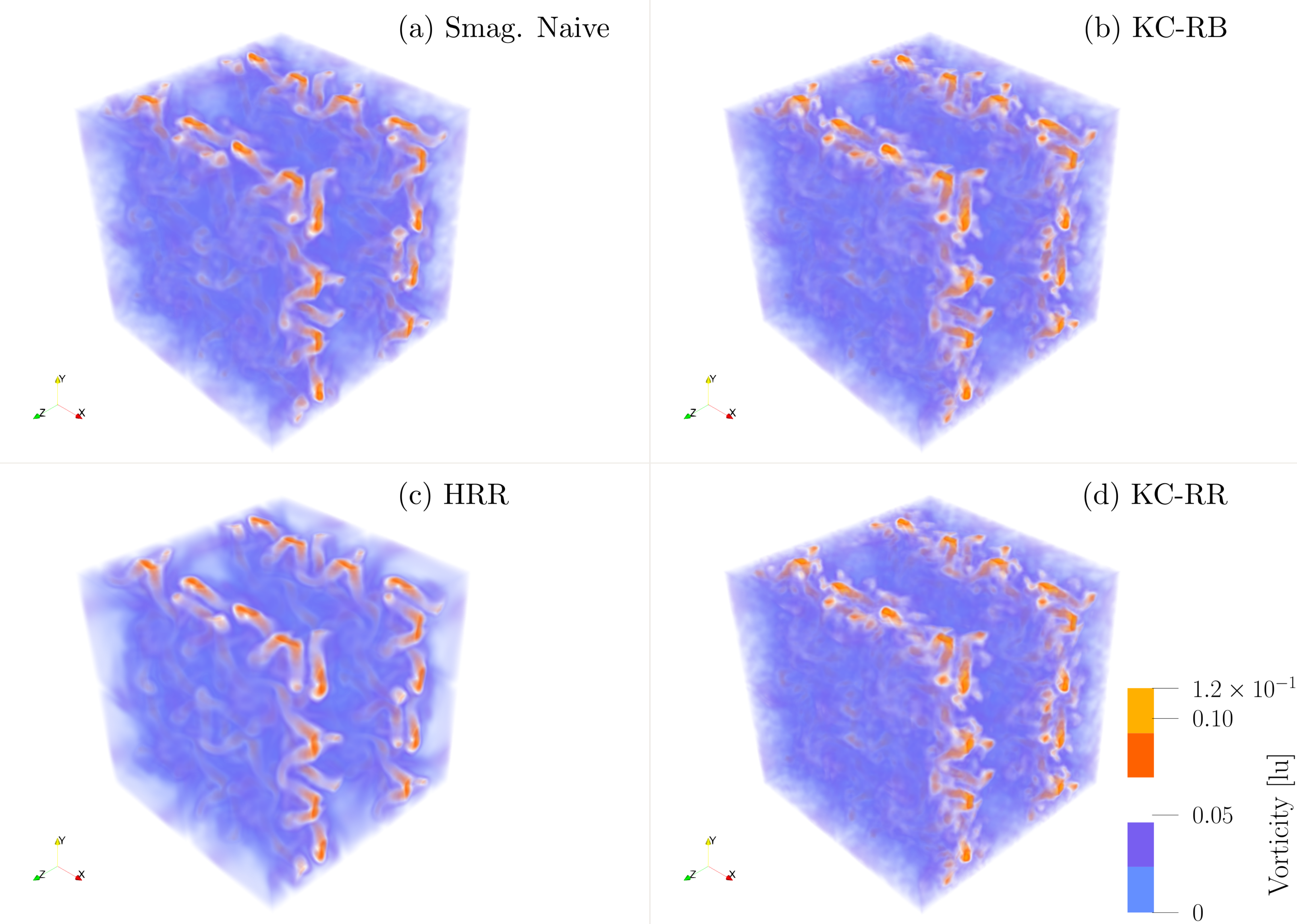}
    \caption{Instantaneous volume renderings of the vorticity magnitude for the Taylor--Green vortex at $\Re = 1600$ on the $N=64$ lattice at $t u_0/L_{\rm TGV} \approx 10$. The viewpoint is placed close to the $(1,1,1)$ diagonal and uses a $30^\circ$ perspective view angle. The common colour bar reports vorticity magnitude in lattice units. Plots (a)--(d) compare Smag.\ Naive, KC-RB, HRR ($\sigma=0.97$), and KC-RR.}
    \label{fig:tgv-comparison}
\end{figure*}
The evolution of the kinetic energy $E_k$ and enstrophy $\mathcal{Z}$ for the Taylor--Green vortex at $\Re = 1600$ and $\Ma = 0.2$ is reported in \cref{fig:tgv-RB,fig:tgv-RR} for various grid resolutions.
The corresponding TKE spectra are shown in \cref{fig:tgv-spectrum-RB,fig:tgv-spectrum-RR} at the nondimensional convective time \(t u_0/L_{\rm TGV}=10\), close to the dissipation peak.
The reference solution is a high-resolution resolved-BGK simulation at $N=512$, computed with the same numerical solver and setup as the comparison runs and used here as an internal numerical reference.

The KC-RB curves use the raw reconstruction in \cref{eq:kcrb_operational_raw_fone}, rather than the Hermite-consistent form in \cref{eq:kcrb_operational_fone}. In weakly compressible LBM, the raw form retains an isotropic divergence contribution that is relaxed at \(\hat\omega_{\sgs}\) rather than \(\hat\omega\), producing a collision mass defect that can modify acoustic activity (Appendix~\ref{app:kcrb_operational}).

To evaluate the residual-based closures, \cref{fig:tgv-RB} contrasts the KC-RB and KC-MP models against the classical Smagorinsky approach (both in its naive local form and with finite-difference-based strain rates).
At the coarsest resolution in \cref{fig:tgv-RB} ($N=32$), Smag.\ Naive loses kinetic energy earliest and gives the lowest smooth enstrophy peak.
Smag.\ FD develops a non-smooth late enstrophy rise during transition.
This behavior indicates operation close to its stability limit.
KC-RB and KC-MP give smoother curves and delay the energy decay relative to Smag.\ Naive, but their enstrophy peaks remain far below the $N=512$ reference.
At $N=64$, the kinetic-energy curves cluster more tightly; KC-RB and KC-MP give higher and better-timed enstrophy peaks than both Smagorinsky variants, although they still underpredict the reference peak.
At $N=128$, the kinetic-energy curves are nearly indistinguishable because the limiter \(\min(\omega^\sharp,\omega_t^\sharp)\) in \cref{eq:full_eff_bgk} largely clips the SGS relaxation.
In the discrete update, this corresponds to the trapezoidally shifted lattice rates \(\hat\omega_{\sgs}=\hat\omega\), defined in Appendix~\ref{app:omega_t_operational}.
In this limit, KC-RB and KC-MP recombine their resolved and SGS components into the standard BGK nonequilibrium relaxation.

The turbulent kinetic energy (TKE) spectra at \(t u_0/L_{\rm TGV}=10\), displayed in \cref{fig:tgv-spectrum-RB}, provide a complementary comparison near the dissipation peak. For the chosen parameters, both the Smagorinsky models and the kinetic closures remain close to the reference over most of the resolved range and exhibit, over part of that range, a slope compatible with the Kolmogorov $k^{-5/3}$ scaling. The visible differences are concentrated near the high-wavenumber end of the spectra.

For the chosen parameter sets, KC-MP ($C_\nu=0.035$) and KC-RB ($C_\nu=0.02$) give similar kinetic-energy, enstrophy, and spectral trends.
This is consistent with their different distribution of SGS dissipation: KC-RB relaxes its population-space estimate of $f_\sgs$, including higher-order nonequilibrium content, whereas KC-MP acts on the second-order Hermite reconstruction of $f_\sgs$.
The larger \(C_\nu\) used by KC-MP may therefore compensate, at least in this case, for its narrower dissipative support.

\Cref{fig:tgv-RR,fig:tgv-spectrum-RR} compares the KC-RR model with the standard recursive regularized (RR) and hybrid recursive regularized (HRR) collision operators, with KC-RB also shown for cross-comparison with \cref{fig:tgv-RB,fig:tgv-spectrum-RB}. Over the resolutions considered here, RR and HRR show faster kinetic-energy decay, lower enstrophy peaks, and stronger high-wavenumber damping than KC-RR. We interpret these lower enstrophy levels and more dissipative spectra as a consequence of the RR/HRR projection mechanism. In a fully resolved flow, RR removes components outside its CE manifold, which are predominantly ghost modes and higher-order numerical errors. In the present under-resolved calculation, the same reconstruction also replaces higher-order departures of the filtered distribution that carry SGS turbulent fluctuations. HRR combines the local population-based second-order stress with a finite-difference strain-rate estimate before the same recursive reconstruction. Smag.\ Naive does not apply this reconstruction. It retains the raw population content and damps high-wavenumber modes dynamically through its eddy-viscosity collision. At $N=64$, KC-RR is slightly less dissipative than legacy KC-RB and retains slightly more high-wavenumber energy. Conversely, appendix~\ref{app:kcrb_reconstruction_sensitivity} shows that Hermite-consistent KC-RB remains close to KC-RR. Therefore, in the matched TGV calculation, the weak-compressibility defect introduced by the raw reconstruction accounts for most of the visible KC-RB/KC-RR separation. KC-RB and KC-RR nevertheless differ in their SGS carriers and high-order reconstruction. KC-RB relaxes the population-space estimate of \(f_\sgs\) defined in \cref{eq:f_sgs_KCRB}, whereas KC-RR relaxes the deviation \(a^{\sgs}=a^{\cNeq}-a^{\rm rr}\) from the RR manifold, as in \cref{eq:kcrr_sgs}. Controlled comparisons beyond the present diagnostic are required to separate the effects of these differences.
At $N=128$, KC-RR overlaps with the BGK enstrophy curve for the same clipping mechanism described above for KC-RB and KC-MP.
When the limiter sets \(\hat\omega_{\sgs}=\hat\omega\), the regularized and SGS Hermite parts recombine, \(a^{\rm rr}+a^{\sgs}=a^{\cNeq}\), and KC-RR approaches a BGK relaxation of the complete retained D3Q27 representation of the coarse-level nonequilibrium. The hats in the \(\hat\omega_{\sgs}=\hat\omega\) notation indicate the shifted lattice rates defined in Appendix~\ref{app:omega_t_operational}.
This also explains why KC-RR and KC-RB collapse onto the same spectrum at $N=128$.

The instantaneous vorticity renderings in \cref{fig:tgv-comparison} show model-dependent differences in the retained vortical structures: at the same coarse resolution, KC-RB and KC-RR retain sharper transitional structures than HRR, while Smag.\ Naive remains intermediate. %

\subsection{Flow in a lid-driven cavity}
\label{ldc}
The lid-driven cavity is a standard test for wall-bounded recirculating flow \cite{leriche_direct_2000,Hegele_2018}. The moving lid injects momentum through the top wall and generates a primary vortex, secondary corner eddies, and strong wall shear layers. At high Reynolds number, the flow is unsteady and the redistribution of energy is controlled by confinement and near-wall transport. This case is therefore useful for assessing how the closures behave near solid boundaries.

\begin{figure*}
    \centering
    \subfloat[$N=64$.]{\includegraphics[width=\textwidth]{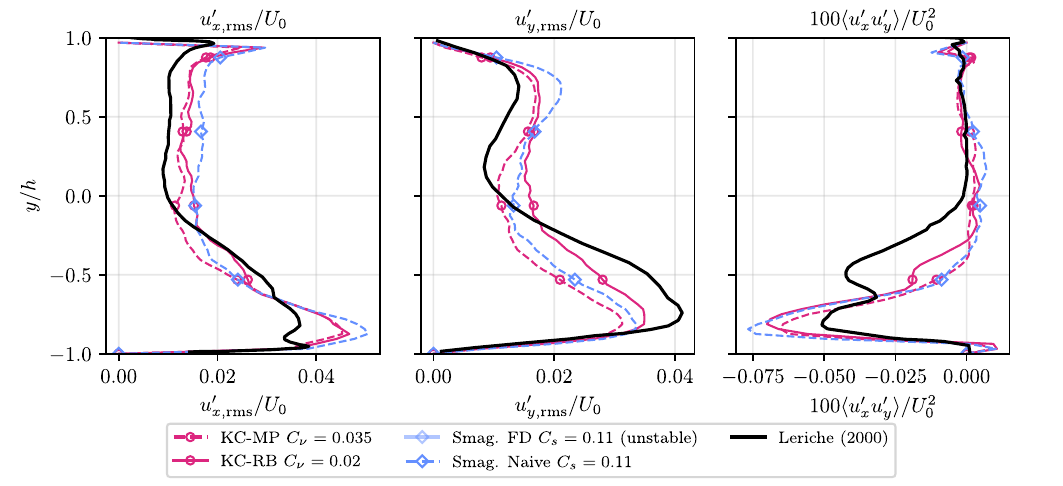}}\\[0.75em]
    \subfloat[$N=128$.]{\includegraphics[width=\textwidth]{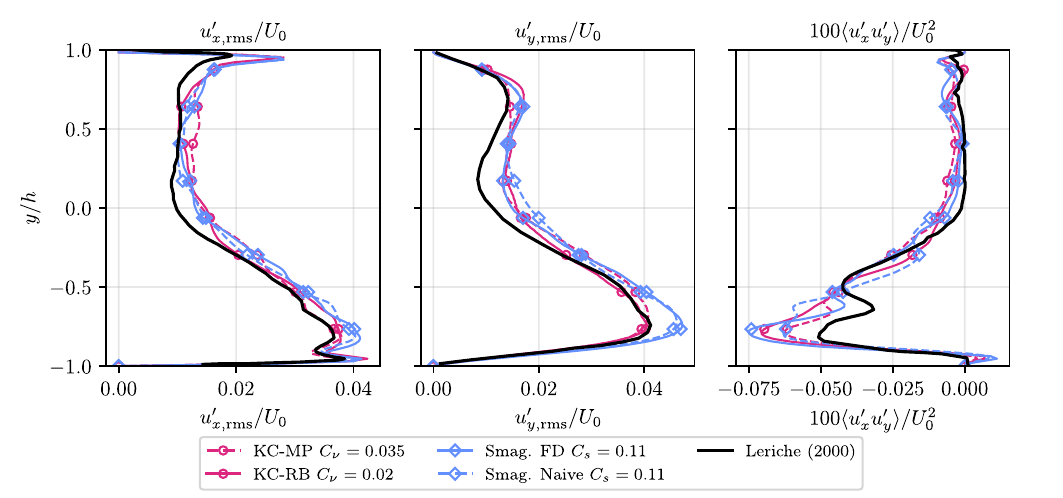}}
    \caption{Lid-driven-cavity root-mean-square statistics at $\Re = 12000$ for the residual-based family. Profiles show $u'_{x,\mathrm{rms}}/u_0$, $u'_{y,\mathrm{rms}}/u_0$, and $100\,\langle u'_x u'_y \rangle/u_0^2$ versus $y/h$. KC-RB ($C_\nu = 0.02$) and KC-MP ($C_\nu = 0.035$) are compared with the Smagorinsky models and the data of Leriche and Gavrilakis~\cite{leriche_direct_2000}. Top: $N=64$; bottom: $N=128$. The finite-difference Smagorinsky variant is marked unstable at $N=64$ because it does not reach the averaging window.}
    \label{fig:ldc-RB}
\end{figure*}

\begin{figure*}
    \centering
    \subfloat[$N=64$.]{
        \includegraphics[width=\textwidth]{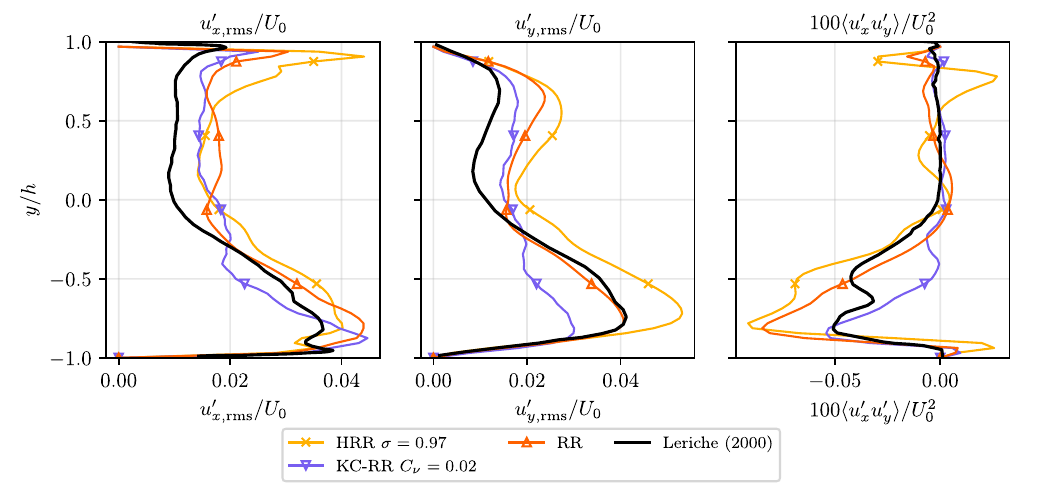}
        }\\[0.75em]
    \subfloat[$N=128$.]{
        \includegraphics[width=\textwidth]{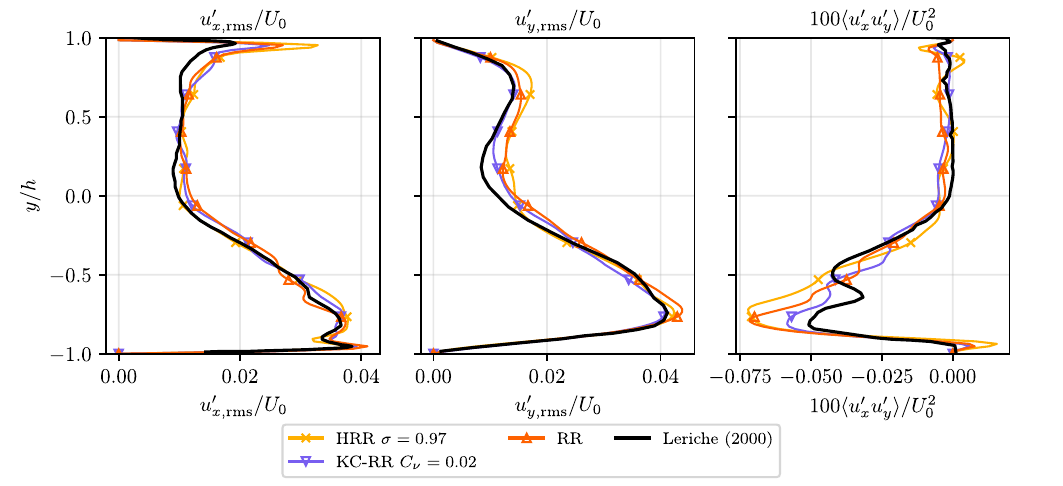}
        }
    \caption{Lid-driven-cavity velocity-fluctuation statistics at $\Re = 12000$ for KC-RR, RR, and HRR. Profiles show $u'_{x,\mathrm{rms}}/u_0$, $u'_{y,\mathrm{rms}}/u_0$, and $100\,\langle u'_x u'_y \rangle/u_0^2$ versus $y/h$. KC-RR ($C_\nu = 0.02$) is compared with Recursive Regularization (RR) and Hybrid Recursive Regularization (HRR, $\sigma = 0.97$). Top: $N=64$; bottom: $N=128$. Leriche and Gavrilakis data~\cite{leriche_direct_2000} are shown as reference.}
    \label{fig:ldc-RR}
\end{figure*}

\begin{table*}[t]
    \caption{Aggregate root-mean-square deviations from the Leriche--Gavrilakis reference profiles~\cite{leriche_direct_2000} for the lid-driven cavity at $\Re = 12000$ on the $N=128$ lattice. The columns report the RMS errors of $u'_{x,\mathrm{rms}}/u_0$, $u'_{y,\mathrm{rms}}/u_0$, and $100\,\langle u'_x u'_y \rangle/u_0^2$ along $y/h$; the last column is the arithmetic mean of the three errors. The table is restricted to $N=128$, where all models complete the stated averaging window.}
    \label{tab:ldc_rms_errors}
    \centering
    \begin{tblr}{
      width=\linewidth,
      colspec={lcccc},
      hline{1,2,9}={solid, 0.08em},
    }
    Model & RMS($u'_{x,\mathrm{rms}}/u_0$) & RMS($u'_{y,\mathrm{rms}}/u_0$) & RMS($100\,\langle u'_x u'_y \rangle/u_0^2$) & Average \\
    \hline
    KC-RR ($C_\nu = 0.02$) & \textbf{0.0048} & \textbf{0.0015} & \textbf{0.0052} & \textbf{0.0038} \\
    KC-MP ($C_\nu = 0.035$) & 0.0055 & 0.0027 & 0.0060 & 0.0048 \\
    RR & 0.0052 & 0.0021 & 0.0072 & 0.0048 \\
    KC-RB ($C_\nu = 0.02$) & 0.0054 & 0.0027 & 0.0080 & 0.0054 \\
    Smag.\ FD ($C_s = 0.11$) & 0.0058 & 0.0038 & 0.0074 & 0.0057 \\
    Smag.\ Naive ($C_s = 0.11$) & 0.0054 & 0.0038 & 0.0091 & 0.0061 \\
    HRR ($\sigma = 0.97$) & 0.0059 & 0.0023 & 0.0103 & 0.0062 \\
    \end{tblr}
\end{table*}

\begin{figure*}
    \includegraphics[width=\textwidth]{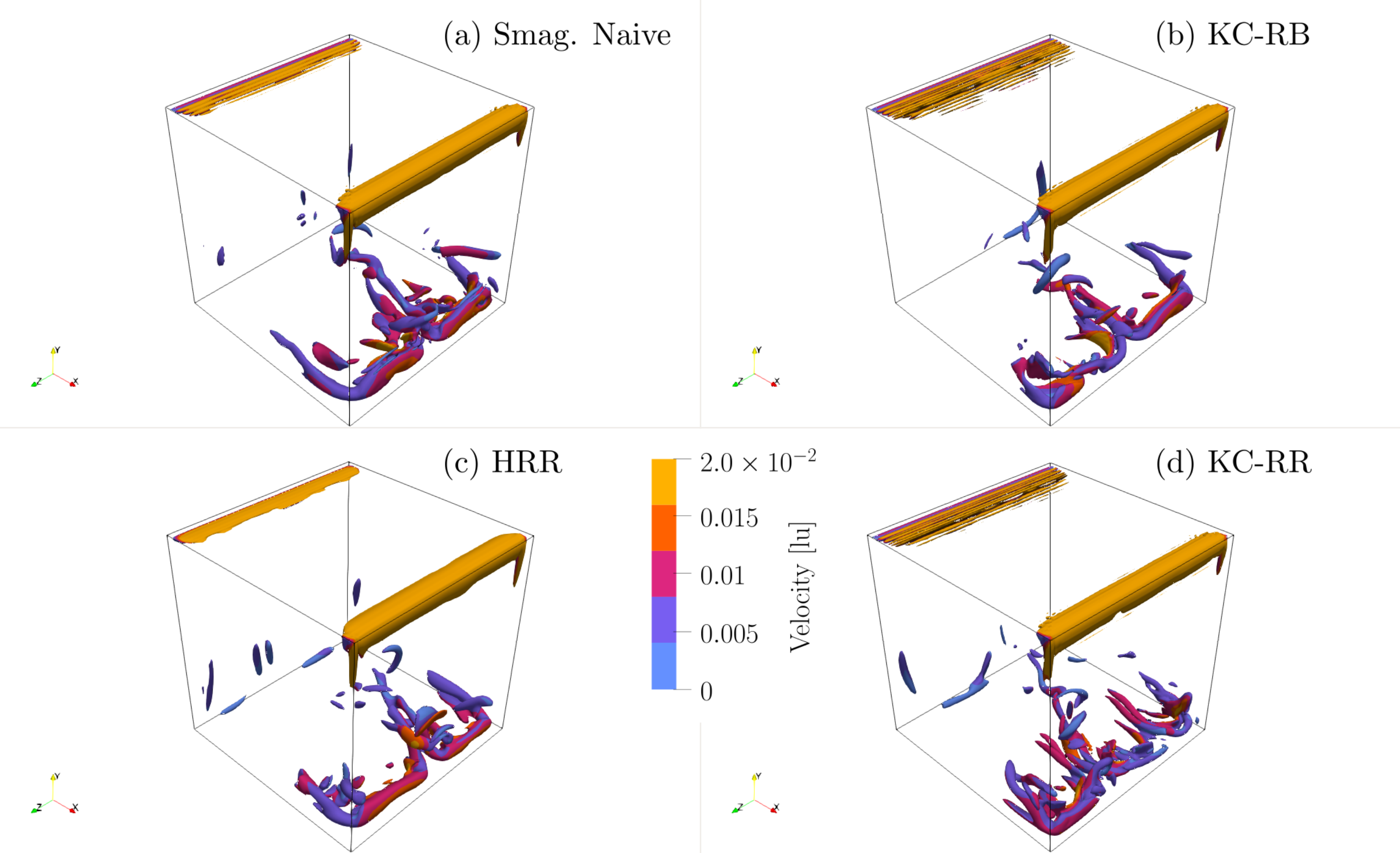}
    \caption{Instantaneous $Q$-criterion isosurfaces for the lid-driven cavity at $\Re=12000$ on the $N=128$ lattice and \(t u_0/(2h)\approx 450\), with \(Qh^2/u_0^2\approx 1.23\) (\(Q=10^{-6}\,\unit{lu}\)), colored by velocity magnitude. The viewpoints are placed close to the $(1,1,1)$ diagonal and use $30^\circ$ perspective view angles. The common colour bar reports velocity magnitude in lattice units. Panels (a)--(d) compare Smag.\ Naive, KC-RB, HRR ($\sigma=0.97$), and KC-RR. The kinetic closures retain a denser field of near-wall vortical structures than Smag.\ Naive and HRR.}
    \label{fig:ldc-128-comparison}
\end{figure*}
We consider $\Re = 12000$ at reference Mach number $\Ma=0.1$ on $N=64$ and $N=128$ grid points. The moving and stationary walls use the Skordos boundary condition~\cite{skordos_initial_1993}, corresponding to BC4 in~\cite{malaspinas_lattice_2009}.
At boundary nodes where velocity gradients are required, we use the same second-order one-sided finite-difference formulas as in Appendix~\ref{app:lbm_discretization}.
These gradients enter both the Skordos boundary condition and the gradient-based collision models.
Normalizing by the lid-velocity scale $u_0$, the reported profiles are $u'_{x,\mathrm{rms}}/u_0 \equiv \sqrt{\langle u'_x u'_x \rangle}/u_0$, $u'_{y,\mathrm{rms}}/u_0 \equiv \sqrt{\langle u'_y u'_y \rangle}/u_0$, and $100\,\langle u'_x u'_y \rangle/u_0^2$ along $y/h$, where $h$ is the cavity half-height and $u'_\alpha=u_\alpha-\langle u_\alpha\rangle$ is the velocity fluctuation about the post-processing time average. The present LBM statistics are averaged over the nominal convective-time window $t u_0/(2h) \in [450,900]$. The comparison curves are plotted using the data coming from the digitization of the plots of the direct numerical simulation of Leriche and Gavrilakis~\cite{leriche_direct_2000}, performed at $\Re = 12000$ on a $129^3$ Chebyshev-collocation grid using a polynomial lid profile with squared factors, whereas our simulations use the corresponding unsquared factors (see \cite{leriche_direct_2000} for the details). The DNS mean-velocity Reynolds number is $\Re_m \approx 10200$. In that DNS, the final-parameter run lasted $1900h/U_0$, and the reported statistics were computed from the last $1000h/U_0$ of sampled data. The present nominal $900h/u_0$ interval is therefore shorter than the reference database and is not presented as a formal statistical-convergence demonstration.

\Cref{fig:ldc-RB} compares KC-RB and KC-MP with the Smagorinsky family. At $N=64$, the finite-difference Smagorinsky variant is unstable and is therefore not available as a converged profile. Among the remaining models, none reproduces all three reference profiles over the full cavity height. For $u'_{x,\mathrm{rms}}/u_0$, KC-RB and KC-MP remain closer to the reference than Smag.\ Naive over the plotted height. For $u'_{y,\mathrm{rms}}/u_0$, they improve the upper part of the profile and remain comparable to Smag.\ Naive in the lower part. For $100\,\langle u'_x u'_y \rangle/u_0^2$, the three converged models show similar errors, and all remain too negative near the lower wall. At $N=128$, the finite-difference Smagorinsky variant is stable and can be compared directly. The model curves then form a tighter group. KC-RB and KC-MP still improve the lower part of $u'_{y,\mathrm{rms}}/u_0$, where they nearly overlap the reference, and they are slightly closer in the lower part of $u'_{x,\mathrm{rms}}/u_0$. The Reynolds-shear-stress profiles remain comparable across the models, with visible residual errors near the lower wall. The aggregated RMS errors reported in \cref{tab:ldc_rms_errors} are consistent with this comparison: KC-MP improves on KC-RB mainly through a smaller shear-stress error (0.0060 against 0.0080), while both residual-based closures remain more accurate on average than the two Smagorinsky variants.

\Cref{fig:ldc-RR} extends the comparison to KC-RR against the RR and HRR operators. At $N=64$, RR gives the closest visual match among these three models. KC-RR follows RR in $u'_{x,\mathrm{rms}}/u_0$ and in the Reynolds shear stress, but underestimates $u'_{y,\mathrm{rms}}/u_0$ below mid-height ($y/h < 0$).
HRR shows the largest departures: it largely overestimates $u'_{x,\mathrm{rms}}/u_0$ in the upper part of the cavity, overestimates $u'_{y,\mathrm{rms}}/u_0$ in the lower half, and deviates strongly in $100\,\langle u'_x u'_y \rangle/u_0^2$ near the lower wall.
At $N=128$, the three profiles move much closer together. The plot shows only modest visual separation, mainly in the lower-wall shear-stress region, so the fine-resolution ordering is better assessed from the RMS errors in \cref{tab:ldc_rms_errors}.

\Cref{tab:ldc_rms_errors} gives the quantitative ranking at $N=128$, where all models complete the stated averaging window. KC-RR has the lowest RMS error in all three statistics and the smallest average error (0.0038). KC-MP and RR share the second-best average error (0.0048), but with different distributions: RR is closer in $u'_{y,\mathrm{rms}}/u_0$, whereas KC-MP is closer in the Reynolds shear stress. KC-RB follows with an average error of 0.0054. The Smagorinsky variants are less accurate on this aggregate measure, with averages of 0.0057 for Smag.\ FD and 0.0061 for Smag.\ Naive. HRR has the largest average error (0.0062), mainly because of its larger shear-stress error.

The instantaneous iso-$Q$ surfaces in \cref{fig:ldc-128-comparison} show that, on the same $N=128$ lattice, the compared closures retain visibly different levels of coherent small-scale structure beneath the moving lid and inside the main recirculation region.
KC-RB and KC-RR produce a richer vortical field than Smag.\ Naive and HRR. The kinetic closures retain an explicit SGS carrier, whereas RR and HRR replace higher-order departures of the filtered distribution with a fully resolved CE reconstruction. Smag.\ Naive instead damps raw population content through its eddy-viscosity collision.

Taken together, the cavity results show that the proposed kinetic closures remain stable and competitive with the Smagorinsky and regularization baselines while retaining resolved second-order and vortical content in a wall-bounded flow. The test therefore extends their assessment beyond periodic decay to a configuration with strong wall shear and recirculation. Because no grid-convergence or parameter-sensitivity study is performed, the quantitative ordering reported above remains specific to the present resolutions, boundary treatment, averaging interval, and model coefficients.

\subsection{Flow past a circular cylinder}\label{fac}
\begin{figure*}
    \includegraphics[width=\textwidth]{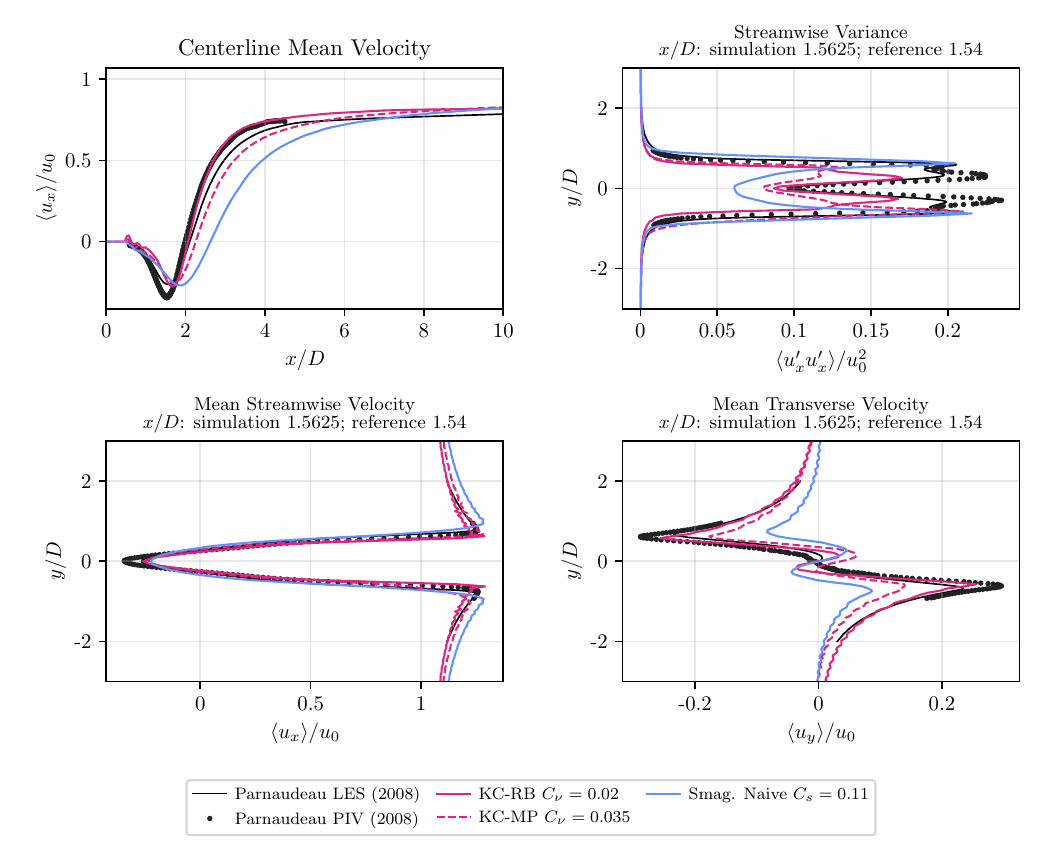}
    \caption{Flow-past-cylinder time-averaged statistics at $\Re = 3900$ for the residual-based kinetic closures on the $480 \times 192 \times 76$ grid. KC-RB and KC-MP are compared with the Smag.\ Naive model, the high-resolution LES reference data of Parnaudeau \emph{et al.}~\cite{parnaudeau_experimental_2008} shown as solid black curves, and digitized PIV measurements shown as black markers.}
    \label{fig:fac-RB}
\end{figure*}
\begin{figure*}
    \includegraphics[width=\textwidth]{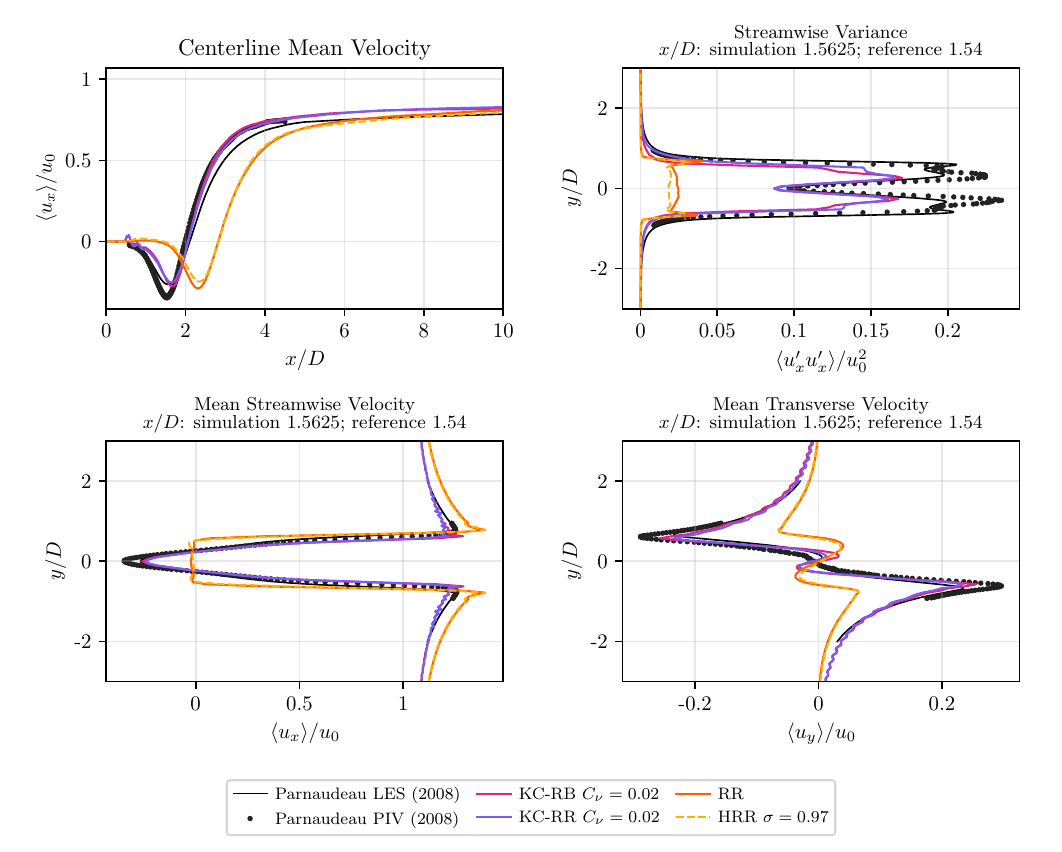}
    \caption{Flow-past-cylinder time-averaged statistics at $\Re = 3900$ for KC-RR and regularized collision models on the $480 \times 192 \times 76$ grid. KC-RR ($C_\nu = 0.02$) is compared with RR, HRR ($\sigma = 0.97$), KC-RB, the Parnaudeau \emph{et al.} LES reference shown as solid black curves~\cite{parnaudeau_experimental_2008}, and digitized PIV measurements shown as black markers.}
    \label{fig:fac-RR}
\end{figure*}
The circular-cylinder wake tests the closures in a separated flow with strong adverse pressure gradients and shear-layer instabilities~\cite{williamson_vortex_1996}. We consider $\Re = 3900$, based on the cylinder diameter $D$ and free-stream velocity $u_0$, at reference Mach number $\Ma = 0.2$. The domain extends approximately $25D$ in the streamwise direction, $10D$ in the transverse direction, and $4D$ in the spanwise direction, and is discretized by $480 \times 192 \times 76$ lattice nodes. Periodic boundary conditions are imposed in the transverse ($y$) and spanwise ($z$) directions. The cylinder has radius $R=9.6$ lattice units, so $D=19.2$ lattice units, and is centered at $(X_c,Y_c)=(120,95)$, i.e. $6.25D$ downstream of the inlet.

The inlet at $x=0$ uses an equilibrium velocity boundary condition. The imposed streamwise velocity is ramped to $u_0$ over the first five convective time units, and the inlet density is extrapolated from the average density of the $x=2$ plane. The outlet uses a non-reflecting convective boundary condition~\cite{izquierdo_characteristic_2008,guo_comparative_2008}. The cylinder surface is treated with the local Filippova--Haenel curved boundary condition~\cite{filippova_lattice-boltzmann_1997,filippova1998grid}, rather than the non-local Mei--Shyy--Luo variant~\cite{mei_accurate_1999}.
During the initial transient, we apply a small cylinder-boundary disturbance to trigger the three-dimensional wake. The disturbance is active only up to $t u_0/D=30$, whereas statistics are collected over $t u_0/D \in [150,400]$. Parnaudeau \emph{et al.}~\cite{parnaudeau_experimental_2008} also discard an initial $150D/U_c$ transient and sample their high-resolution LES over the following $250D/U_c$, approximately $52$ vortex-shedding cycles. The present sampling window follows that reference protocol, but does not by itself demonstrate statistical convergence of the present coarse-grid calculation.
This single-node choice suits the present GPU-oriented Bailey AA-pattern implementation~\cite{bailey_accelerating_2009} and remains practical at higher Reynolds numbers~\cite{marson_directional_2022,ginzburg_unified_2023}. More elaborate ELI-type variants with pressure correction (denoted K4 in~\cite{marson_directional_2022,ginzburg_unified_2023}) may improve wall accuracy~\cite{marson_enhanced_2021,marson_directional_2022,ginzburg_unified_2023}; their combination with the present kinetic-closure framework, possibly with wall models, is left for future work.

This configuration remains a coarse LES. The cylinder is resolved by $D=19.2$ lattice spacings, which is not sufficient to establish grid convergence for a curved-wall wake at this Reynolds number. The case is therefore used as a consistency test for the closures in separated flow, not as a definitive accuracy benchmark for the cylinder wake. Time-averaged profile statistics are collected over $t u_0/D \in [150,400]$ and are compared with the high-resolution LES reference data of Parnaudeau~\emph{et al.}~\cite{parnaudeau_experimental_2008}. The figures also include black markers obtained by digitizing the PIV measurements reported by Parnaudeau~\emph{et al.}~\cite{parnaudeau_experimental_2008}. The markers cover the centerline mean streamwise velocity and the near-wake mean-velocity and streamwise-variance profiles shown in the figures. Because the PIV values were also extracted from published plots, they provide an additional visual consistency reference rather than a separate quantitative error benchmark. The centerline profiles are sampled on the lattice node line $Y=95$, passing through the cylinder center. We assess the closures through numerical stability, profile consistency, and qualitative wake structure.

\begin{figure*}
    \includegraphics[width=\textwidth]{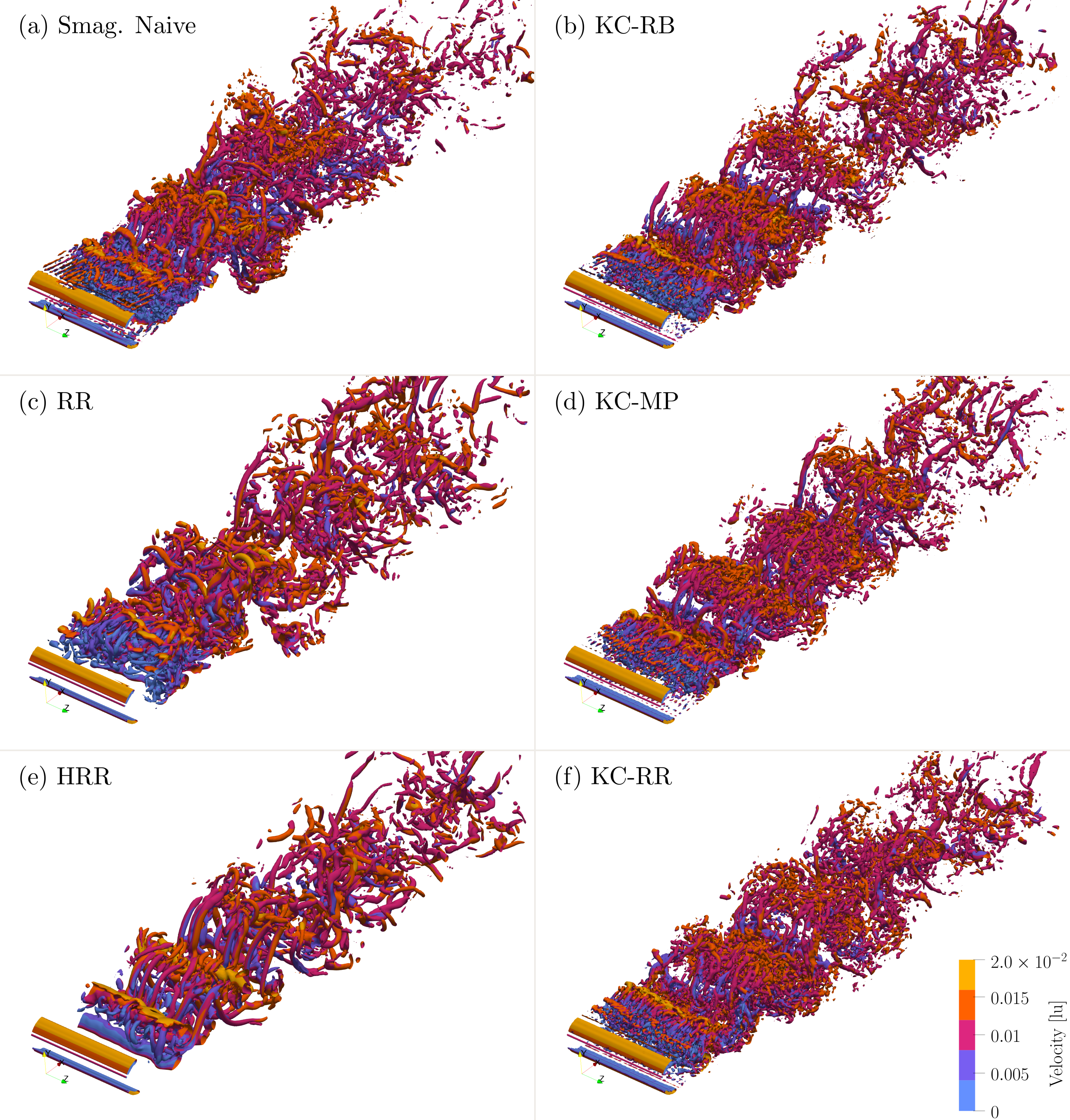}
    \caption{Instantaneous $Q$-criterion isosurfaces for the cylinder wake at $t u_0/D \approx 150$, with $Q D^2/u_0^2 \approx 2.76$ ($Q=10^{-4}\,\unit{lu}$), colored by velocity magnitude. All panels use a common axonometric view under parallel projection. The viewpoint direction from focal point to viewpoint is $\bm d_{\rm cam}\approx(-0.584,0.582,0.566)$, with parallel scale $109.58$. The common colour bar reports velocity magnitude in lattice units. Panels (a)--(f) show Smag.\ Naive, KC-RB, RR, KC-MP, HRR ($\sigma=0.97$), and KC-RR, respectively. %
    }
    \label{fig:fac-comparison}
\end{figure*}

\Cref{fig:fac-RB,fig:fac-RR} compare the stable cylinder runs included in this study. Smag.\ FD becomes unstable at $t u_0/D\simeq55.33$, before the averaging interval. Smag.\ Naive reaches the averaging interval and is retained as the local-moment Smagorinsky reference.

The profile comparisons show that the kinetic closures remain stable and recover the main wake trends in the mean velocity and streamwise-variance profiles. However, we do not assign a quantitative closure ranking from this case. At the present wall resolution, profile-error values and apparent model rankings are sensitive to cylinder placement and wall resolution. We interpret this sensitivity mainly as a consequence of the limited geometric resolution of the cylinder and the corresponding curved-boundary discretization, rather than as a closure-accuracy effect.

Thus the cylinder case is used here as a consistency test for separated-flow behavior. The results indicate that the proposed closures can be applied to the cylinder wake without the immediate instability observed for Smag.\ FD, and that they produce mean and variance profiles broadly comparable with the reference profiles. A grid-converged accuracy assessment of the drag, shedding frequency, recirculation length, and profile errors would require a systematic study of wall resolution, curved-boundary placement, and averaging-window length. This higher-resolution validation is left for future work.

The instantaneous wake shape in \cref{fig:fac-comparison} remains useful as a qualitative diagnostic. The kinetic closures retain a more broken wake with more numerous small vortical structures than the more strongly smoothed cases. This interpretation is consistent with the higher-resolution vorticity plots of Parnaudeau \emph{et al.}~\cite{parnaudeau_experimental_2008}, where the wake contains smaller and less organized structures.

\section{Concluding remarks}

The kinetic closure of turbulence is built on the observation that, in the filtered Boltzmann description, SGS advective information is not lost by transport but is carried by the filtered distribution itself~\cite{marson_kinetic_2025}.
The kinetic turbulence closure problem is therefore structurally different from its macroscopic filtered Navier--Stokes counterpart: it is not the reconstruction of missing SGS transport, but the modeling of the collision-covariance source term generated by the unresolved collision-product covariance. The collision-product covariance is distinct from the molecular-chaos defect associated with the \emph{Stosszahlansatz}.
This structural inversion clarifies the long-standing ambiguity that has repeatedly led kinetic closures to be interpreted as kinetic versions of the Boussinesq/Smagorinsky ansatz for the macroscopic SGS stress.

Within this perspective, we show that the main limitation of standard BGK-type models in LES is not due to the breakdown of molecular chaos, nor to the algebraic simplicity of the BGK operator itself, but to the inherited assumption of instantaneous Markovian relaxation from the Boltzmann integral. At LES scales, the kinetic time filter can induce a finite collision-product covariance, and the associated collision-covariance source term can survive within the collision term and break this local memoryless hypothesis because it represents the memory of the subgrid collision dynamics. The collision-side closure problem is therefore dual: one must both distinguish the filtered fine-grained equilibrium from the equilibrium built from filtered moments alone, and model the non-Markovian relaxation process generated at the filter scale.
In this context, the kinetic time filter has a space--time interpretation along particle characteristics.
A consistent LES-type spatial filter has \(\Delta_x \gtrsim c_t\Delta_t\), and the choice \(\Delta_x=O(c_t\Delta_t)\) defines a matched space--time coarse graining.

The kinetic closure of turbulence framework provides a formal route to address this dual collision-side closure problem by introducing a filtered recorrelated BGK--BE that makes the collision-covariance source term explicit. Its Chapman--Enskog analysis shows that the relevant asymptotic parameter is the ratio of kinetic to macroscopic reference timescales, which in the present nondimensionalization is $\Ma^2/\Re_\ell$, rather than any turbulence-specific scale ratio imposed by hand. The derivation is restricted to the classical hydrodynamic regime, with $\Kn\ll1$ and $\epsilon\ll1$, and does not address rarefied-gas dynamics. No additional CE-type separation is assumed among the nonlinearly coupled hydrodynamic turbulence scales. In this setting, the standard resolved BGK model appears as a limiting non-dynamic case of the more general recorrelated collision operator.

Within the present kinetic closure, the Newtonian stress form at RANS/LES scales follows from the adopted consistency ansatz defined by \cref{aneq,msgs}. This ansatz isolates the Newtonian stress-compatible part, while the remaining SGS-stress evolution constraint is assigned to the non-Markovian source-term closure.

On this basis, we proposed a first operational closure in which the unclosed first CE coefficient of the collision-covariance source, $\overline{\Ecc^{\sharp(1)}}$, is represented through the SGS carrier $f_{\sgs}$ and an effective turbulent relaxation frequency $\omega_t^\sharp$~\cite{marson_kinetic_2025}. This structure was then realized through three concrete kinetic closures (KC-RB~\cite{marson_kinetic_2025}, KC-MP, and KC-RR). The corresponding lattice--Boltzmann simulations show improved stability with respect to the finite-difference Smagorinsky model and model-dependent accuracy gains in the reported benchmarks, while the cylinder case remains sensitive to wall representation and force/statistics diagnostics. Across the three instantaneous flow visualizations, the KC models retain richer resolved small-scale vortical structures than the corresponding Smagorinsky and regularization-based comparisons. These results are promising, but dedicated numerical studies are required to establish accuracy benchmarks against existing methods. In parallel, further development of the SGS collision operator and numerical optimization of the present prototype implementations may improve the accuracy and computational efficiency of KC methods.

The present derivation and numerical realization remain limited to isothermal flows. Extending the filtered recorrelated equation and its Chapman--Enskog analysis to the energy-equation level is the natural route to adiabatic and thermal configurations. This extension could also clarify how turbulent fluctuations and TKE dynamics enter the kinetic closure when energy transport is retained.

Two numerical and modeling directions remain for further development.
First, future work could characterize the interaction between discretization errors and physical SGS content in the filtered distribution and investigate whether additional regularization can reduce the former without suppressing the latter.
Second, the scalar model $\omega_t^\sharp f_{\sgs}$ is only the simplest specialization of the turbulent collision operator. A more general MRT-like operator $\Omega_t^\sharp\circ f_{\sgs}$ could assign different relaxation laws to different SGS moments and encode more detailed SGS physics.

Across these extensions, the main open point is the constitutive determination of the turbulent relaxation frequency $\omega_t^\sharp$, or more generally of the turbulent collision operator $\Omega_t^\sharp$, which sets the dissipation timescale of the collision-covariance source term. In the present work, we estimate this remaining ingredient, including the empirical coefficient $C_\nu$, through a simple phenomenological eddy-viscosity closure. This point does not concern the kinetic closure architecture itself, nor the introduction of an additional macroscopic transport equation to carry unresolved information, since that information is already embedded in the filtered kinetic distribution through the exact advection.
Therefore, a natural next step is to determine this relaxation more directly from the temporal statistics of the collision-product covariance and from the relaxation dynamics of $f_{\sgs}$, so that the constitutive model can be refined without abandoning the kinetic structure established here.

\section*{ACKNOWLEDGMENTS}

The authors gratefully acknowledge support from the Swiss National Science Foundation (Grant No. 212882, Advances in turbulence modeling with the lattice Boltzmann method).

\section*{AUTHOR DECLARATIONS}

\subsection*{Conflict of Interest}
The authors have no conflicts to disclose.

\subsection*{Author Contributions}
\begin{description}
    \item[Francesco Marson] Conceptualization; Methodology; Formal analysis; Software; Validation; Visualization; Funding acquisition (support); Writing -- original draft; Writing -- review and editing.
    \item[Orestis Malaspinas] Conceptualization; Supervision; Funding acquisition (lead); Writing -- review and editing.
\end{description}

\subsection*{AI Use Statement}
AI-assisted writing tools were used during manuscript preparation.
The writing workflow consisted of human drafting, AI-assisted text refactoring, human editing, targeted AI fixes, and final human revision.
AI tools were used for language editing, proofreading, LaTeX refactoring, local consistency checks, discussion of wording and presentation of specific concepts, and occasional bibliography-management support together with Zotero; all references and citations were reviewed by the authors.
AI tools were not used for the core conceptualization, methodology, or formal analysis of the work.
AI tools were not used to generate simulation data, alter figures, or fabricate bibliographic sources.
The authors reviewed and edited all AI-assisted text and take responsibility for the integrity, accuracy, and originality of the manuscript.

\subsection*{Ethical Use Statement}
Consistent with the license of the supporting dataset and software release, the authors ask readers to use the manuscript, methods, numerical models, and results only for peaceful and civil purposes, and not for military or defense purposes, whether directly or indirectly.
The authors further ask persons acting on behalf of, employed by, or materially supporting organizations owned, controlled, or directed by individuals subject to a public International Criminal Court arrest warrant, summons to appear, confirmed charges, conviction, or sentence for genocide, crimes against humanity, war crimes, or the crime of aggression to refrain from directly or indirectly using, implementing, or applying the approaches, methods, numerical models, or results described in this article.

\section*{DATA AVAILABILITY}

The data that support the findings of this study are publicly available in Zenodo at \href{https://doi.org/10.5281/zenodo.21861350}{10.5281/zenodo.21861350}~\cite{marson_malaspinas_kinetic_closure_dataset_2026}.
The associated software release is publicly available in Zenodo at \href{https://doi.org/10.5281/zenodo.21260883}{10.5281/zenodo.21260883}~\cite{marson_malaspinas_kinetic_closure_software_2026}.
The dataset and software release are distributed under the End-User Ethical Dataset and Software License Agreement (EEDSLA), Version 1.1 (July 2026), which prohibits direct or indirect military or defense use and includes an exclusion based on International Criminal Court jurisdiction.

\appendix
\section{Collision-product covariance analysis}\label{app:klimontovich_covariance}

This appendix uses the Klimontovich formalism to distinguish the statistical mechanisms associated with hypotheses (i) and (iii) of \cref{sec:filtered_recorrelated_BGK_Boltzmann_equation}. The two mechanisms are the breakdown of microscopic molecular chaos (\emph{Stosszahlansatz}) and the macroscopic collision-product covariance induced by finite temporal filtering. Their distinction has historical context.
Early two-point kinetic theories argued that binary molecular chaos had to be discarded to retain long-range hydrodynamic correlations and connected the growth of these correlations under shear to transition and turbulence~\cite{tsuge_breakdown_1971,tsuge_approach_1974}.
Tsuge, in particular, placed shear-induced hydrodynamic fluctuations and molecular thermodynamic fluctuations within the same ensemble description. The Reynolds-number growth of the hydrodynamic fluctuations was therefore attributed to a two-particle molecular-chaos defect, without introducing a distinct temporal-filtering level.
Related identifications still appear in recent works~\cite{chliamovitch_turbulence_2018,girimaji_coherence_2026}. For example, the recent preprint~\cite{girimaji_coherence_2026} links the onset of macroscopic coherent structures and non-Markovian memory with a loss of microscopic-state decorrelation, explicitly identified with the \emph{Stosszahlansatz}. The same preprint extends the thermodynamic analogy by describing the unresolved SGS field as a ``turbulent bath'' and interpreting local production--dissipation balance as local equilibrium and describing the unresolved bath as ``truly thermal'' at finer resolution. Although the preprint distinguishes turbulent and molecular baths, its invocation of the \emph{Stosszahlansatz} risks conflating turbulent decorrelation with molecular chaos. Neither ensemble averaging (RANS) nor spatial filtering (LES), by itself, justifies equating turbulent decorrelation with molecular chaos. The \emph{Stosszahlansatz} factorizes incoming two-particle statistics at the molecular collision scale. The turbulent SGS fluctuations invoked by the preprint instead belong to the coarse-grained hydrodynamic field, whereas thermodynamic fluctuations originate from molecular degrees of freedom. Viscous conversion of turbulent kinetic energy into heat couples these levels but does not turn the SGS field into a thermodynamic bath.
The thermal analogy may provide an intuitive interpretive picture, but the unresolved SGS fluctuations considered here are ordinary hydrodynamic motions governed by the Navier--Stokes equations. A sufficiently resolved DNS would resolve them as part of the same flow field. Their unresolved status in an LES does not by itself endow them with molecular thermodynamic-bath properties.
The filter-induced non-Markovian memory considered here likewise concerns the coarse-grained evolution of the one-particle field rather than a loss of molecular chaos. Thermodynamic nonequilibrium does not by itself imply a failure of molecular chaos, because the Boltzmann equation describes nonequilibrium evolution while retaining the \emph{Stosszahlansatz}.

The analysis below shows that the breakdown of molecular chaos $\Cchaos$ and the collision-product covariance $\Ccc$ have similar covariance structures but arise from different statistical operations and have different physical meanings. Specifically, $\Cchaos$ is the instantaneous ensemble cross-particle correlation, whereas $\Ccc$ is the covariance of the continuous one-particle density field generated by the finite temporal filter \(\langle\cdot\rangle_{\Delta_t}\).

For the Klimontovich field $N$ and its moments, we use units with particle mass $m=1$.
The microscopic $N_p$-particle field is defined as
\begin{equation}
    N(\boldsymbol{z}, t) = \sum_{i=1}^{N_p} \delta(\boldsymbol{z} - \boldsymbol{z}_i(t)).
\end{equation}
The evolution of the singular field follows the Klimontovich equation~\cite{klimontovich_statistical_1967,klimontovich_silin_1962},
\begin{equation}
    \pp_t N
    + \xi_\alpha \pp_{x_\alpha}N
    + a_{N,\alpha}\pp_{\xi_\alpha}N
    =0,
\end{equation}
where $a_{N,\alpha}$ are the components of the microscopic acceleration generated by the intermolecular potentials, assumed independent of $\boldsymbol{\xi}$. Thus the equation contains no modeled collision integral.

Two distinct averaging operators define the relevant probability spaces:
\begin{enumerate}
    \item The Gibbs ensemble average $\langle \cdot \rangle_{\text{ens}}$, representing an expectation over realizations of initial microscopic configurations.
    \item The macroscopic temporal average $\langle \cdot \rangle_{\Delta_t}$, evaluated over a finite time window $\Delta_t$ within a single realization.
\end{enumerate}

The one-particle distribution function $\mathcal{F}$ corresponds directly to the ensemble average of the exact field:
\begin{equation}
    \mathcal{F}(\boldsymbol{z}, t) = \langle N(\boldsymbol{z}, t) \rangle_{\text{ens}}.
\end{equation}
The ensemble-averaged product of the field decomposes as~\cite{klimontovich_silin_1962,tsuge_new_1975}
\begin{equation}\label{eq:NN_decomposition}
    \langle N(\boldsymbol{z}, t)\, N(\boldsymbol{z}_{i\!i}, t) \rangle_{\text{ens}} = \mathcal{F}_2(\boldsymbol{z}, \boldsymbol{z}_{i\!i}, t) + \delta(\boldsymbol{z} - \boldsymbol{z}_{i\!i})\,\mathcal{F}(\boldsymbol{z}, t),
\end{equation}
where $\mathcal{F}_2$ is the two-particle distribution function and the diagonal term accounts for the self-pair contribution ($i = j$ in the double sum). For distinct phase-space coordinates ($\boldsymbol{z} \neq \boldsymbol{z}_{i\!i}$), this reduces to
\begin{equation}
    \mathcal{F}_2(\boldsymbol{z}, \boldsymbol{z}_{i\!i}, t) = \langle N(\boldsymbol{z}, t)\, N(\boldsymbol{z}_{i\!i}, t) \rangle_{\text{ens}}.
\end{equation}

For distinct phase-space coordinates, or after removing the self-pair term, the error associated with the molecular chaos assumption ($\Cchaos$) evaluates to the covariance of the exact field $N$ under the ensemble operator:
\begin{alignedEq}
    \Cchaos &= \mathcal{F}_2 - \mathcal{F}\mathcal{F}_{i\!i}\\& = \langle N(\boldsymbol{z}) N(\boldsymbol{z}_{i\!i}) \rangle_{\text{ens}} - \langle N(\boldsymbol{z}) \rangle_{\text{ens}} \langle N(\boldsymbol{z}_{i\!i}) \rangle_{\text{ens}}.
\end{alignedEq}
Analogously, the collision-product covariance $\Ccc$ is the covariance of $\mathcal{F}$ under the temporal operator:
\begin{alignedEq}
    \Ccc(\mathcal{F}\mathcal{F}_{i\!i}) &= \langle \mathcal{F}(\boldsymbol{z}) \mathcal{F}(\boldsymbol{z}_{i\!i}) \rangle_{\Delta_t} - \langle \mathcal{F}(\boldsymbol{z}) \rangle_{\Delta_t} \langle \mathcal{F}(\boldsymbol{z}_{i\!i}) \rangle_{\Delta_t}\\
    &= \left\langle
    \langle N(\boldsymbol{z}) \rangle_{\text{ens}}
    \langle N(\boldsymbol{z}_{i\!i}) \rangle_{\text{ens}}
    \right\rangle_{\Delta_t}\\
    &\quad - \left\langle \langle N(\boldsymbol{z}) \rangle_{\text{ens}} \right\rangle_{\Delta_t}
    \left\langle \langle N(\boldsymbol{z}_{i\!i}) \rangle_{\text{ens}} \right\rangle_{\Delta_t}.
\end{alignedEq}
The finite temporal average $\langle \cdot \rangle_{\Delta_t}$ is not, in general, an ensemble average.
The same temporal average is also not necessarily a Reynolds operator.
Therefore the covariance $\Ccc$ cannot be expected to reduce directly to a single Reynolds-like contribution.
To make the simplest LES analogy explicit, we now restrict $\langle \cdot \rangle_{\Delta_t}$ to a fixed linear temporal filter that commutes with the derivatives used in the filtered kinetic equation~\cite{sagaut_large_2006}.
The fixed-filter assumption is already a simplification.
A more general finite-time average could also introduce commutation errors.
Let $F \equiv \langle\mathcal{F}\rangle_{\Delta_t}$ and $F_{\rm cc} \equiv \mathcal{F} - F$.
In \cref{fcc_decomp}, the brackets acting on products containing $F$ denote a repeated application of the same temporal filter.
For a finite non-Reynolds filter, repeated filtering need not reduce to the identity.
Since the collision product $\mathcal{F}\mathcal{F}_{i\!i}$ is quadratic, substituting $\mathcal{F}=F+F_{\rm cc}$ gives
\begin{alignedEq}\label{fcc_decomp}
    \Ccc\left(\mathcal{F}\mathcal{F}_{i\!i}\right) &\equiv \langle \mathcal{F}\mathcal{F}_{i\!i} \rangle_{\Delta_t} - F F_{i\!i} \\&=
    \underbrace{\langle F_{\rm cc} F_{{\rm cc}, i\!i} \rangle_{\Delta_t}}_{\CRey}
    + \underbrace{(\langle F F_{i\!i} \rangle_{\Delta_t} - F F_{i\!i})}_{\CLeo}\\
    &\quad + \underbrace{(\langle F F_{{\rm cc}, i\!i} \rangle_{\Delta_t} + \langle F_{\rm cc} F_{i\!i} \rangle_{\Delta_t})}_{\CCross}.
\end{alignedEq}
\Cref{fcc_decomp} is the temporal analogue of a Leonard-type SGS decomposition.
Here $\CRey$ is the Reynolds-like fast-time residual product, $\CLeo$ is a Leonard-like term, and $\CCross$ collects the cross terms.
The first term becomes a true covariance only under the Reynolds-average reduction used below.
Thus $\Ccc$ is not, in general, only $\CRey$.
Even under the simplified linear-filter assumption, the finite temporal filter produces additional terms that have no counterpart in the ensemble covariance $\Cchaos$ when the ensemble average is used as a Reynolds operator.

To isolate the physical meaning of the remaining covariance, we now make a stronger reduction.
We either assume that the temporal filter behaves as a Reynolds average over the relevant window, or we discard the Leonard-like and cross terms as a modeling approximation.
Then $\Ccc(\mathcal{F}\mathcal{F}_{i\!i})\approx\CRey$, and the comparison focuses on the origin of this surviving Reynolds-like contribution.
The term $\Cchaos$ measures microscopic particle correlations under an ensemble average.
In the Reynolds-average limit, $\CRey$ measures temporal covariance of the ensemble-reduced one-particle distribution along one continuum realization, not a covariance of the Klimontovich field within that realization.

Under this Reynolds-average reduction, the two quantities share a covariance form but describe different mechanisms.
Define the microscopic statistical fluctuation as $\delta N = N - \langle N \rangle_{\text{ens}}$.
The fluctuation $\delta N$ measures the deviation of the exact particle field from the one-particle PDF $\mathcal{F}$.
By contrast, the temporal fluctuation $F_{\rm cc}$ defined above measures the deviation of the one-particle distribution from its time-filtered value.
The microscopic covariance and the Reynolds-like temporal contribution are
\begin{align}
    \Cchaos &= \langle \delta N(\boldsymbol{z}) \delta N(\boldsymbol{z}_{i\!i}) \rangle_{\text{ens}}, \\
    \CRey &= \langle F_{\rm cc}(\boldsymbol{z}) F_{\rm cc}(\boldsymbol{z}_{i\!i}) \rangle_{\Delta_t}.
\end{align}

The distinction between these quantities can be illustrated through two limiting physical regimes.
First, consider the steady laminar flow of a dense liquid.
Because the fluid is dense, short-range intermolecular potentials induce persistent spatial correlations between particle positions.
The pair distribution is therefore not factorizable, $\mathcal{F}_2 \neq \mathcal{F}\mathcal{F}_{i\!i}$, and $\Cchaos \neq 0$.
Conversely, because the flow is steady, the one-particle PDF $\mathcal{F}(\boldsymbol{z}, t)$ is constant over the filter window $\Delta_t$.
The temporal fluctuation $F_{\rm cc}$ vanishes identically, yielding $\CRey = 0$.

Alternatively, consider a turbulent dilute gas in a regime where the Boltzmann molecular-chaos assumption is valid.
The mean free path is then large compared with the molecular interaction range, and pre-collision phase-space correlations are negligible.
The pair distribution factorizes and $\Cchaos \to 0$, recovering the Boltzmann framework.
However, if $\Delta_t$ spans unresolved macroscopic fluctuations, the one-particle PDF $\mathcal{F}(\boldsymbol{z}, t)$ can still fluctuate within the filter window.
Thus $\CRey$ can remain finite even when molecular chaos is valid.

Both errors have the structure of a covariance generated by an averaging operator, but their physical origins differ.
The term $\Cchaos$ quantifies phase-space correlations of particle pairs, including correlations generated by intermolecular potentials.
Within the restricted approximation $\Ccc(\mathcal{F}\mathcal{F}_{i\!i})\approx\CRey$, the term $\CRey$ quantifies the temporal covariance of the one-particle PDF introduced by unresolved macroscopic fluctuations.
Conflating these two operations would incorrectly mix the modeling of dense-gas thermodynamic interactions with the modeling of macroscopic turbulent transport.

\section{Collision-side-only scaling assumptions in filter-based kinetic closures}
\label{app:ansumali_scaling}

This appendix examines the filtered formulation of Ansumali
et al.~\cite{ansumali_kinetic_2004} and the ADM formulation of Sagaut
et al.~\cite{sagaut_toward_2010,malaspinas_advanced_2011}.
They are treated together because their limitations relative to the present framework have the same structural origin.
Both approaches rely on the
fact that a homogeneous spatial filter commutes with BE linear streaming, and
both then treat the unresolved distribution residual as a small correction acting
only through collision, although their specific smallness assumptions differ: Ansumali uses a shrinking-filter ordering, whereas ADM assumes a small deconvolution error.
However, this ordering removes the $O(1)$ SGS momentum flux
from the filtered Euler limit that then leads to the laminar form of the filtered Euler equation.
Furthermore, both approaches address only the part of the collision problem that concerns the reconstruction of the fine-grained equilibrium, namely \compref{tab:comparison:d-be1} in \cref{tab:comparison}, but do not consider the non-Markovian collision-covariance source term \compref{tab:comparison:d-be2}.

\subsection{Collision-side-only scaling in a filtered formulation}

The filtered formulation of~\cite{ansumali_kinetic_2004} gives the following hydrodynamic limit (their Eq.\ (41)), converted here from acoustic to the present convective variables, while distinguishing the source's viscosity-based Knudsen number, $\Kn_A=\nu/(\sqrt{\theta_R}\mathcal L)=\Ma/\Re$, from the molecular $\Kn$ used in the present CE expansion:
\begin{equation}\label{ansumali}
\begin{aligned}
    0 &=
    \pp_t     \MT[\tilde u^\flat][1]
        +
        \T[\pp][1][1][]P_A^\flat
    +
\T[\pp][2][2][] \left(
        \,\MT[\tilde u^\flat][1] \MT[\tilde u^\flat][2]\right)
\\
&\quad-  2\,{\frac{1}{\Re}}\,
     \MT[\pp^{(1)}][2]
   \MT[\tilde S^\flat][1][2]
\\
&\quad+\frac{\Ma}{12\Re}
\MT[\partial][2]
\big[
(\MT[\tilde S^\flat][1][3]+\MT[\mathcal{W}^\flat][1][3])(\MT[\tilde S^\flat][3][2]-\MT[\mathcal{W}^\flat][3][2])
\big]\mathrlap{,}
\end{aligned}
\end{equation}
where $P_A^\flat\equiv\lim_{\Ma\to0}[(\bar\rho-\rho_R)\tilde\theta^\sharp/(\rho_R\Ma^2)]$ is the incompressibility pressure in the isothermal low-Mach limit, with $\bar\rho-\rho_R=O(\Ma^2)$ and constant $\tilde\theta^\sharp$.
Equation~(41) of Ref.~\cite{ansumali_kinetic_2004} uses acoustic velocity $u^\sharp=\Ma\,u^\flat$ with macroscopic coordinates $x^\flat=x/\mathcal L$. Converting to convective velocity and time gives the viscous and SGS coefficients $2\Kn_A/\Ma=2/\Re$ and $\Kn_A/12=\Ma/(12\Re)$, respectively, in \cref{ansumali}.

The resolved strain and rotation rate tensors are defined as
\begin{align}
    \tilde S_{\alpha_1 \alpha_2}^\flat &= \frac{1}{2}(\pp_{\alpha_2} \tilde u_{\alpha_1} + \pp_{\alpha_1} \tilde u_{\alpha_2})\,,\\
    \mathcal{W}_{\alpha_1 \alpha_2}^\flat &= \frac{1}{2}(\pp_{\alpha_2} \tilde u_{\alpha_1} - \pp_{\alpha_1} \tilde u_{\alpha_2})\,.
\end{align}

The leading ${O}(1)$ terms in \cref{ansumali} form an unfiltered Euler or a filtered-laminar Euler system, so the momentum flux contains no SGS contribution. In the present fixed-filter LES ordering, the filtered Euler equations contain the ${O}(1)$ SGS stress tensor $\MT[m^{\sgs}][1][2]$ because inertial energy transfer enters through the advective flux. By placing the non-linear SGS terms at ${O}(\Kn_A)$, \cref{ansumali} removes them from the Euler-level macroscopic transport and reintroduces them only as a viscous-order correction. The shrinking-filter scaling is therefore not a necessity of filtering itself, but follows from the implicit choice $\MT[m^{\sgs\zero}][1][2]=0$ together with the requirement that the first nonzero SGS correction enter at viscous order, as detailed below.

The decisive additional assumption enters through the asymptotic matching. In the present raw-moment notation, Eq.\ (32) of~\cite{ansumali_kinetic_2004} amounts to imposing
\begin{equation}\label{eq:ansumali_nonconserved_moment_matching}
\begin{aligned}
    \overline{\T[m][1][n]}
    &=
    \T[m^\zero][1][n]
    +\Kn_A\,\overline{\T[m^\one][1][n]}\\
    &\quad
    +O(\Kn_A^2),
    \qquad n\ge 2 .
\end{aligned}
\end{equation}
In \cref{eq:ansumali_nonconserved_moment_matching} only, $\overline{\T[m^\one][1][n]}$ denotes the source's first-order coefficient in $\Kn_A$. Here $\T[m^\zero][1][n]$ is the resolved equilibrium moment computed from the filtered variables, i.e. from $f^\zero(\overline f)$. By contrast, the present filtered-equilibrium decomposition gives
\begin{equation}\label{eq:filtered_equilibrium_moment_decomposition}
\begin{aligned}
    \overline{\T[m^\zero][1][n]}
    &=
    \T[m^\zero][1][n]
    +\T[m^{\sgs\zero}][1][n]+O(\epsilon),
    \qquad n\ge 2 ,
\end{aligned}
\end{equation}
and therefore
\begin{equation}\label{eq:present_filtered_moment_expansion}
\begin{aligned}
    \overline{\T[m][1][n]}
    &=
    \T[m^\zero][1][n]
    +\T[m^{\sgs\zero}][1][n]\\
    &\quad
    +\epsilon\overline{\T[m^\one][1][n]}
    +\epsilon\T[m^{\sgs\one}][1][n]+O(\epsilon^2),
    \quad n\ge 2 .
\end{aligned}
\end{equation}
Thus the matching used in Eq.\ (32) of~\cite{ansumali_kinetic_2004} corresponds, in the notation of the present manuscript, to setting
\begin{equation}\label{eq:ansumali_zero_sgs_moment_condition}
\begin{aligned}
    \T[m^{\sgs\zero}][1][n]=0,
    \qquad n\ge 2,
\end{aligned}
\end{equation}
and in particular $\MT[m^{\sgs\zero}][1][2]=0$. Condition~\cref{eq:ansumali_zero_sgs_moment_condition} is not a consequence of homogeneous filtering. It is implicit in choosing the resolved equilibrium, reconstructed solely from the filtered conserved variables, as the complete $O(1)$ part of every filtered nonconserved moment. This structural choice removes all $O(1)$ SGS stresses from the macroscopic momentum flux: the resulting $O(1)$ hydrodynamic limit is the resolved Euler system with no SGS stress, as shown explicitly by their Eq.\ (33). The filter-induced SGS contribution is left to enter only through the next-order relaxation terms.

Consequently, the relevant filter-induced non-linearity is isolated within the local equilibrium moments of the relaxation operator. The Taylor expansion of the filter in Eq.\ (25) of~\cite{ansumali_kinetic_2004} shows that filtering these nonlinear equilibrium moments generates $O(\Delta_x^2/L^2)$ corrections, where $L=\mathcal L$ in the present notation. In the non-dimensionalized kinetic equation, their Eq.\ (29) places the filtered-equilibrium correction on the right-hand side with scaling $\frac{1}{\Kn_A}(\Delta_x/L)^2$.

For an independent macroscopic filter ($\Delta_x/L \sim O(1)$), this term contributes to the leading collision balance as $\Kn_A \to 0$. To retain a singularly perturbed hierarchy with SGS terms at the viscous timescale, the formulation then chooses, through their Eq.\ (30), $\Delta_x/L \sim \sqrt{\Kn_A}$, or equivalently $\Delta_x\sim(\ell_A L)^{1/2}$ with $\ell_A=\Kn_A L$. This restriction shifts the filtered-equilibrium correction to first order in the Knudsen expansion, yielding the tensor-diffusivity contribution in their Eqs.\ (36)--(41). In the present notation, it is the analogue of the collision-side ordering $f_\sgs=\epsilon f_\sgs^\one$, with no independent $O(1)$ SGS stress in the filtered Euler flux. Therefore, the filter-width relation should not be interpreted as a general physical restriction on the LES filter, but as a conditional scaling required to place the filter correction at viscous order within this collision-side-only hierarchy encoded by~\cref{eq:ansumali_zero_sgs_moment_condition}.

\subsection{Approximate Deconvolution Method}

Consider the BE for the unfiltered distribution $f$:
\begin{equation}\label{eq:adm_be_full}
    \pp_t  f + \xi_{\alpha}\pp_{\alpha} f = \mathcal{Q}(f)\,.
\end{equation}
In the ADM formulation~\cite{sagaut_toward_2010,malaspinas_advanced_2011}, a deconvolved distribution $f^*$ is introduced to approximate the unfiltered state, with the reconstruction error $f' \equiv f - f^*$. At the level of their unresolved advective moments, $f'$ is analogous, but not identical, to the SGS equilibrium commutation residual $f_\sgs$ in the present filtered hierarchy. Substituting $f = f^* + f'$ into \cref{eq:adm_be_full} and using the linearity of streaming:
\begin{equation}\label{eq:adm_decomposed}
    \pp_t  f^* + \xi_{\alpha}\pp_{\alpha} f^*
    + \pp_t  f' + \xi_{\alpha}\pp_{\alpha} f'
    = \mathcal{Q}(f^* + f')\,.
\end{equation}
We denote the ADM spatial filter by \(\overline{(\cdot)} \equiv \mathcal{G}*(\cdot)\), where \(\mathcal{G}*\) is the notation used in~\cite{sagaut_toward_2010,malaspinas_advanced_2011}.
This filter commutes exactly with the linear streaming operator, so filtering \cref{eq:adm_decomposed} gives
\begin{equation}\label{eq:adm_filtered}
\begin{aligned}
    \overline{\pp_t f^* + \xi_{\alpha}\pp_{\alpha} f^*}
    &+
    \underbrace{\overline{\pp_t f' + \xi_{\alpha}\pp_{\alpha} f'}}_{\text{advective streaming of } f'}\\
    &=
    \overline{\mathcal{Q}(f^* + f')}\,.
\end{aligned}
\end{equation}
At this stage, the decomposition reflects the structured hierarchy of our current investigation. The distribution \( f^* \) is responsible for generating the resolved, or deconvolved, moments, while the flow of \( f' \) represents the advective deconvolution error. Additionally, there is a residual term produced by \( \mathcal{Q}(f^* + f') - \mathcal{Q}(f^*) \), which accounts for the collision-side deconvolution error.

At the physical level, deconvolution reconstructs part of the information smoothed by the filter, but a residual remains (namely \(f'\), which is analogous to the SGS equilibrium commutation residual \(f_\sgs\) in the filtered case). The deconvolved equations are therefore more resolved, but not structurally different from the filtered equations. The residual has both an advective projection, carried by the streaming of \(f'\), and a collision-side projection, carried by \(\mathcal{Q}(f^*+f')-\mathcal{Q}(f^*)\). For a finite LES filter, the advective projection is the leading macroscopic contribution because its second moment enters the Euler-level momentum flux. The collision-side projection can affect the Navier--Stokes-order dissipative terms. However, it remains a collision-side error, not an Euler-level transport flux, and therefore cannot replace at leading order the advective SGS stress carried by the streamed \(f'\). In the near-DNS limit, where the deconvolution error itself becomes asymptotically small, both projections can be neglected consistently.

However, the central approximation (Eq.~(7) of \cite{malaspinas_advanced_2011}) is $|f'|=|f-f^*| \ll 1$: the deconvolved distribution is assumed to be close to the true one.
Based on this, the ADM formulation drops the underbraced advective streaming of $f'$ from \cref{eq:adm_filtered} as an independent transport term while retaining $f'$ through the collision-side residual \(\mathcal{R}_2=\overline{\mathcal{Q}(f^*+f')-\mathcal{Q}(f^*)}\).
For a finite LES filter, this step is not justified by the smallness of \(f'\) alone and structurally parallels the collision-side-only ordering encoded by \cref{eq:ansumali_zero_sgs_moment_condition} in~\cite{ansumali_kinetic_2004}.
It retains the collision-side error but assumes that the advective deconvolution error is negligible, although its filtered-hierarchy counterpart is the leading SGS advective momentum flux.
This yields Eq.~(9) of \cite{malaspinas_advanced_2011}:
\begin{equation}\label{eq:adm_eq9}
    \overline{\pp_t f^* + \xi_{\alpha}\pp_{\alpha} f^* - \mathcal{Q}(f^*)}
    =
    \mathcal{R}_2\,,
\end{equation}
where $\mathcal{R}_2$ denotes the Taylor-closed ADM collision-deconvolution residual. Comparing \cref{eq:adm_filtered} with \cref{eq:adm_eq9}, the dropped term is precisely the filtered advective streaming of $f'$. Its first velocity moment is
\[
    \pp_t\int_\Xi \xi_{\alpha_1}\overline{f'}\,\dd\bm\xi
    +
    \pp_{\alpha_2}\int_\Xi
    \xi_{\alpha_1}\xi_{\alpha_2}\overline{f'}\,\dd\bm\xi ,
\]
so it contains the divergence of a second reconstruction-error moment. This is the ADM structural counterpart of the SGS momentum-flux contribution $\pp_{\alpha_2}\MT[m^{\sgs}][1][2]$ in the present filtered hierarchy.

The consequence is immediate. When taking the first velocity moment (Eq.\ (21) of \cite{malaspinas_advanced_2011}), the momentum flux is expressed in terms of the deconvolved moments and the pressure tensor $P^*$. The Euler-level equation obtained in Eq.\ (25) of \cite{malaspinas_advanced_2011},
\[
    \pp_t (\rho^* \MT[u^*][1])
    + \pp_{\alpha_2}(\rho^* \MT[u^*][1]\MT[u^*][2])
    = -\pp_{\alpha_1}p^*,
\]
therefore contains no explicit SGS stress tensor. The simplified model (Eq.\ 28 of \cite{malaspinas_advanced_2011}) yields the standard Navier--Stokes form for the deconvolved variables, with the ADM action retained through the explicit filtering/deconvolution step rather than through an Euler-level SGS flux.
Even in the full ADM model (Eq.~(27) of \cite{malaspinas_advanced_2011}), the correction terms are deconvolution corrections at the Navier--Stokes level. The issue is therefore not that ADM lacks an SGS mechanism, but where that mechanism is placed: it models the collision-side deconvolution residual while neglecting the advective counterpart that structurally mirrors the Euler-level SGS tensor $\MT[m^{\sgs}][1][2]$ of the present filtered hierarchy, as displayed in \cref{momentum_moments}.

In conclusion, the shortcoming of the ADM macroscopic equations is structurally analogous to that in Ansumali's approach~\cite{ansumali_kinetic_2004}: it neglects the emergence of an explicit SGS advective stress tensor in the Euler-level macroscopic equation.
As discussed in the previous section, however, even if SGS advection were included in the solution, this model would still lack an explicit model for the non-Markovian term considered here.

\section{A relative-frame formulation}
\label{app:girimaji}
In the present framework, the central kinetic modeling problem is not to reconstruct missing advective transport. Because filtering commutes with the linear Boltzmann streaming operator, the unresolved momentum transport is already carried by the filtered distribution and appears, at Euler level, through the SGS equilibrium commutation residual $f_{\sgs} \equiv \overline{f^{(0)}} - f^{(0)}$. The genuinely unresolved kinetic physics enters instead through the collision side, where filtering generates a non-Markovian collision-product covariance.

The relative-frame formulation for RANS and LES introduced by Girimaji \cite{girimaji_boltzmann_2007} targets a different representation: it rewrites the moment-level advective commutation structure into explicit transformed transport and acceleration terms by changing variables to a moving velocity frame. This appendix analyzes what that representation changes in the filtered hierarchy, first at Euler level and then in the modeled kinetic equation used for LES.

\subsection{Non-orthogonality of relative phase-space coordinates}
\label{app:geometric_obliquity}
In the standard kinetic description, the one-particle phase space is the product of physical space and velocity space, $\Gamma = \boldsymbol{x}^\flat \times \boldsymbol{\xi}^\flat$.
Here $\boldsymbol{x}^\flat$ and $\boldsymbol{\xi}^\flat$ are the convectively nondimensional spatial and velocity coordinates.
The coordinate directions are defined by the partial derivatives
\begin{equation}
    \mathbf{e}_{x_\alpha}^\flat \equiv \cpartial_\alpha \big|_{\boldsymbol{\xi}^\flat}, \quad \quad \mathbf{e}_{\xi_\beta}^{\flat} \equiv \pp_{\xi_\beta}^{\flat} \big|_{\boldsymbol{x}^\flat} \,.
\end{equation}
With the usual Euclidean product metric on $\Gamma$, these directions are orthogonal:
\begin{equation}
    \mathbf{e}_{x_\alpha}^\flat \cdot \mathbf{e}_{\xi_\beta}^{\flat} = 0 .
\end{equation}

Relative-frame formulations~\cite{girimaji_boltzmann_2007} introduce the relative velocity
\begin{equation}
    \boldsymbol{\eta}^\flat
    =
    \boldsymbol{\xi}^\flat - \boldsymbol{u}'^\flat(\boldsymbol{x}^\flat,t),
    \qquad
    \boldsymbol{u}'^\flat
    \equiv
    \boldsymbol{u}^\flat-\tilde{\boldsymbol{u}}^\flat .
\end{equation}
This is a smooth change of variables, but it mixes spatial and velocity coordinates because $\boldsymbol{u}'^\flat$ depends on $\boldsymbol{x}^\flat$.
For \(g(\boldsymbol{x}^\flat,\boldsymbol{\eta}^\flat,t)=f(\boldsymbol{x}^\flat,\boldsymbol{\xi}^\flat,t)\), the chain rule gives
\begin{align}
    \cpartial_t g \big|_{\boldsymbol{\eta}^\flat} &= \cpartial_t f \big|_{\boldsymbol{\xi}^\flat} + \cpartial_t u'^\flat_\gamma \pp_{\xi_\gamma}^{\flat} f \big|_{\boldsymbol{x}^\flat} \,, \label{eq:chain_rule_t} \\
    \cpartial_\alpha g \big|_{\boldsymbol{\eta}^\flat} &= \cpartial_\alpha f \big|_{\boldsymbol{\xi}^\flat} + \cpartial_\alpha u'^\flat_\gamma \pp_{\xi_\gamma}^{\flat} f \big|_{\boldsymbol{x}^\flat} \,, \label{eq:chain_rule_x} \\
    \pp_{\eta_\beta}^{\flat} g \big|_{\boldsymbol{x}^\flat} &= \pp_{\xi_\beta}^{\flat} f \big|_{\boldsymbol{x}^\flat} \,.
\end{align}
Thus the relative-velocity direction remains unchanged,
\(\mathbf{E}_{\eta_\beta}^{\flat} \equiv \pp_{\eta_\beta}^{\flat} = \mathbf{e}_{\xi_\beta}^{\flat}\).
In contrast, the spatial direction at fixed \(\boldsymbol{\eta}^\flat\) becomes
\begin{equation}
    \mathbf{E}_{x_\alpha}^\flat \equiv \cpartial_\alpha \big|_{\boldsymbol{\eta}^\flat} = \mathbf{e}_{x_\alpha}^\flat + \cpartial_\alpha u'^\flat_\gamma \mathbf{e}_{\xi_\gamma}^{\flat} \,.
\end{equation}
Therefore the relative coordinate system is not orthogonal in general.
The mixed inner product is
\begin{equation}
    \mathbf{E}_{x_\alpha}^\flat \cdot \mathbf{E}_{\eta_\beta}^{\flat} = \cpartial_\alpha u'^\flat_\beta \neq 0 \,.
\end{equation}
Thus spatial and relative-velocity directions become coupled whenever $\cpartial_\alpha u'^\flat_\beta \neq 0$.

\subsection{The relative-frame BGK equation}
When rewriting the inertial streaming operator in the moving, non-inertial coordinates, the chain rule generates explicitly $\boldsymbol{\eta}^\flat$-dependent transport terms, including a force-like velocity-space derivative. Applying the filter operator over the transformed variables yields the following relative-frame kinetic equation reported in \cite{girimaji_boltzmann_2007}:
\begin{equation}\label{girimajiBE}
    \cpartial_t \overline{g} + \eta^\flat_{\alpha_1} \cpartial_{\alpha_1} \overline{g} + \cpartial_{\alpha_1} \overline{(g u'^\flat_{\alpha_1})} - \pp_{\eta_{\alpha_2}}^{\flat} \overline{(g \mathcal{A}^\flat_{\alpha_2})} = -\frac{\omega^\sharp}{\epsilon}(\overline{g}-g^{(0)}),
\end{equation}
where $\mathcal{A}^\flat_{\alpha_2} \equiv \cpartial_t u'^\flat_{\alpha_2} + \eta^\flat_{\alpha_3} \cpartial_{\alpha_3} u'^\flat_{\alpha_2} + u'^\flat_{\alpha_3} \cpartial_{\alpha_3} u'^\flat_{\alpha_2}$ is the induced phase-space acceleration, \(\overline{g^{(0)}}=g^{(0)}\), and the right-hand side is the BGK collision term \(-\frac{\omega^\sharp}{\epsilon}(\overline{g}-g^{(0)})\) in the relative-frame representation.

The transformed equation therefore contains both an additional physical-space transport term, $\cpartial_{\alpha_1} \overline{(g u'^\flat_{\alpha_1})}$, and a force-like velocity-space term, $\pp_{\eta_{\alpha_2}}^{\flat} \overline{(g \mathcal{A}^\flat_{\alpha_2})}$, arising from the chosen moving velocity coordinate. In this sense, the relative-frame formulation makes explicit terms that are induced by the coordinate transformation rather than by any change in the underlying inertial-frame physics.

In~\cite{girimaji_boltzmann_2007} \cref{girimajiBE} is then simplified by assuming $\overline{(g u'^\flat_{\alpha_1})} \approx \overline{g}\,\overline{u'^\flat_{\alpha_1}} = 0$:
\begin{equation}\label{girimajiBE2}
    \cpartial_t \overline{g} + \eta^\flat_{\alpha_1} \cpartial_{\alpha_1} \overline{g}  - \pp_{\eta_{\alpha_2}}^{\flat} \overline{(g \mathcal{A}^\flat_{\alpha_2})} = -\frac{\omega^\sharp}{\epsilon}(\overline{g}-g^{(0)}).
\end{equation}

\subsection{Euler limit}

In the inertial-frame Euler limit already derived in \cref{momentum_moments} the SGS stress appears at leading order through the second raw moment of $\overline{f^{(0)}} = f^{(0)} + f_{\sgs}$.
In contrast, the Euler level component of \cref{girimajiBE2} reads
\begin{equation}\label{eq:girimaji_euler}
    \begin{aligned}
        \partial_t^{(1)} \overline{g^{(0)}}
        + \eta^\flat_{\alpha_1}\partial_{\alpha_1}^{(1)} \overline{g^{(0)}}
        - \pp_{\eta_{\alpha_2}}^{\flat}
          \overline{(g^{(0)}\mathcal{A}_{\alpha_2}^{\flat(1)})}
        = -\omega^\sharp \overline{g^{(1)}}\,.
    \end{aligned}
\end{equation}
This equation shows the effect of Girimaji's closure at Euler order. In the inertial filtered formulation, the SGS tensor is carried by the second raw moment of $\overline{f^{(0)}}=f^{(0)}+f_{\sgs}$ and enters the Euler flux in \cref{momentum_moments}. After the relative-frame equilibrium replacement, $\overline{g^{(0)}}$ contains only the resolved advective and pressure moments. The tensor $\MT[m^{\sgs}][1][2]$ is therefore no longer transported as part of the one-particle equilibrium distribution; it must be supplied through the force-like coefficient $\overline{(g^{(0)}\mathcal{A}_{\alpha_2}^{\flat(1)})}$.

Taking the momentum moment $\int \eta_{\alpha_1}^\flat (\cdot)\,\dd\bm\eta^\flat$ of the unclosed transformed equation, assuming Reynolds averaging rules for the density-weighted moments, yields three contributions.
The first $\boldsymbol{\eta}^\flat$-moment of the additional transport term $\cpartial_{\alpha_1} \overline{(g u'^\flat_{\alpha_1})}$ does not contribute under these Favre identities, so the remaining terms are:
\begin{enumerate}
    \item \textit{Time derivative:} $\int \eta_{\alpha_1}^\flat \partial_t^{(1)} \overline{g^{(0)}} \dd\bm\eta^\flat = \partial_t^{(1)} (\bar{\rho} \MT[\tilde{u}^\flat][1])$.
    \item \textit{Relative streaming flux:} Under this assumed equilibrium closure, the second moment of the relative equilibrium,
    $\T[m^{\zero}][1][2] = \int \eta_{\alpha_1}^\flat \eta_{\alpha_2}^\flat \overline{g^{(0)}} \dd\bm\eta^\flat$, contains only the resolved advective part and the thermal pressure:
    \begin{equation}
        \begin{aligned}
            \T[m^{\zero}][1][2] &= \int \eta_{\alpha_1}^\flat \eta_{\alpha_2}^\flat \overline{g^{(0)}} \dd\bm\eta^\flat \\
            &= \bar{\rho} \MT[\tilde u^\flat][1] \MT[\tilde u^\flat][2] + \bar{p}^\flat \MT[\delta][1][2] \,.
        \end{aligned}
    \end{equation}
    Thus, the unresolved covariance $\MT[m^{\sgs}][1][2]$ is absent from this equilibrium contribution.
    \item \textit{Force-like moment:} The acceleration term $\mathcal{A}_{\alpha_2}^\flat$ contributes through the velocity-space derivative. Integration by parts gives
    \begin{equation}
        - \int \eta_{\alpha_1}^\flat \pp_{\eta_{\alpha_2}}^{\flat} \overline{\big(g^{(0)}  \mathcal{A}_{\alpha_2}^{\flat(1)}\big)} \dd\bm\eta^\flat
        = \int \overline{g^{(0)}  \mathcal{A}_{\alpha_1}^{\flat(1)}} \dd\bm\eta^\flat \,,
    \end{equation}
    where $\mathcal{A}_{\alpha_2}^{\flat(1)} \equiv \partial_t^{(1)} u'^\flat_{\alpha_2} + \eta^\flat_{\alpha_3} \partial_{\alpha_3}^{(1)} u'^\flat_{\alpha_2} + u'^\flat_{\alpha_3} \partial_{\alpha_3}^{(1)} u'^\flat_{\alpha_2}$ denotes the induced phase-space acceleration at the convective scale. Under the same incompressible moment identities used in \cite{girimaji_boltzmann_2007} to obtain Eq.~(17), this force-like contribution precisely supplies the SGS-stress divergence that is absent from the equilibrium second moment.
\end{enumerate}

Accordingly, the transformed first-moment balance recovers the same Euler-level momentum equation obtained in the inertial frame (\cref{momentum_moments}):
\begin{equation}
    \partial_t^{(1)} (\bar{\rho} \MT[\tilde u^\flat][1]) + \partial_{\alpha_2}^{(1)} (\bar{\rho} \MT[\tilde u^\flat][1] \MT[\tilde u^\flat][2] + \bar{p}^\flat \MT[\delta][1][2] + \MT[m^{\sgs}][1][2]) = 0 \,.
\end{equation}
At the Euler level, the coordinate transformation therefore changes only the representation of the unresolved stress: in the inertial frame it appears explicitly through the SGS part of the filtered equilibrium moment, whereas in the relative frame it re-enters through the transformed force-like contribution. Under the present Euler-level analysis, the same SGS second-order stress content remains.

In this respect, the closed relative-frame equation is closer to the FNSE structure in \cref{FNSE}: the advective SGS stress requires a separate closure instead of being carried by $f_{\sgs}$. However, in light of the developments discussed in this article, we find that this approach results in the loss of a fundamental advantage of kinetic modeling, which is precisely that of obtaining the transport of the SGS component ``for free''.

\subsection{Navier--Stokes-order limit}
The work~\cite{girimaji_boltzmann_2007} does not perform a CE but only a change of variables. A CE expansion could in principle be carried out in the relative variables by using \cref{eq:girimaji_euler} to express $\overline{g^{(1)}}$ in the ${O}(\epsilon^2)$ equation to see the effect of the collision-relaxation in the relative reference frame (rhs of \cref{girimajiBE}).
\Cref{eq:girimaji_euler} shows that the transformed advective couplings remain explicitly present in the nonequilibrium correction consistently with the expansion in the inertial frame \cref{CE}.
However, given that the work in~\cite{girimaji_boltzmann_2007} does not perform this Navier--Stokes-order analysis, no hydrodynamic limit (in the CE sense) of \cref{girimajiBE2} is derived in~\cite{girimaji_boltzmann_2007}.

\subsection{LES stress model}

With the notation used here and without the CE-order decomposition, the closed equation actually advanced in \cite{girimaji_boltzmann_2007} uses
\begin{align}
    \overline{(g u'^\flat_{\alpha_1})} &\approx 0, \\
    \overline{(g \mathcal{A}^\flat_{\alpha_1})} &\approx
    \overline{g}\,\bar{\rho}^{-1}\cpartial_{\alpha_2}\MT[m^{\sgs}][1][2].
\end{align}
Thus Eq.~(19) of \cite{girimaji_boltzmann_2007} corresponds to
\begin{equation}
    \cpartial_t \overline{g}
    + \eta^\flat_{\alpha_1} \cpartial_{\alpha_1} \overline{g}
    - \bar{\rho}^{-1}\cpartial_{\alpha_2}\MT[m^{\sgs}][1][2]\,
      \pp_{\eta_{\alpha_1}}^{\flat}\overline{g}
    =
    -\frac{\omega^\sharp}{\epsilon}(\overline{g}-g^{(0)})\,.
\end{equation}

In~\cite{girimaji_boltzmann_2007}, $\MT[m^{\sgs}][1][2]$ is not computed from the relative-frame kinetic equation but is prescribed by a Smagorinsky SGS model.

\subsection{Conclusions}
The relative-frame transformation does not remove the SGS influence from the kinetic dissipative problem; it rewrites the transport-side contribution while leaving the collision-side closure unaddressed. At Euler level it recovers the same unclosed second-order stress content already carried by $f_{\sgs}$ in the inertial filtered formulation, but represents it through transformed transport and acceleration terms that are explicitly modeled by a Smagorinsky SGS model.

In other words, the relative-frame framework externalizes the advective SGS content carried implicitly by the inertial filtered distribution and closes it through a Smagorinsky model of the SGS stress, as in the FNSE. In this sense, the claim of reconciliation with macroscopic SGS modeling is correct: the SGS cascade is modeled through an external transport closure, so a separate collision-covariance dissipation model is no longer needed within that route. However, this bypasses rather than solves the non-Markovian collision-covariance closure problem of the filtered kinetic equation.

Therefore, the physical role of the relative frame remains limited. It recasts unresolved momentum transfer, but it neither alters the Euler-level SGS content nor supplies an independent kinetic closure nor a Chapman--Enskog constitutive derivation of the dissipative stresses. More importantly, it discards the fundamental advantage of kinetic modeling: the linear transport of the SGS terms.

\section{Klimontovich turbulence formulation of Chen et al. and its relation to the collision-covariance source term}
\label{app:chen_comparison}

This appendix develops a comparison between the Klimontovich-based turbulence formulation introduced by Chen et al. \cite{chen_average_2023,chen_average_2024} and the source-term framework of the present work. The collision term introduced in Chen's formulation plays the same structural role as the filtered collision-covariance source term $\overline{\Ecc^\sharp}$, but the two approaches differ in how that unresolved term is derived and closed. For readability, the original notation of \cite{chen_average_2023,chen_average_2024} is written into the conventions of the present article wherever possible.
Chen et al. denote by angle brackets a Reynolds ensemble average over fluid-element realizations.
For the present comparison, we denote Chen's average by an overline.

\subsection{Top-down kinetic lifting of the Navier--Stokes equations}

Chen's approach~\cite{chen_average_2023} defines a probability density function $\psi(\boldsymbol{x},\boldsymbol{\xi},t)$ for an ensemble of infinitesimal fluid elements.
In its incompressible form, this PDF is normalized as $\int \psi\,\mathrm{d}\bm{\xi} = 1$, so its moments give the hydrodynamic velocity directly rather than density-weighted kinetic moments. The evolution of $\psi$ obeys a Klimontovich-type equation with a body-force acceleration $A_\alpha(\boldsymbol{x},t)$:
\begin{equation}\label{eq:chen_klimontovich}
    \partial_t \psi + \xi_\alpha\,\partial_\alpha \psi + A_\alpha\,\partial_{\xi_\alpha} \psi = 0.
\end{equation}
In the unfiltered case, $\psi$ is a monokinetic distribution, i.e. a Dirac delta in velocity space centered at $\boldsymbol{u}(\boldsymbol{x},t)$. This zero-temperature constraint gives
\begin{equation}\label{eq:chen_zero_temp}
    \int \xi_{\alpha_1}\xi_{\alpha_2}\, \psi\, \mathrm{d}\bm{\xi} = u_{\alpha_1} u_{\alpha_2},
\end{equation}
where $u_\alpha(\boldsymbol{x},t)$ is the (unaveraged) Navier--Stokes velocity field. Taking the first moment of \cref{eq:chen_klimontovich} and using \cref{eq:chen_zero_temp},
\begin{equation}
    \partial_t u_{\alpha_1} + \partial_{\alpha_2}(u_{\alpha_1} u_{\alpha_2}) = A_{\alpha_1}.
\end{equation}
Thus, choosing $A_\alpha \equiv -\partial_\alpha p + \nu\,\partial_\beta\partial_\beta u_\alpha$ recovers the exact incompressible Navier--Stokes equations. The kinetic equation \cref{eq:chen_klimontovich} is therefore not an independent physical postulate but a top-down kinetic lifting of the NSE into a Klimontovich representation.

\subsection{Averaging and the emergence of the collision term}

Here and below, \(\overline{(\cdot)}\) denotes Chen's fluid-element average, whereas \(\langle\cdot\rangle_{\rm ens}\) remains reserved for the molecular ensemble average that defines \(\mathcal F\). With this convention, the fluctuations are defined by
\begin{equation}\label{eq:chen_avg}
    \begin{aligned}
    \delta\psi&\equiv \psi-\overline{\psi},
    &
    \delta A_\alpha&\equiv A_\alpha-\overline{A}_\alpha .
    \end{aligned}
\end{equation}
The following first-moment argument shows why these two averaging operations cannot be identified.

\paragraph{First-moment distinction.} In Chen's formulation, Eq.~(5) defines the hydrodynamic fluid-element velocity by \(\int_\Xi \xi_\alpha\psi\,\mathrm{d}\bm{\xi}=u_\alpha\).
In the present framework, \cref{rho_u_theta} identifies the same hydrodynamic velocity through the time-filtered molecular distribution \(F\).
Thus the local first-moment comparison gives
\begin{equation}\label{eq:chen_1st}
    \frac{1}{\varrho}
    \int_\Xi \xi_\alpha F\,\mathrm{d}\bm{\xi}
    =
    \int_\Xi \xi_\alpha \psi\,\mathrm{d}\bm{\xi}.
\end{equation}
Chen's Eq.~(10) then defines the hydrodynamically averaged velocity \(\overline{u}_\alpha\) by
\(
    \int_\Xi \xi_\alpha \overline{\psi}\,\mathrm{d}\bm{\xi}
    =
    \overline{u}_\alpha.
\)
Therefore the Chen fluctuation split gives
\begin{equation}
    \int_\Xi \xi_\alpha \psi\,\mathrm{d}\bm{\xi}
    =
    \int_\Xi \xi_\alpha \overline{\psi}\,\mathrm{d}\bm{\xi}
    +
    \int_\Xi \xi_\alpha
    \delta\psi\,\mathrm{d}\bm{\xi}.
\end{equation}
To make this distinction explicit, suppose temporarily that the \(\overline{(\cdot)}\) average in \cref{eq:chen_avg} is actually the molecular ensemble average \(\langle(\cdot)\rangle_{\rm ens}\).
Then, for a fixed linear time filter, the molecular ensemble average commutes with the time filter and acts as a Reynolds operator.
Therefore, the left-hand side of \cref{eq:chen_1st} reads
\begin{equation}\label{eq:chen_lhs}
    \begin{aligned}
    \left\langle
    \frac{1}{\varrho}
    \int_\Xi \xi_\alpha
    \langle\langle N\rangle_{\rm ens}\rangle_{\Delta_t}
    \,\mathrm{d}\bm{\xi}
    \right\rangle_{\rm ens}
    &=
    \frac{1}{\varrho}
    \int_\Xi \xi_\alpha
    \langle\langle N\rangle_{\rm ens}\rangle_{\Delta_t}
    \,\mathrm{d}\bm{\xi}\\
    &=
    \frac{1}{\varrho}
    \int_\Xi \xi_\alpha
    F
    \,\mathrm{d}\bm{\xi} = u_\alpha.
    \end{aligned}
\end{equation}
The same ensemble operation would remove the fluctuation part from the right-hand side of \cref{eq:chen_1st},
\begin{equation}\label{eq:chen_rhs}
    \left\langle\int_\Xi \xi_\alpha \overline{\psi}\,\mathrm{d}\bm{\xi}\right\rangle_{\rm ens}+
    \cancel{\left\langle
    \int_\Xi \xi_\alpha
    \delta\psi
    \,\mathrm{d}\bm{\xi}
    \right\rangle_{\rm ens}}
    =
    U_\alpha,
\end{equation}
where, under this counterfactual identification, \(U_\alpha = \langle u_\alpha \rangle_{\rm ens}\).
Combining \cref{eq:chen_lhs,eq:chen_rhs} leads to
\begin{equation}
    u_\alpha = U_\alpha
\end{equation}
which would collapse the local and averaged hydrodynamic velocities.
Thus, in the present notation, Chen's average is the operation that maps \(\psi\) to \(\overline\psi\), not the molecular average that maps \(N\) to \(\mathcal F\).

The word ensemble can still be retained when referring to statistically equivalent fluid-element samples.
Homogeneity and stationarity permit equivalent spatial and temporal statistics, consistent with the Lagrangian correlations considered in~\cite{chen_average_2024}.

\paragraph{Chen's equation and the collision term.} Averaging \cref{eq:chen_klimontovich} then yields
\begin{equation}\label{eq:chen_averaged}
    \partial_t \overline{\psi}
    + \xi_\alpha\,\partial_\alpha \overline{\psi}
    + \overline{A}_\alpha\,\partial_{\xi_\alpha} \overline{\psi}
    = C,
\end{equation}
where
\begin{equation}\label{eq:chen_collision}
    C \equiv -\partial_{\xi_\alpha}
    \overline{\delta A_\alpha\,\delta\psi}.
\end{equation}
The term \(C\) is a velocity-space divergence built from \(\delta A_\alpha\delta\psi\).
Since \(\delta A_\alpha\) is independent of \(\boldsymbol{\xi}\) and \(\int \delta\psi\,\mathrm{d}\bm{\xi}=0\), its zeroth and first velocity moments vanish, so \(C\) conserves mass and momentum, consistently with Eq.~(21) of~\cite{chen_average_2023}.
However, it does not conserve \(\mathcal{K}\), the second central moment of \(\overline{\psi}\).
Because the zeroth and first moments of \(C\) vanish, the raw and central second moments of the collision contribution coincide.
The second-moment balance of the averaged kinetic equation, Eq.~(32) of~\cite{chen_average_2023}, can then be compared with the turbulent-kinetic-energy balance obtained from the averaged NSE, Eq.~(25) of~\cite{chen_average_2023}.
This comparison gives Eq.~(33) of~\cite{chen_average_2023}, written here in central-moment form as
\begin{equation}\label{eq:chen_C_energy_identity}
    \int_\Xi \frac{1}{2}(\xi_\alpha-\overline u_\alpha)(\xi_\alpha-\overline u_\alpha)\, C\,\mathrm{d}\bm{\xi}
    =
    -\partial_\beta W_\beta+\nu\,\partial_\beta\partial_\beta \mathcal{K}-\varepsilon,
\end{equation}
where \(W_\beta\) is Chen's pressure-work flux.
The reduction to \(-\varepsilon\) holds only after homogeneous/global averaging, or when the flux and diffusion terms vanish.

In the present work (\cref{sec:filtered_recorrelated_BGK_Boltzmann_equation}), the dimensional collision-covariance source term $\Ecc$ arises from the time-filtering of the binary collision integral.
After nondimensionalization and homogeneous LES filtering, this term enters the modeled equation as $\overline{\Ecc^\sharp}$.

Both $C$ and $\overline{\Ecc^\sharp}$ play the same structural role as unresolved covariance-driven source terms, although they arise from different derivations. The structural parallel and the closure-level differences are summarized in \cref{tab:chen_structural}.

\begin{table*}[t]
\caption{Comparison between the collision term $C$ of~\cite{chen_average_2023} and the present filtered collision-covariance source term $\overline{\Ecc^\sharp}$. Here $\mathcal{K}$, $\varepsilon$, and $\varnu_T$ denote Chen's TKE, externally supplied dissipation rate, and modeled turbulent viscosity.}
\label{tab:chen_structural}
\begin{tblr}{
  width=\linewidth,
  colspec={lXX},
  hline{1,2,9}={solid, 0.08em},
}
 & Chen's formulation (RANS) & Kinetic closure (present LES formulation) \\
\hline
Origin of unresolved source & Fluctuations of NSE stresses ($\delta A_\alpha$, $\delta\psi$) & Collision-product covariance ($\Ccc(\mathcal{F}\mathcal{F}_{i\!i})$)\\
Conservation / dissipation & Mass, momentum; does not conserve TKE & Exact elastic covariance: conserves mass, momentum, and energy. Isothermal scalar closure: conserves mass and momentum and removes SGS energy \\
Effect & \SetCell[c=2]{c} Unresolved covariance-driven source acting on the velocity distribution & \\
Equilibrium target & $\overline{\psi}^{\Eq}(\mathcal{K}^{\Eq})$ (turbulent Gaussian) & $0$ (null SGS carrier; no turbulent equilibrium state) \\
Relaxation rate & $1/\vartau_t$ with $\vartau_t = \frac{6}{7}\mathcal{K}/\varepsilon$ & $\omega_t^\sharp$ (phenomenological SGS relaxation rate) \\
Turbulent temperature & Yes ($T_{\rm turb} = 2\mathcal{K}/3$) & No \\
Example operational realization & $\varnu_T = C_\mu \mathcal{K}^2/\varepsilon$, $C_\mu \approx 0.0816$ (needs separate $\varepsilon$-equation) & e.g.\ $\omega_t^\sharp=\tilde\theta^\sharp/(C_\nu\Delta_x^\sharp\sqrt{k_\sgs^\sharp})$  \\
\end{tblr}
\end{table*}

\subsection{The difference in closure strategy}

The energy identity \cref{eq:chen_C_energy_identity} constrains the second moment of \(C\), but it does not determine the full collision operator.
The closure adopted in \cite{chen_average_2023,chen_average_2024} then assumes that the covariance-driven interaction relaxes \(\overline{\psi}\) toward a Gaussian target.
The target is centered at \(\overline u_\alpha\) to preserve momentum.
Its width is set by a reduced equilibrium turbulent kinetic energy, \(\mathcal{K}^{\Eq}=\mathcal{K}-\vartau_t\varepsilon\), so that the relaxation accounts for the energy dissipated during one relaxation time. This sink-only target omits the pressure-work and diffusion terms in \cref{eq:chen_C_energy_identity}.
This gives the BGK form
\begin{equation}\label{eq:chen_closure}
    \begin{aligned}
    C &\approx -\frac{1}{\vartau_t}
    \left(\overline{\psi} - \overline{\psi}^{\Eq}\right),\\
    \overline{\psi}^{\Eq}
    &= W\exp\left[-\frac{3(\xi_\alpha - \overline u_\alpha)^2}{4\mathcal{K}^{\Eq}}\right],
    \qquad
    \mathcal{K}^{\Eq}=\mathcal{K}-\vartau_t\varepsilon .
    \end{aligned}
\end{equation}
Here \(\mathcal{K} = \frac{1}{2}\overline{\delta u_\alpha\delta u_\alpha}\), \(\varepsilon\) is the macroscopic turbulent dissipation rate, and \(\vartau_t = \vartau_t(\boldsymbol{x},t)\) is the turbulent relaxation time. The factor \(W\) normalizes the PDF to unity.
Chen interprets \(\mathcal{K}\) as defining the turbulent-temperature analogue, \(T_{\rm turb}=2\mathcal{K}/3\), while the BGK target uses the reduced equilibrium value \(\mathcal{K}^{\Eq}\).
The turbulent viscosity follows from the Chapman--Enskog expansion of the BGK equation as $\varnu_T = \frac{2}{3}\vartau_t \mathcal{K}^{\Eq}$.

It is important to note that while $\mathcal{K}$ is self-consistently transported by the kinetic formulation (as the second moment of \(\overline{\psi}\)), the turbulent dissipation rate $\varepsilon$ is not natively determined. The framework relies on $\varepsilon$ as an external macroscopic input to define the relaxation time $\vartau_t$ and the equilibrium state \(\overline{\psi}^{\Eq}\). Consequently, the value of $\varepsilon$ must be supplied independently (e.g., from a separate macroscopic dissipation transport equation) to close the kinetic model.

By comparison, in the present framework, the analogous source term is modeled as (\cref{Ebgk_approx})
\begin{equation}
    \overline{\Ecc^\sharp} \approx -\omega_t^\sharp f_{\sgs}.
\end{equation}
The modeled term preserves mass and momentum because the SGS carrier $f_{\sgs}$ has vanishing zeroth and first velocity moments by construction, while its second and higher moments remain non-zero and encode the unresolved stress content. The closure therefore dissipates SGS second-moment content without introducing a turbulent equilibrium state or turbulent temperature as an additional constitutive variable. The present approach retains the molecular thermodynamic equilibrium (the Maxwell-Boltzmann distribution at the physical temperature $\theta_R$) and relaxes the SGS equilibrium commutation residual $f_{\sgs}$ as the SGS carrier at $\omega_t^\sharp$ (\cref{tab:chen_structural}).

\subsection{Summary}

Chen's approach and the present one~\cite{marson_kinetic_2025} both go beyond the simple calculation of fine-grained equilibrium or the modeling of the advective SGS term, and instead focus on modeling unresolved covariance-driven source terms.
However, Chen's Klimontovich framework and the present collision-covariance
framework diverge at the closure level, as summarized in
\cref{tab:chen_structural}:
Chen's Klimontovich formulation introduces a turbulent equilibrium state and determines the relaxation time $\vartau_t$ from a fluctuation-dissipation argument involving the external macroscopic dissipation rate $\varepsilon$, whereas the present kinetic framework represents the collision-covariance source term through the SGS carrier $f_{\sgs}$ and the turbulent relaxation frequency $\omega_t^\sharp$, without introducing a turbulent temperature, a separate equilibrium state, or a secondary macroscopic transport equation for SGS information that is already carried by the filtered distribution under the linear streaming operator.
Furthermore, the present model is implemented within the LBM update without an additional macroscopic transport equation.

\section{Hermite representation}
\label{app:hermite_representation}

This appendix gives the nondimensional form of Grad's fixed-reference Hermite expansion used in Sections~\ref{subsec:KCRB_theory}--\ref{subsec:KCRR_theory}~\cite{grad_note_1949,shan_2006_hydrodynamics-navier-stokes}. The particle velocity variable $\bm{\xi}$ enters through two nondimensional representations: the diffusive $\bm{\xi}^\sharp=\bm{\xi}/\sqrt{\theta_R}$ and the convective $\bm{\xi}^\flat=\bm{\xi}/\mcU$. Independently of the nondimensionalization, $\bm{\xi}$ is inherently mixed-scale:
\begin{align}
    \xi_\alpha^\flat &= u_\alpha^\flat + \Ma^{-1}\zeta_\alpha^\sharp \,,\\
    \xi_\alpha^\sharp &= \Ma\, u_\alpha^\flat + \zeta_\alpha^\sharp \,.
\end{align}
Thus, powers of \(\Ma^{-1}\) enter in the description through the thermal part of the particle velocity.

At the reference state, $\bm u_R=\bm 0$, so $\bm\xi^\sharp=\Ma\bm\xi^\flat$ and $d\bm\xi^\sharp=\Ma^3d\bm\xi^\flat$. Hence the same normalized Gaussian takes the forms
\begin{align}\label{w}
    w^\sharp(\bm\xi^\sharp)
    &= \frac{\rho_R}{(2\pi \theta_R^\sharp)^{3/2}}
    \exp\left(-\frac{\xi^{\sharp 2}}{2\theta_R^\sharp}\right),
    \\ \label{wflat}
    w^\flat(\bm\xi^\flat)
    &= \Ma^3 w^\sharp(\Ma\bm\xi^\flat)
    = \frac{\rho_R\Ma^3}{(2\pi \theta_R^\sharp)^{3/2}}
    \exp\left(-\frac{\Ma^2\xi^{\flat 2}}{2\theta_R^\sharp}\right)
    ,
\end{align}
Equivalently, $w^\flat=f^\zero(\rho_R,\bm 0,\theta_R^\sharp)$, where $f^\zero$ is the convective velocity-space-density Maxwellian associated with $f\equiv f^\flat$; its arguments specify the reference thermodynamic state rather than the velocity-space measure. Since $w^\flat\,d\bm\xi^\flat=w^\sharp\,d\bm\xi^\sharp$, both reference weights have the same zeroth moment,
\begin{equation}
    \int_\Xi w^\flat\,d\bm\xi^\flat
    = \int_\Xi w^\sharp\,d\bm\xi^\sharp
    = \rho_R
    \equiv \frac{\varrho_R}{\varrho_R}
    = 1.
\end{equation}
The reference temperature is likewise normalized in diffusive units as $\theta_R^\sharp\equiv\theta_R/\theta_R=1$. We retain $\rho_R$ and $\theta_R^\sharp$ in the formulas below to keep the reference normalization explicit.
The fixed-reference Hermite polynomials follow from the scaled Rodrigues formula
\begin{equation}
    H^\sharp_{\alpha_1 \alpha_2 \cdots \alpha_n}
    = \frac{(-\theta_R^\sharp)^n}{w^\sharp}\,
    \partial_{\xi^\sharp_{\alpha_1}}\partial_{\xi^\sharp_{\alpha_2}}\cdots
    \partial_{\xi^\sharp_{\alpha_n}}w^\sharp .
\end{equation}
Thus the Hermite basis is defined in diffusive ($\sharp$) variables, while the expansion of $f\equiv f^\flat$ below uses the convective ($\flat$) density weight $w^\flat$. This mixed representation is deliberate: the Hermite basis in diffusive variables preserves the fixed thermal Gaussian structure, whereas the density and coefficients in convective variables keep the hydrodynamic moments in convective units.
The first polynomials are
\begin{align}
    H^\sharp_0 &= 1 ,  \\
    H^\sharp_{\alpha_1} &= \xi^\sharp_{\alpha_1} = \Ma\,\xi^\flat_{\alpha_1} ,  \\
    H^\sharp_{\alpha_1 \alpha_2} &= \xi^\sharp_{\alpha_1}\xi^\sharp_{\alpha_2} - \theta_R^\sharp\delta_{\alpha_1 \alpha_2}
    = \Ma^2\xi^\flat_{\alpha_1} \xi^\flat_{\alpha_2} - \theta_R^\sharp\delta_{\alpha_1 \alpha_2} , \\
    H^\sharp_{\alpha_1 \alpha_2 \alpha_3} &= \xi^\sharp_{\alpha_1}\xi^\sharp_{\alpha_2}\xi^\sharp_{\alpha_3} \nonumber\\
    &\quad - \theta_R^\sharp\left( \delta_{\alpha_1 \alpha_2} \xi^\sharp_{\alpha_3}
    + \delta_{\alpha_1 \alpha_3} \xi^\sharp_{\alpha_2}
    + \delta_{\alpha_2 \alpha_3} \xi^\sharp_{\alpha_1} \right)\nonumber\\
    &= \Ma^3\xi^\flat_{\alpha_1} \xi^\flat_{\alpha_2} \xi^\flat_{\alpha_3} \nonumber\\
    &\quad - \Ma\,\theta_R^\sharp\left( \delta_{\alpha_1 \alpha_2} \xi^\flat_{\alpha_3}
    + \delta_{\alpha_1 \alpha_3} \xi^\flat_{\alpha_2}
    + \delta_{\alpha_2 \alpha_3} \xi^\flat_{\alpha_1} \right).
\end{align}
The convectively scaled polynomials $H^\flat_{\alpha_1\dots\alpha_n}\equiv\Ma^{-n}H^\sharp_{\alpha_1\dots\alpha_n}$ read
\begin{align} \label{eq:H_advective}
    H^\flat_0 &= 1 ,  \\
    H^\flat_{\alpha_1} &= \xi^\flat_{\alpha_1} ,  \\
    H^\flat_{\alpha_1 \alpha_2} &= \xi^\flat_{\alpha_1} \xi^\flat_{\alpha_2}
    - \Ma^{-2}\theta_R^\sharp\delta_{\alpha_1 \alpha_2} , \\
    H^\flat_{\alpha_1 \alpha_2 \alpha_3}
    &= \xi^\flat_{\alpha_1} \xi^\flat_{\alpha_2} \xi^\flat_{\alpha_3}\nonumber\\
    &\quad - \Ma^{-2}\theta_R^\sharp\left(\delta_{\alpha_1 \alpha_2} \xi^\flat_{\alpha_3}
    + \delta_{\alpha_1 \alpha_3} \xi^\flat_{\alpha_2}\right.\nonumber\\
    &\qquad\qquad\qquad\left.
    + \delta_{\alpha_2 \alpha_3} \xi^\flat_{\alpha_1}\right).
\end{align}
With the same convective scaling for the Hermite coefficients, $a_{\alpha_1 \dots \alpha_n} \equiv \Ma^{-n} a^\sharp_{\alpha_1 \dots \alpha_n}$, the expansion has three equivalent forms:
\begin{equation}\label{hermite-expansion}
\begin{aligned}
    f
    &= \frac{w^\flat}{\rho_R} \sum_{n=0}^{\infty} \frac{1}{n! \theta_R^{\sharp n}}\,
    H^\sharp_{\alpha_1 \dots \alpha_n} a^\sharp_{\alpha_1 \dots \alpha_n}.
\end{aligned}
\end{equation}
\begin{equation}
\begin{aligned}
    f
    &= \frac{w^\flat}{\rho_R} \sum_{n=0}^{\infty} \frac{\Ma^{n}}{n! \theta_R^{\sharp n}}\,
    H^\sharp_{\alpha_1 \dots \alpha_n} a_{\alpha_1 \dots \alpha_n}.
\end{aligned}
\end{equation}
\begin{equation}
\begin{aligned}
    f
    &= \frac{w^\flat}{\rho_R} \sum_{n=0}^{\infty} \frac{\Ma^{2n}}{n! \theta_R^{\sharp n}}\,
    H^\flat_{\alpha_1 \dots \alpha_n} a_{\alpha_1 \dots \alpha_n}.
\end{aligned}
\end{equation}
Here $H^\flat_{\alpha_1 \dots \alpha_n}$ are the convectively scaled Hermite polynomials and $a_{\alpha_1 \dots \alpha_n}\equiv a_{\alpha_1 \dots \alpha_n}^\flat$ are the corresponding convectively scaled Hermite moments~\cite{shan_2006_hydrodynamics-navier-stokes, malaspinas_lattice_2009, malaspinas_increasing_2015}.
The first equation is the fixed-reference diffusive-coefficient ($\sharp$) form for $f \equiv f^\flat$. The second equation uses the same fixed-reference polynomials but the convectively scaled coefficients $a_{\alpha_1\dots\alpha_n}=\Ma^{-n}a^\sharp_{\alpha_1\dots\alpha_n}$; this is the form used in the closure formulas below because it bridges the Hermite basis in diffusive variables with convectively scaled ($\flat$) hydrodynamic moments. The third equation rewrites the same expansion with the convectively scaled polynomials. Since $\rho_R=\theta_R^\sharp=1$ here, the retained reference factors only record the fixed-density and fixed-temperature normalization.

The CE reconstruction uses $\epsilon \overline{f^\one}$, with $\epsilon=\Ma \, \Kn$. Because $\overline{f^\one}$ has zero density and momentum moments, its Hermite expansion starts from $n=2$:
\begin{equation}\label{f1}
    \epsilon \overline{f^\one}
    = \frac{w^\flat}{\rho_R} \sum_{n=2}^{\infty} \frac{\Ma^{n}}{n! \theta_R^{\sharp n}}\,H^\sharp_{\alpha_1\dots\alpha_n}\,
      \epsilon \overline{a^{\one}_{\alpha_1\dots\alpha_n}}.
\end{equation}
Since $\int_\Xi \overline{f^\one} \dd\bm{\xi}^\flat = 0$, the isotropic subtraction in $H^\flat_{\alpha_1\alpha_2}$ vanishes after projection. The filtered raw second-order $O(\epsilon)$ moment
\[
    \overline{m^\one_{\alpha_1 \alpha_2}}
    \equiv \int_\Xi \overline{f^\one} \xi^\flat_{\alpha_1} \xi^\flat_{\alpha_2} \dd\bm{\xi}^\flat
\]
therefore coincides with the filtered second-order convectively scaled Hermite coefficient $\overline{a^\one_{\alpha_1 \alpha_2}}$.
Truncating \cref{f1} at second order therefore gives
\begin{equation}\label{eq:second_order_ce_reconstruction}
    \epsilon \overline{f^\one}
    \approx \frac{\Ma^2 w^\flat}{2\rho_R\theta_R^{\sharp 2}}\,H^\sharp_{\alpha_1\alpha_2}\,\epsilon\overline{m^\one_{\alpha_1\alpha_2}}.
\end{equation}
Using the convectively scaled Hermite polynomials of \cref{eq:H_advective}, the corresponding equilibrium Hermite coefficients are
\begin{align}
    a^{\zero} &\equiv \int_\Xi f^\zero H^\flat_0 \dd\bm\xi^\flat = \bar \rho , \\
    \MT[a^{\zero}][1] &\equiv \int_\Xi f^\zero H^\flat_{\alpha_1} \dd\bm\xi^\flat = \bar \rho \MT[\tilde u^\flat][1] , \\
    \MT[a^{\zero}][1][2] &\equiv \int_\Xi f^\zero H^\flat_{\alpha_1\alpha_2} \dd\bm\xi^\flat \nonumber \\&= \bar \rho \left(\MT[\tilde u^\flat][1]\MT[\tilde u^\flat][2] + \frac{(\tilde\theta^\sharp-\theta_R^\sharp)\MT[\delta][1][2]}{\Ma^2}\right) , \\
    \MT[a^{\zero}][1][2][3] &\equiv \int_\Xi f^\zero H^\flat_{\alpha_1\alpha_2\alpha_3} \dd\bm\xi^\flat \nonumber\\
    &= \bar \rho \Big[\MT[\tilde u^\flat][1]\MT[\tilde u^\flat][2]\MT[\tilde u^\flat][3]
    +\frac{\tilde\theta^\sharp-\theta_R^\sharp}{\Ma^2}\times\nonumber\\& \quad\left(\MT[\delta][1][2]\MT[\tilde u^\flat][3] + \MT[\delta][2][3]\MT[\tilde u^\flat][1] + \MT[\delta][3][1]\MT[\tilde u^\flat][2]\right)\Big] .
\end{align}
In the isothermal lattice--Boltzmann setting, $\tilde\theta^\sharp=1$, so these reduce to the standard equilibrium coefficients used by recursive regularization.

Compared with the raw moments, the Hermite coefficients subtract the isotropic reference contribution already built into the basis polynomials.
For completeness, the first raw moments of the resolved equilibrium $f^\zero$ are
\begin{align}
    m^\zero &\equiv \int_\Xi f^\zero \dd\bm\xi^\flat = \bar \rho , \\
    \MT[m^\zero][1] &\equiv \int_\Xi \xi^\flat_{\alpha_1} f^\zero \dd\bm\xi^\flat = \bar \rho \MT[\tilde u^\flat][1] , \\
    \MT[m^\zero][1][2] &\equiv \int_\Xi \xi^\flat_{\alpha_1}\xi^\flat_{\alpha_2} f^\zero \dd\bm\xi^\flat \nonumber \\&= \bar \rho \left(\MT[\tilde u^\flat][1] \MT[\tilde u^\flat][2] + \frac{\tilde\theta^\sharp \MT[\delta][1][2]}{\Ma^2}\right) , \\
    \MT[m^\zero][1][2][3] &\equiv \int_\Xi \xi^\flat_{\alpha_1}\xi^\flat_{\alpha_2}\xi^\flat_{\alpha_3} f^\zero \dd\bm\xi^\flat \nonumber\\
    &= \bar \rho \Big[\MT[\tilde u^\flat][1]\MT[\tilde u^\flat][2]\MT[\tilde u^\flat][3]
    + \frac{\tilde\theta^\sharp}{\Ma^2}\times\nonumber\\& \quad\left(\MT[\delta][1][2]\MT[\tilde u^\flat][3] + \MT[\delta][2][3]\MT[\tilde u^\flat][1] + \MT[\delta][3][1]\MT[\tilde u^\flat][2]\right)\Big] .
\end{align}

\section{Lattice--Boltzmann discretization}
\label{app:lbm_discretization}

This appendix summarizes the $\mathrm{D3Q27}$ lattice--Boltzmann numerical realization of the kinetic closures formulated in \cref{sec:kinetic_closures}.

\paragraph{Velocity space discretization}
The continuous velocity space is discretized using the standard $\mathrm{D3Q27}$ lattice set $\{\bm{\xi}_i\}_{i=0}^{26}$, which is the Cartesian product of the one-dimensional $\mathrm{D1Q3}$ Gauss--Hermite rule~\cite{kruger_lattice_2017}.
Here $i=0,\ldots,26$ labels a discrete velocity and $\xi_{i\alpha}$ denotes its Cartesian component.
Accordingly, the D3Q27 rule integrates Gaussian-weighted tensor-product monomials exactly when each one-dimensional degree is at most five~\cite{kruger_lattice_2017}. This quadrature exactness does not imply independent representation of all moments up to that order: powers larger than two in one Cartesian direction reduce on the node set, e.g.\ \(\xi_{i\alpha}^3=(\Delta x/\Delta t)^2\xi_{i\alpha}\) and \(\xi_{i\alpha}^4=(\Delta x/\Delta t)^2\xi_{i\alpha}^2\). Thus moments with repeated powers in one Cartesian direction alias onto lower-order moments.
Thus a mixed moment such as \(\xi_{ix}^2\xi_{iy}^2\xi_{iz}^2\) is independently represented, whereas one-direction moments such as \(\xi_{ix}^3\), \(\xi_{ix}^4\), and \(\xi_{ix}^5\) are not independent.

The continuous filtered distribution $\overline{f}$ is projected onto discrete populations $\overline{\mathrm{f}}_i(\bm{x},t)$ associated with lattice weights $w_i$.
The lattice spacing $\Delta x$ and time step $\Delta t$ define the reference units of the discrete solver. The discrete velocities satisfy $\xi_{i\alpha}\in\{-1,0,1\}\Delta x/\Delta t$. The isothermal equilibrium is parametrized by the fixed reference lattice temperature \(\theta_R \equiv c_l^2 = \frac{1}{3}(\Delta x/\Delta t)^2\), where \(c_l\) denotes the lattice speed of sound~\cite{kruger_lattice_2017,malaspinas_lattice_2009}.
In the isothermal LBM formulation, the equilibrium distribution is computed using $\theta=\theta_R$, which corresponds to setting $\theta^\sharp = 1$.

\paragraph{The filtered lattice Boltzmann equation} The discrete equation is intended as a realization of the modeled FRBGK--BE \eqref{eq:full_eff_bgk} with the kinetic time-filter width and the homogeneous spatial-filter width identified with the lattice spacings, namely \(\Delta_t=\Delta t\) and \(\Delta_x=\Delta x\).
Since \(c_t=\sqrt{3\theta_R}=\Delta x/\Delta t\), this choice gives \(c_t\Delta t=\Delta x\) for the RMS thermal-speed estimate in \cref{eq:kinetic_filter_width}.
Thus the discrete equation can be interpreted as acting on a space--time filtered distribution whose spatial and temporal coarse-graining scales are connected by the RMS thermal speed.
With the identification \(\Delta_t=\Delta t\) and \(\Delta_x=\Delta x\), the discrete resolution therefore obeys the same scale matching as the continuum filtering argument.
The kinetic evolution equation is then integrated along characteristics using a second-order trapezoidal rule:
\begin{multline}
    \overline{\mathrm{f}}_i(\bm{x} + \bm{\xi}_i\Delta t, t + \Delta t) - \overline{\mathrm{f}}_i(\bm{x}, t) =\\ \frac{\Delta t}{2} \left[ \mathcal{Q}_i(\bm{x}, t) + \mathcal{Q}_i(\bm{x} + \bm{\xi}_i\Delta t, t + \Delta t) \right],
\end{multline}
where $\mathcal{Q}_i$ represents the discrete, fully modeled collision operator, including the SGS collision contribution, and the shift $\bm{x}\mapsto\bm{x}+\bm{\xi}_i\Delta t$ moves a population by one lattice spacing in each active Cartesian direction.
To remove the future-time collision term introduced by the trapezoidal
quadrature and express the resulting second-order update as a local explicit
collision step, we introduce the classical time-shifted populations
\cite{kruger_lattice_2017}:
\begin{equation}\label{eq:shifted_population_definition}
    \overline{f}_i(\bm{x},t) \equiv \overline{\mathrm{f}}_i(\bm{x},t) - \frac{\Delta t}{2}\mathcal{Q}_i(\bm{x}, t).
\end{equation}
The substitution cancels the future-time collision term and gives
\begin{equation}\label{eq:shifted_lbe_implicit_collision}
    \overline f_i(\bm{x}+\bm{\xi}_i\Delta t,t+\Delta t)
    =
    \overline f_i(\bm{x},t)
    +
    \Delta t\,\mathcal{Q}_i(\overline{\mathrm f})(\bm{x},t),
\end{equation}
with \(\overline{\mathrm f}_i=\overline f_i+\frac{\Delta t}{2}\mathcal{Q}_i(\overline{\mathrm f})\).
For a BGK collision, \(f_i^\zero\) is unchanged by the shift because the collision conserves the fields that define the equilibrium.
The standard trapezoidal inversion expresses the BGK collision explicitly in
terms of the shifted populations:
\begin{equation}
    \mathcal{Q}_i^{\rm BGK}(\overline{\mathrm f})
    =
    -\left(\frac{\nu}{\theta_R}+\frac{\Delta t}{2}\right)^{-1}
    \left(\overline f_i-f_i^\zero\right).
\end{equation}
For the present two-rate closures, Appendix~\ref{app:trapezoidal_reconstruction} shows how the same half-step shift acts on the first-order CE contribution and the SGS carrier.
For the locally evaluated relaxation rates, held fixed during the inversion, the shifted form gives the explicit two-rate collision exactly, provided that the first-order CE contribution is written with the shifted relaxation time.
The unshifted lattice relaxation times are
\begin{equation}\labelAndRemember{eq:tau_tausgs_viscosities}{
    \tau=\frac{\nu}{\theta_R},
    \qquad
    \tau_{\sgs}=\frac{\nu_t}{\theta_R}.}
\end{equation}
We therefore use
\begin{equation}\label{eq:omega_t_discrete}
    \begin{aligned}
    \hat\omega
    = \left(\tau+\frac{\Delta t}{2}\right)^{-1}\!\!, \quad
    \hat\omega_{\sgs}
    = \left(\tau_{\sgs}+\frac{\Delta t}{2}\right)^{-1}.
    \end{aligned}
\end{equation}
With these shifted rates, the operational two-rate update corresponding to \cref{eq:full_eff_bgk} reads
\begin{equation}\label{eq:operational_two_rate_update}
    \begin{aligned}
    \overline{f}_i(\bm{x}+\bm{\xi}_i\Delta t,t+\Delta t)
    &=
    \overline{f}_i(\bm{x},t)
    \\
    &\quad
    -
    \Delta t\,\hat\omega\left(\overline{f}_i-\overline{f_i^\zero}\right)
    -
    \Delta t\,\hat\omega_{\sgs} f_{\sgs,i}.
    \end{aligned}
\end{equation}
Here $f_{\sgs,i}$ denotes the closure-specific discrete SGS carrier reconstructed locally from the shifted populations \(\overline f_i\), and \(\overline{f_i^\zero}\equiv f_i^\zero+f_{\sgs,i}\) denotes the modeled target.
For KC-RB, this carrier is the shifted SGS-carrier estimate
\begin{equation}\labelAndRemember{eq:kcrb_operational_fsgs}{
    f_{\sgs,i}
    \equiv
    \overline f_i-f_i^\zero-\overline{f_i^\one}.}
\end{equation}
When $\hat\omega_{\sgs}=\hat\omega$, the two-rate update reduces exactly to the standard resolved BGK update.
Operationally, the update is applied as a local, filtered-density and filtered-momentum conserving, collision step (except for the legacy raw KC-RB mass defect discussed in Appendix~\ref{app:kcrb_operational}):
\begin{equation}
    f_i^{\rm post}(\bm{x},t)
    =
    \overline{f}_i(\bm{x},t)
    -
    \Delta t\,\hat\omega\left(\overline{f}_i-\overline{f_i^\zero}\right)
    -
    \Delta t\,\hat\omega_{\sgs} f_{\sgs,i},
\end{equation}
followed by streaming,
\begin{equation}\label{eq:streaming}
    \overline{f}_i(\bm{x}+\bm{\xi}_i\Delta t,t+\Delta t)
    =
    f_i^{\rm post}(\bm{x},t).
\end{equation}
Thus all closure terms enter the local collision; streaming only shifts the post-collision populations along the lattice links.

\paragraph{Macroscopic fields and isothermal discrete equilibrium}
The macroscopic conserved variables can then be evaluated directly from the shifted populations without implicit coupling:
\begin{align}
    \bar\rho &= \sum_i \overline{f}_i, \\
    \bar\rho\tilde u_{\alpha_1} &= \sum_i \overline{f}_i \xi_{i\alpha_1}.
\end{align}
The corresponding discrete equilibrium is then constructed from these conserved moments, so that
\begin{align}
    \sum_i f_i^\zero &= \bar\rho, \\
    \sum_i f_i^\zero \xi_{i\alpha_1} &= \bar\rho\tilde u_{\alpha_1}.
\end{align}

Rather than using either the globally truncated second-order polynomial expansion in $\bm{\xi}_i\cdot\tilde{\bm{u}}$ or a higher-order global Hermite expansion, the discrete equilibrium is taken in the complete product form induced by the Cartesian product of the one-dimensional $\mathrm{D1Q3}$ equilibrium truncation~\cite{karlin_factorization_2010}.
A global higher-order expansion would try to impose directional modes such as \(H_{xxx}\) or \(H_{yyyy}\), which alias on $\mathrm{D3Q27}$, whereas the product form keeps only modes with degree at most two in each Cartesian direction. With $w_i=\prod_{\alpha=1}^{3}w^{(1D)}_{\xi_{i\alpha}}$, \(w^{(1D)}_0=2/3\), and \(w^{(1D)}_{\pm \Delta x/\Delta t}=1/6\), the equilibrium populations are
\begin{equation}\label{eq:discrete_equilibrium}
    f_i^\zero = w_i \bar\rho\prod_{\alpha=1}^{3}
    \left(
        1 + \frac{\xi_{i\alpha}\tilde u_\alpha}{\theta_R}
        + \frac{(\xi_{i\alpha}^2-\theta_R)\tilde u_\alpha^2}{2\theta_R^2}
    \right).
\end{equation}
Each Cartesian factor is precisely the $\mathrm{D1Q3}$ quadratic truncation, while their product retains all mixed velocity terms admissible on the $\mathrm{D3Q27}$ tensor-product lattice.

\paragraph{Finite-difference strain-rate tensor}
The Chapman--Enskog viscous contribution $\overline{f_i^\one}$ (including the CE prefactor) used in KC-RB, KC-MP, and KC-RR is reconstructed from macroscopic velocity gradients.
In the implementation considered in \cref{sec:test_cases} these gradients are evaluated by coordinate-wise central finite differences:
\begin{equation}\label{eq:fd}
    \partial_{\alpha_1}\tilde u_{\alpha_2}(\bm{x})
    \approx
    \frac{\tilde u_{\alpha_2}(\bm{x}+\bm e_{\alpha_1}\Delta x)
    -\tilde u_{\alpha_2}(\bm{x}-\bm e_{\alpha_1}\Delta x)}
    {2\Delta x},
\end{equation}
and the strain-rate tensor is computed as
\begin{equation}\label{eq:strain_rate_discrete}
    S_{\alpha_1\alpha_2}
    \equiv
    \frac{1}{2}\left(\partial_{\alpha_2}\tilde u_{\alpha_1}+\partial_{\alpha_1}\tilde u_{\alpha_2}\right).
\end{equation}
To avoid read--write races during parallel collision, the implementation used for the numerical tests evaluates these gradients from the previous time step's velocity field, introducing an \(O(\Delta t)\) temporal lag error.
At boundary nodes, the implementation uses the corresponding one-sided second-order formulas.
Let \(\tilde u_{\alpha_2}^{b,\pm k}\equiv
\tilde u_{\alpha_2}(\bm{x}_b\pm k\bm e_{\alpha_1}\Delta x)\), with
\(\tilde u_{\alpha_2}^{b,0}\equiv\tilde u_{\alpha_2}(\bm{x}_b)\).
For a lower boundary normal to \(\bm e_{\alpha_1}\),
\begin{equation}\label{eq:fd_boundary_lower}
    \partial_{\alpha_1}\tilde u_{\alpha_2}(\bm{x}_b)
    \approx
    \frac{-3\tilde u_{\alpha_2}^{b,0}
    +4\tilde u_{\alpha_2}^{b,+1}
    -\tilde u_{\alpha_2}^{b,+2}}{2\Delta x}.
\end{equation}
For an upper boundary normal to \(\bm e_{\alpha_1}\),
\begin{equation}\label{eq:fd_boundary_upper}
    \partial_{\alpha_1}\tilde u_{\alpha_2}(\bm{x}_b)
    \approx
    \frac{3\tilde u_{\alpha_2}^{b,0}
    -4\tilde u_{\alpha_2}^{b,-1}
    +\tilde u_{\alpha_2}^{b,-2}}{2\Delta x}.
\end{equation}
We used \cref{eq:fd,eq:fd_boundary_lower,eq:fd_boundary_upper} for the validations in \cref{sec:test_cases}.
A weighted lattice-link stencil could improve rotational isotropy, but we leave this alternative for future work~\cite{shan_analysis_2006}.

The same finite-difference velocity gradients are also used by the Skordos boundary condition~\cite{skordos_initial_1993} in the lid-driven cavity test case (\cref{ldc}).

\subsection{Common operational algorithm}
\label{app:omega_t_operational}
The operational update used in the simulations follows the standard collide--stream sequence. At each fluid node and before collision (after streaming):
\begin{enumerate}
    \item compute the local macroscopic fields from the shifted populations,
    \begin{equation}
        \bar\rho=\sum_i \overline f_i,
        \qquad
        \bar\rho\,\tilde u_\alpha=\sum_i \overline f_i \xi_{i\alpha};
    \end{equation}
    \item construct the local equilibrium \(f_i^\zero=f_i^\zero(\bar\rho,\tilde{\bm u})\) from \cref{eq:discrete_equilibrium};
    \item evaluate the velocity gradients and \(S_{\alpha_1\alpha_2}\) from \cref{eq:fd,eq:strain_rate_discrete} (or use a weighted lattice-link stencil~\cite{shan_analysis_2006}, not tested herein);
    \item compute the closure-specific SGS trace \(m^{\sgs}_{\alpha_1\alpha_1}\) used below, from \cref{eq:kcrb_operational_msgstrace} for KC-RB and \cref{eq:kcmp_operational_msgstrace} for KC-MP and KC-RR;
    \item evaluate the SGS amplitude and the clipped turbulent relaxation frequency through
\begin{equation}\label{eq:omega_sgs_operational}
    \begin{aligned}
        k_\sgs &\approx \frac{1}{2\rho_0}\left|m^{\sgs}_{\alpha_1\alpha_1}\right|,
        \qquad \rho_0=1\,\unit{lu}, \\
        \nu_t^{\rm eff} &= \max\!\left(\nu, C_\nu \Delta \sqrt{k_\sgs}\right), \\
        \hat\omega_{\sgs} &= \left(\frac{\nu_t^{\rm eff}}{\theta_R}+\frac{\Delta t}{2}\right)^{-1} \\
        &= \min\!\left[
        \hat\omega,\left(\frac{C_\nu \Delta \sqrt{k_\sgs}}{\theta_R}+\frac{\Delta t}{2}\right)^{-1}
        \right];
    \end{aligned}
\end{equation}
Here \(k_\sgs\), \(\nu_t^{\rm eff}\), and \(\Delta\) are the lattice-unit counterparts of \(k_\sgs^\sharp\), \(\nu_t^\sharp\), and \(\Delta_x^\sharp\), after the clipping by the molecular viscosity. The quantity \(\rho_0\) is the fixed reference density used by the discrete implementation. The simulations reported here set \(\rho_0=1\,\unit{lu}\) (with $\unit{lu}$ being the density lattice unit), rather than dividing the SGS trace by the local density \(\bar\rho\).
In the operational reconstructions below, \(\hat\tau\equiv1/\hat\omega=\nu/\theta_R+\Delta t/2\) denotes the shifted molecular relaxation time.
    \item apply the closure-specific collision rule: \cref{eq:kcrb_operational_collision} for KC-RB, \cref{eq:kcmp_operational_collision} for KC-MP, and \cref{eq:kcrr_operational_second_order_collision,eq:kcrr_operational_high_order_collision,eq:kcrr_operational_post_population} for KC-RR;
    \item stream the post-collision populations.
\end{enumerate}
For the uniform lattices considered here, \(\Delta=\Delta x\).

\subsection{Operational implementation of KC-RB}
\label{app:kcrb_operational}
After the common pre-collision steps of Appendix~\ref{app:omega_t_operational}, the model-specific part enters through steps 4 and 6.

\paragraph*{Step 4: SGS trace}
The SGS trace is obtained as follows:
\begin{enumerate}
    \item approximate the filtered first-order CE contribution,
	\begin{equation}\label{eq:kcrb_operational_fone}
	    \overline{f_i^\one}
	    \approx
	    -\frac{w_i\bar\rho\hat\tau}{\theta_R}\,
    H_{i\alpha_1\alpha_2}S_{\alpha_1\alpha_2},
\end{equation}
where
\begin{equation}\label{eq:discrete_second_order_hermite}
    H_{i\alpha_1\alpha_2} \equiv \xi_{i\alpha_1}\xi_{i\alpha_2}-\theta_R\delta_{\alpha_1\alpha_2};
\end{equation}
    \item form the KC-RB estimate of the SGS carrier,
\[
    \recallEqAndTag{eq:kcrb_operational_fsgs}
\]
    \item compute the SGS trace used in the turbulent relaxation model,
\begin{equation}\label{eq:kcrb_operational_msgstrace}
    m^{\sgs}_{\alpha_1\alpha_1}
    =
    \sum_i f_{\sgs,i} \xi_i^2;
\end{equation}
\end{enumerate}

\paragraph*{Step 6: collision rule}
The post-collision populations are
\begin{equation}\label{eq:kcrb_operational_collision}
    f_i^{\rm post}
    =
    \overline f_i
    -
    \Delta t\,\hat\omega\,\overline{f_i^\one}
    -
    \Delta t\,\hat\omega_{\sgs}f_{\sgs,i}.
\end{equation}

The simulations reported in the present paper did not use this Hermite-consistent form. To remain aligned with the original incompressible formulation proposed in~\cite{marson_kinetic_2025}, they used the simplified first-order CE reconstruction
\begin{equation}\label{eq:kcrb_operational_raw_fone}
    \overline{f_{i,{\xi\xi}}^\one}
    \equiv
    -\frac{w_i\bar\rho\hat\tau}{\theta_R}\,
    \xi_{i\alpha_1}\xi_{i\alpha_2}S_{\alpha_1\alpha_2}.
\end{equation}
At second order, a raw moment can represent a Hermite coefficient only when it is evaluated from a massless distribution, because the isotropic Hermite subtraction then vanishes. Conversely, a distribution reconstructed from a second-order coefficient must retain the Hermite polynomial: replacing it with \(\xi_{i\alpha_1}\xi_{i\alpha_2}\) generally gives a nonzero zeroth moment.
The two realizations satisfy
\begin{equation}
    \overline{f_{i,{\xi\xi}}^\one}
    =
    \overline{f_i^\one}
    -
    w_i\bar\rho\hat\tau\,\partial_{\alpha_1}u_{\alpha_1}.
\end{equation}
The simplified \(\xi\xi\) form therefore introduces an isotropic defect proportional to \(\partial_{\alpha_1}u_{\alpha_1}\) in the operative LBM implementation of KC-RB.
For this realization, evaluating the conserved fields directly from \(\overline f_i\) is exact when \(\partial_{\alpha_1}u_{\alpha_1}=0\) or \(\hat\omega_{\sgs}=\hat\omega\). In the weakly compressible LBM regime, this mass defect can modify the acoustic activity because the extra isotropic contribution is relaxed with \(\hat\omega_{\sgs}\) instead of \(\hat\omega\). This may improve robustness, but it remains a consistency error specific to the present KC-RB realization and can make the solution more sensitive to the chosen Mach number. This issue is absent from the KC-MP and KC-RR implementations below, which retain the Hermite reconstruction.
Appendix~\ref{app:kcrb_reconstruction_sensitivity} compares \cref{eq:kcrb_operational_raw_fone,eq:kcrb_operational_fone} in an otherwise fixed Taylor--Green calculation. Replacing the raw form changes both the CE reference and the SGS carrier entering the turbulent relaxation. The resulting difference is therefore not equivalent to a fixed additional bulk viscosity.

\subsection{Operational implementation of KC-MP}
\label{app:kcmp_operational}
After the common pre-collision steps of Appendix~\ref{app:omega_t_operational}, the model-specific part enters through steps 4 and 6.

\paragraph*{Step 4: SGS trace}
The SGS trace is obtained as follows:
\begin{enumerate}
    \item reconstruct the common second-order CE reference stress,
\begin{equation}
    m^{\rm CE}_{\alpha_1\alpha_2}
    \equiv
    -2\bar\rho\theta_R\hat\tau S_{\alpha_1\alpha_2};
\end{equation}
    \item compute the second-order moment of the coarse-level nonequilibrium,
\begin{equation}
    m^{\cNeq}_{\alpha_1\alpha_2}
    =
    \sum_i(\overline f_i-f_i^\zero)\xi_{i\alpha_1}\xi_{i\alpha_2};
\end{equation}
Because \(f^\cNeq\) has zero zeroth moment, \(m^{\cNeq}_{\alpha_1\alpha_2}=a^{\cNeq}_{\alpha_1\alpha_2}\); thus this raw moment is also the second-order Hermite coefficient used by KC-RR.
    \item subtract the CE reference stress to obtain the closure-specific SGS second moment,
\begin{equation}
    m^{\sgs}_{\alpha_1\alpha_2}
    =
    m^{\cNeq}_{\alpha_1\alpha_2}
    -
    m^{\rm CE}_{\alpha_1\alpha_2};
\end{equation}
    \item take the trace used in the turbulent relaxation model,
\begin{equation}\label{eq:kcmp_operational_msgstrace}
    m^{\sgs}_{\alpha_1\alpha_1}
    =
    m^{\cNeq}_{\alpha_1\alpha_1}
    -
    m^{\rm CE}_{\alpha_1\alpha_1}.
\end{equation}
\end{enumerate}

\paragraph*{Step 6: collision rule}
The collision rule is applied as follows:
\begin{enumerate}
    \item reconstruct the SGS carrier,
\begin{equation}
    f_{\sgs,i}
    =
    \frac{w_i}{2\theta_R^2}\,
    H_{i\alpha_1\alpha_2}m^{\sgs}_{\alpha_1\alpha_2};
\end{equation}
    \item identify the operational estimate of the filtered first-order CE contribution,
\begin{equation}
    \overline{f_i^\one}
    \approx
    (\overline f_i-f_i^\zero)-f_{\sgs,i};
\end{equation}
    \item apply the split collision rule,
\begin{equation}\label{eq:kcmp_operational_collision}
    f_i^{\rm post}
    =
    \overline f_i
    -
    \Delta t\,\hat\omega\,\overline{f_i^\one}
    -
    \Delta t\,\hat\omega_{\sgs}f_{\sgs,i};
\end{equation}
\end{enumerate}
Because the reconstruction is performed in the same Hermite basis, no additional compressibility error analogous to the KC-RB \(\xi\xi\)-approximation is introduced here.

\subsection{Operational implementation of KC-RR}
\label{app:kcrr_operational}
KC-RR operates directly on the discrete Hermite coefficients of the coarse-level nonequilibrium \(f_i^\cNeq=\overline f_i-f_i^\zero\), using the Hermite basis introduced in \cref{eq:discrete_second_order_hermite}.
After the common pre-collision steps of Appendix~\ref{app:omega_t_operational}, the model-specific part enters through steps 4 and 6.

\paragraph*{Step 4: SGS trace}
KC-RR uses the same second-order SGS-trace construction as KC-MP in Appendix~\ref{app:kcmp_operational}. In the coefficient notation required for recursive regularization, its second-order seed is the same CE reference stress,
\begin{equation}\label{eq:kcrr_operational_rr_seed}
    a^{\rm rr}_{\alpha_1\alpha_2}
    =
    m^{\rm CE}_{\alpha_1\alpha_2}.
\end{equation}
Since \(f^\cNeq\) is massless, \(a^{\cNeq}_{\alpha_1\alpha_2}=m^{\cNeq}_{\alpha_1\alpha_2}\), and
\begin{equation}
    a^{\sgs}_{\alpha_1\alpha_2}
    =
    a^{\cNeq}_{\alpha_1\alpha_2}
    -
    a^{\rm rr}_{\alpha_1\alpha_2}
    =
    m^{\sgs}_{\alpha_1\alpha_2}.
\end{equation}
The trace in \cref{eq:kcmp_operational_msgstrace} therefore applies directly. The coefficient \(a^{\rm rr}_{\alpha_1\alpha_2}\) also seeds the higher-order RR recursion in Step~6.

\paragraph*{Step 6: collision rule}
For the higher-order projection, D3Q27 provides only three independent lattice functions in each Cartesian direction:
\(1\), \(\xi_{i\alpha}\), and \(\xi_{i\alpha}^2-\theta_R\).
Thus higher powers are not independent on the lattice.
For example, \(\xi_{i\alpha}^3=(\Delta x/\Delta t)^2\xi_{i\alpha}\).
Consequently, components with degree larger than two in one Cartesian direction alias onto lower-order lattice modes.
Removing the density mode, the three momentum modes, and the six second-order modes leaves \(3^3-1-3-6=17\) independent higher-order components:
\[
    \mathcal{B}_{\rm high}
    \equiv
    \left\{
    \begin{aligned}
    &xxy,xxz,yyz,xyy,xzz,yzz,xyz,\\
    &xxyy,xxzz,yyzz,xxyz,xyyz,xyzz,\\
    &xxyyz,xxyzz,xyyzz,xxyyzz
    \end{aligned}
    \right\}.
\]
This basis spans the remaining high-order phase space without overlapping the conserved or SGS-stress sectors.
The collision rule is then applied as follows:
\begin{enumerate}
    \item relax the second-order coefficients,
\begin{equation}\label{eq:kcrr_operational_second_order_collision}
    \begin{aligned}
        a_{xx}^{\rm post} &= (1-\Delta t\,\hat\omega)a_{xx}^{\rm rr} + (1-\Delta t\,\hat\omega_{\sgs})a_{xx}^{\sgs}, \\
        a_{yy}^{\rm post} &= (1-\Delta t\,\hat\omega)a_{yy}^{\rm rr} + (1-\Delta t\,\hat\omega_{\sgs})a_{yy}^{\sgs}, \\
        a_{zz}^{\rm post} &= (1-\Delta t\,\hat\omega)a_{zz}^{\rm rr} + (1-\Delta t\,\hat\omega_{\sgs})a_{zz}^{\sgs}, \\
        a_{xy}^{\rm post} &= (1-\Delta t\,\hat\omega)a_{xy}^{\rm rr} + (1-\Delta t\,\hat\omega_{\sgs})a_{xy}^{\sgs}, \\
        a_{xz}^{\rm post} &= (1-\Delta t\,\hat\omega)a_{xz}^{\rm rr} + (1-\Delta t\,\hat\omega_{\sgs})a_{xz}^{\sgs}, \\
        a_{yz}^{\rm post} &= (1-\Delta t\,\hat\omega)a_{yz}^{\rm rr} + (1-\Delta t\,\hat\omega_{\sgs})a_{yz}^{\sgs};
    \end{aligned}
\end{equation}
    \item compute the high-order D3Q27 Hermite coefficients,
\begin{equation}
    a^{\cNeq}_{\alpha_1\dots\alpha_n}
    =
    \sum_i(\overline f_i-f_i^\zero)H_{i,\alpha_1\dots\alpha_n},
    \quad
    (\alpha_1\dots\alpha_n)\in\mathcal{B}_{\rm high},
\end{equation}
where no component in \(\mathcal{B}_{\rm high}\) contains a directional power above two.
    \item compute the regularized higher-order coefficients from \(a^{\rm rr}_{\alpha_1\alpha_2}\) with the RR recursion~\cite{malaspinas_increasing_2015},
\[
    \recallEqAndTag{eq:kcrr_rr};
\]
    \item define the SGS high-order coefficients,
\begin{equation}\label{eq:kcrr_operational_sgs_high}
    a^{\sgs}_{\alpha_1\dots\alpha_n}
    =
    a^{\cNeq}_{\alpha_1\dots\alpha_n}
    -
    a^{\rm rr}_{\alpha_1\dots\alpha_n},
    \quad
    (\alpha_1\dots\alpha_n)\in\mathcal{B}_{\rm high};
\end{equation}
    \item relax the higher-order coefficients,
\begin{equation}\label{eq:kcrr_operational_high_order_collision}
    \begin{aligned}
    a^{\rm post}_{\alpha_1\dots\alpha_n}
    &= (1-\Delta t\,\hat\omega)\,a^{\rm rr}_{\alpha_1\dots\alpha_n} \\
    &\quad + (1-\Delta t\,\hat\omega_{\sgs})\,a^{\sgs}_{\alpha_1\dots\alpha_n},
    \quad
    (\alpha_1\dots\alpha_n)\in\mathcal{B}_{\rm high};
    \end{aligned}
\end{equation}
    \item define the directional discrete Hermite factors,
\begin{equation}
    \begin{aligned}
        H_{x,i} &\equiv \xi_{ix},
        & H_{xx,i} &\equiv \xi_{ix}^2-\theta_R, \\
        H_{y,i} &\equiv \xi_{iy},
        & H_{yy,i} &\equiv \xi_{iy}^2-\theta_R, \\
        H_{z,i} &\equiv \xi_{iz},
        & H_{zz,i} &\equiv \xi_{iz}^2-\theta_R,
    \end{aligned}
\end{equation}
    \item reconstruct the post-collision populations,
\begin{equation}\label{eq:kcrr_operational_post_population}
    f_i^{\rm post}
    =
    f_i^\zero
    +
    w_i\sum_{m=2}^{6}T_m,
\end{equation}
where
{\allowdisplaybreaks
\begin{align*}
        T_2 &= \frac{1}{2\theta_R^2}\Big[H_{xx,i}a_{xx}^{\rm post}+H_{yy,i}a_{yy}^{\rm post}+H_{zz,i}a_{zz}^{\rm post}\Big] \\
        &\quad + \frac{1}{\theta_R^2}\Big[H_{x,i}H_{y,i}a_{xy}^{\rm post}+H_{x,i}H_{z,i}a_{xz}^{\rm post} \\
        &\qquad + H_{y,i}H_{z,i}a_{yz}^{\rm post}\Big], \\[0.8ex]
        T_3 &= \frac{1}{2\theta_R^3}\Big[H_{xx,i}H_{y,i}a_{xxy}^{\rm post}+H_{xx,i}H_{z,i}a_{xxz}^{\rm post} \\
        &\qquad + H_{yy,i}H_{z,i}a_{yyz}^{\rm post}\Big] \\
        &\quad + \frac{1}{2\theta_R^3}\Big[H_{x,i}H_{yy,i}a_{xyy}^{\rm post}+H_{x,i}H_{zz,i}a_{xzz}^{\rm post} \\
        &\qquad + H_{y,i}H_{zz,i}a_{yzz}^{\rm post}\Big] \\
        &\quad+\frac{1}{\theta_R^3}H_{x,i}H_{y,i}H_{z,i}a_{xyz}^{\rm post}, \\[0.8ex]
        T_4 &= \frac{1}{4\theta_R^4}\Big[H_{xx,i}H_{yy,i}a_{xxyy}^{\rm post}+H_{xx,i}H_{zz,i}a_{xxzz}^{\rm post} \\
        &\quad\quad + H_{yy,i}H_{zz,i}a_{yyzz}^{\rm post}\Big] \\
        &\quad + \frac{1}{2\theta_R^4}\Big[H_{xx,i}H_{y,i}H_{z,i}a_{xxyz}^{\rm post} \\
        &\qquad + H_{x,i}H_{yy,i}H_{z,i}a_{xyyz}^{\rm post} \\
        &\qquad + H_{x,i}H_{y,i}H_{zz,i}a_{xyzz}^{\rm post}\Big], \\[0.8ex]
        T_5 &= \frac{1}{4\theta_R^5}\Big[H_{xx,i}H_{yy,i}H_{z,i}a_{xxyyz}^{\rm post} \\
        &\quad\quad + H_{xx,i}H_{y,i}H_{zz,i}a_{xxyzz}^{\rm post} \\
        &\quad\quad + H_{x,i}H_{yy,i}H_{zz,i}a_{xyyzz}^{\rm post}\Big], \\[0.8ex]
        T_6 &= \frac{1}{8\theta_R^6}H_{xx,i}H_{yy,i}H_{zz,i}a_{xxyyzz}^{\rm post};
\end{align*}}
\end{enumerate}
In the present implementation, the same pair \((\hat\omega,\hat\omega_{\sgs})\) is used for the third-order and higher-order ghost sectors. As in KC-MP, the Hermite formulation avoids the additional compressibility inconsistency discussed above for the simplified KC-RB realization.

\subsection{Trapezoidal reconstruction of the two-rate collision}
\label{app:trapezoidal_reconstruction}
The explicit two-rate update in \cref{eq:operational_two_rate_update}
can be derived directly from the Hermite-consistent KC-RB split in the transformed equation \cref{eq:shifted_lbe_implicit_collision}, with the local relaxation times held fixed during the collision step.
The update is explicit only after expressing \(\mathcal{Q}_i(\overline{\mathrm f})\) in terms of \(\overline f_i\).
For clarity, first recall the unshifted relaxation times from \cref{eq:tau_tausgs_viscosities} and keep them fixed during the local algebra,
\[
    \recallEqAndTag{eq:tau_tausgs_viscosities}
\]
The continuous KC-RB decomposition, after velocity discretization, applies to the unshifted trapezoidal population,
\begin{equation}\label{eq:physical_discrete_sgs_carrier}
    \mathrm{f}_{\sgs,i}
    =
    \overline{\mathrm{f}}_i
    -
    \mathrm{f}_i^\zero
    -
    \overline{\mathrm{f}_i^\one},
\end{equation}
where the upright symbols denote the discrete projections of the corresponding continuous quantities, with the SGS carrier reconstructed as above.
Since the collision has zero conserved moments, \(\mathrm{f}_i^\zero=f_i^\zero\).
The first-order CE contribution \(\overline{\mathrm{f}_i^\one}\) is reconstructed from the continuum CE expansion in \cref{eq:f1_KCRB}, here written in dimensional lattice units as
\begin{equation}
    \overline{\mathrm{f}_i^\one}
    \approx
    -\frac{w_i\bar\rho\,\tau}{\theta_R}\,
    H_{i\alpha_1\alpha_2}S_{\alpha_1\alpha_2}.
\end{equation}
Thus \(\overline{\mathrm{f}_i^\one}\) depends on the continuum \(\tau=\nu/\theta_R\), not on the shifted operational \(\hat\tau=1/\hat\omega\).
The operational reconstruction in \cref{eq:kcrb_operational_fone} uses the same Hermite projection but the shifted relaxation time. Therefore, for these reconstructed CE contributions,
\begin{equation}
    \overline{\mathrm{f}_i^\one}
    =
    \frac{\tau}{\hat\tau}\overline{f_i^\one}.
\end{equation}
The trapezoidal shift leaves the conserved fields unchanged, so the same \(\bar\rho\), \(\tilde{\bm u}\), and \(S_{\alpha_1\alpha_2}\) enter both CE reconstructions.
However, the shift changes the relaxation time multiplying the reconstructed CE contribution: the unshifted contribution uses \(\tau\), while the operational shifted contribution uses \(\hat\tau\).
Thus the following algebra rewrites the unshifted SGS split in terms of shifted populations and shifted CE contributions.
The two-rate collision then reads
\begin{equation}\label{eq:correct_Qi}
    \begin{aligned}
    \mathcal{Q}_i(\overline{\mathrm{f}})
    &=
    -\frac{1}{\tau}\overline{\mathrm{f}_i^\one}
    -
    \frac{1}{\tau_{\sgs}}
    \left(
    \overline{\mathrm{f}}_i
    -
    f_i^\zero
    -
    \overline{\mathrm{f}_i^\one}
    \right)\\
    &\approx
    -\frac{1}{\hat\tau}\overline{f_i^\one}
    -
    \frac{1}{\tau_{\sgs}}
    \left(
    \overline{\mathrm{f}}_i
    -
    f_i^\zero
    -
    \frac{\tau}{\hat\tau}\overline{f_i^\one}
    \right).
    \end{aligned}
\end{equation}
Substituting the inverse shift
\(\overline{\mathrm f}_i=\overline f_i+\Delta t\,\mathcal{Q}_i(\overline{\mathrm f})/2\)
from \cref{eq:shifted_population_definition} gives the implicit local relation
\begin{equation}
    \begin{aligned}
    \mathcal{Q}_i(\overline{\mathrm{f}})
    &\approx
    -\frac{1}{\hat\tau}\overline{f_i^\one}
    -
    \frac{1}{\tau_{\sgs}}
    \left(
    \overline f_i
    +
    \frac{\Delta t}{2}\mathcal{Q}_i(\overline{\mathrm f})
    -
    f_i^\zero
    -
    \frac{\tau}{\hat\tau}\overline{f_i^\one}
    \right).
    \end{aligned}
\end{equation}
Solving this scalar linear relation for the collision gives
\begin{equation}
    \begin{aligned}
    \mathcal{Q}_i(\overline{\mathrm{f}})
    &\approx
    -\hat\omega_{\sgs}\left(\overline f_i-f_i^\zero\right)
    +
    \left(\hat\omega_{\sgs}-\hat\omega\right)\overline{f_i^\one}\\
    &=
    -\hat\omega\,\overline{f_i^\one}
    -
    \hat\omega_{\sgs}
    \left(
    \overline f_i
    -
    f_i^\zero
    -
    \overline{f_i^\one}
    \right),
    \end{aligned}
\end{equation}
where
\(\hat\omega=(\tau+\Delta t/2)^{-1}\) and
\(\hat\omega_{\sgs}=(\tau_{\sgs}+\Delta t/2)^{-1}\).
Thus the shifted trapezoidal inversion recovers the operational two-rate collision in \cref{eq:operational_two_rate_update}, with
the KC-RB carrier defined in \cref{eq:kcrb_operational_fsgs}. The turbulent relaxation rate is evaluated from the shifted SGS trace according to \cref{eq:omega_sgs_operational}.

\section{KC-RB reconstruction sensitivity}
\label{app:kcrb_reconstruction_sensitivity}

\begin{figure*}[t]
    \centering
    \includegraphics[width=0.9\textwidth]{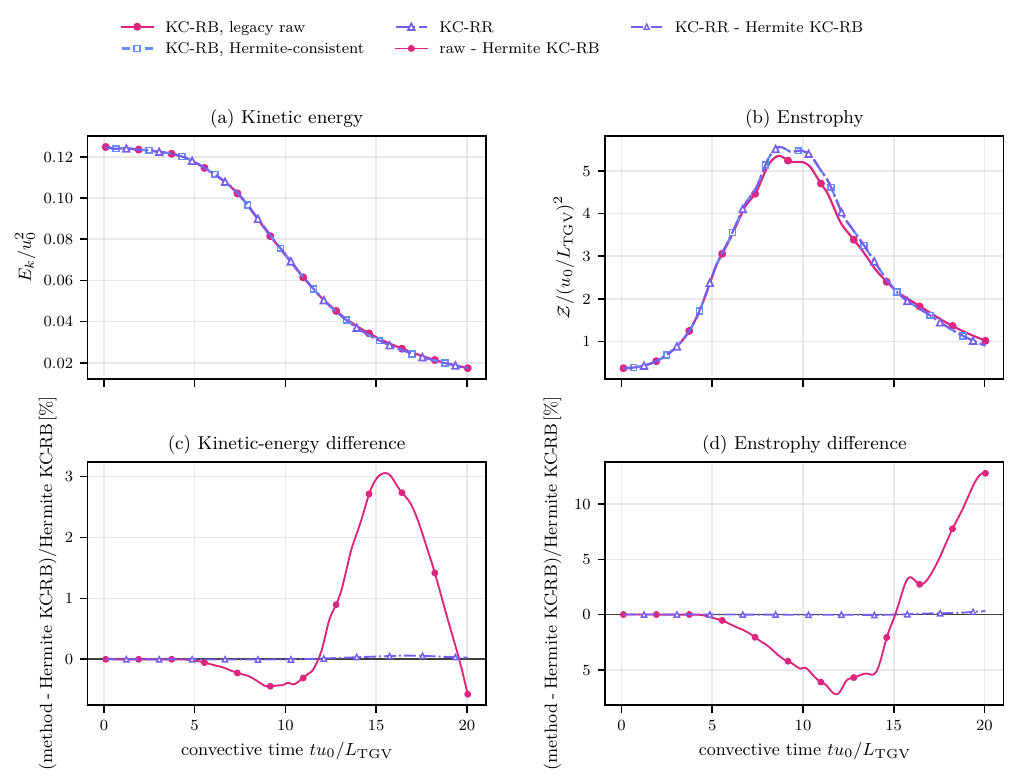}
    \caption{Controlled sensitivity of KC-RB to the first-order CE reconstruction in the \(N=64\) Taylor--Green vortex. The legacy raw form uses \(\xi_{i\alpha}\xi_{i\beta}\), whereas the Hermite-consistent form uses \(H_{i\alpha\beta}=\xi_{i\alpha}\xi_{i\beta}-\theta_R\delta_{\alpha\beta}\). The upper panels use the reference-scale normalization of the main TGV figures, \(E_k/u_0^2\) and \(\mathcal{Z}/(u_0/L_{\rm TGV})^2\). The lower panels show deviations from Hermite KC-RB. KC-RR provides a contextual Hermite-consistent reference, but its recursive projection and SGS carrier also differ from KC-RB.}
    \label{fig:kcrb_raw_vs_hermite_n64}
\end{figure*}
This appendix isolates the effect of using the raw reconstruction in \cref{eq:kcrb_operational_raw_fone} instead of the Hermite-consistent form in \cref{eq:kcrb_operational_fone}. We repeat the periodic D3Q27 Taylor--Green calculation at \(N=64\), \(\Re=1600\), \(\Ma=0.2\), and \(C_\nu=0.02\). Both calculations use the product equilibrium and second-order finite differences. They differ only in the construction of the first-order CE distribution.

\Cref{fig:kcrb_raw_vs_hermite_n64} shows maximum relative differences of \(3.06\%\) in \(E_k/u_0^2\) and \(12.82\%\) in \(\mathcal{Z}/(u_0/L_{\rm TGV})^2\) between the two KC-RB variants. Both differences change sign. Hence the trajectories have no uniform dissipation ordering. Replacing the raw form changes both the CE contribution and the SGS carrier used for local relaxation. The comparison identifies an implementation sensitivity, not an accuracy ranking.

KC-RR is included as a contextual Hermite-consistent reference. Its maximum differences from Hermite KC-RB are \(0.061\%\) in \(E_k/u_0^2\) and \(0.314\%\) in \(\mathcal{Z}/(u_0/L_{\rm TGV})^2\) over the displayed interval. These deviations are much smaller than the raw-versus-Hermite KC-RB differences. KC-RR also uses a recursive projection and a different SGS carrier. Therefore, the KC-RR comparison does not isolate the first-order reconstruction. However, the much smaller deviations indicate that the raw reconstruction dominates the legacy KC-RB/KC-RR separation in the matched \(N=64\) TGV test.

\end{document}